\documentclass[aps,prd,twocolumn,floatfix,superscriptaddress,groupedaddress,nofootinbib,preprintnumbers]{revtex4-2}
\usepackage[utf8]{inputenc}

\pdfoutput=1
\usepackage{graphicx}
\usepackage{latexsym} 
\usepackage{amsfonts}
\usepackage{amssymb}
\usepackage{amsmath}
\usepackage{xcolor}
\usepackage{cancel}
\usepackage{multirow}
\usepackage[hidelinks]{hyperref}

\definecolor{SeaGreen}{rgb}{0.13,0.55,0.33}

\newcommand{\flowone}{\textsc{Flow-1}}
\newcommand{\flowtwo}{\textsc{Flow-2}}
\newcommand{\flowthree}{\textsc{Flow-3}}
\newcommand\geant{\textsc{Geant}4}
\newcommand{\icalo}{i\textsc{CaloFlow}}
\newcommand{\cf}{\textsc{CaloFlow}}
\newcommand{\LtoLF}{\textsc{L2LFlows}}
\newcommand\atlfast{\textsc{AtlFast3}}
\begin{document}
\title{Inductive Simulation of Calorimeter Showers with Normalizing Flows}

\author{Matthew R.~Buckley}
\affiliation{NHETC, Dept. of Physics and Astronomy, Rutgers University, Piscataway, NJ 08854, USA}

\author{Claudius Krause}
\affiliation{NHETC, Dept. of Physics and Astronomy, Rutgers University, Piscataway, NJ 08854, USA}
\affiliation{Institut f\"ur Theoretische Physik, Universit\"at Heidelberg, Germany }
\author{Ian Pang}
\affiliation{NHETC, Dept. of Physics and Astronomy, Rutgers University, Piscataway, NJ 08854, USA}

\author{David Shih}
\affiliation{NHETC, Dept. of Physics and Astronomy, Rutgers University, Piscataway, NJ 08854, USA}

\date{\today}

\begin{abstract}

Simulating particle detector response is the single most expensive step in the Large Hadron Collider computational pipeline. Recently it was shown that normalizing flows can accelerate this process while achieving unprecedented levels of accuracy, but scaling this approach up to higher resolutions relevant for future detector upgrades leads to prohibitive memory constraints. To overcome this problem, we introduce  {\it Inductive CaloFlow} (\icalo), a framework for fast detector simulation based on an inductive series of normalizing flows trained on the pattern of energy depositions in pairs of consecutive calorimeter layers. We further use a teacher-student distillation to increase sampling speed without loss of expressivity. As we demonstrate with Datasets 2 and 3 of the {\it CaloChallenge2022}, \icalo\
can realize the potential of normalizing flows in performing fast, high-fidelity simulation on detector geometries that are $\sim 10$ -- $100$ times higher granularity than previously considered.

\end{abstract}
\maketitle

\section{Introduction} \label{sec:intro}

The computational resources required by the physics program of the Large Hadron Collider (LHC) are immense. In addition to the formidable demands set by the acquisition, reconstruction, and analysis of the physics events themselves, the physics goals of the LHC call for detailed and accurate simulations of these events. Such simulation requires even greater computer resources than the data acquisition and analysis itself (see \cite{Bruning:2015dfu, Calafiura:2729668, Software:2815292, Collaboration:2802918, HEPSoftwareFoundation:2020daq} for recent reviews of the current status and future plans for LHC computing). 

The most significant bottleneck (see e.g., Figure 1 of \cite{Calafiura:2729668}) in simulating LHC events is modeling the response of the detector --- and in particular that of the calorimeter --- to incident particles using \geant\ \cite{Agostinelli:2002hh,1610988,ALLISON2016186}. As the high-luminosity runs of the LHC progress, the demand for efficient simulation of collider events will only grow more pressing. 

In recent years, a wide variety of deep generative models --- including Generative Adversarial Networks (GANs), Variational AutoEncoder (VAE)-based models, normalizing flows, and diffusion models  \cite{Paganini:2017hrr,Paganini_2018,deOliveira:2017rwa,Erdmann:2018kuh,Erdmann:2018jxd,Belayneh:2019vyx,Buhmann:2020pmy,ATL-SOFT-PUB-2020-006,Krause:2021ilc,Krause:2021wez,Buhmann:2021lxj,buhmann2021fast,Buhmann:2021caf,ATLAS:2021pzo,Mikuni:2022xry,ATLAS:2022jhk,Adelmann:2022ozp,Krause:2022jna,Cresswell:2022tof,Diefenbacher:2023vsw,Buhmann:2023bwk,Hashemi:2023ruu}
--- have demonstrated their potential for fast and accurate surrogate modeling of \geant.
The probability distribution of energy depositions within a calorimeter can be learned by neural networks trained on collections of \geant\ events, 
and then new simulated events can be produced from these networks much faster than running \geant\ itself. This approach has gone beyond proof-of-concept, with \atlfast\ \cite{ATLAS:2021pzo} (the current official fast-simulation framework of the ATLAS collaboration) adopting a GAN for a portion of its calorimeter simulation.

To spur new solutions \cite{Mikuni:2022xry,Krause:2022jna,Cresswell:2022tof} to the problem of fast calorimeter simulation, the Fast Calorimeter Simulation Challenge 2022 ({\it CaloChallenge2022})~\cite{calochallenge} presents three datasets~\cite{CaloChallenge_ds1,CaloChallenge_ds2,CaloChallenge_ds3}, each with increasing numbers of detector segments. 
Previous work \cite{Krause:2022jna} showed that the \cf{} method of \cite{Krause:2021ilc,Krause:2021wez}, based on normalizing flows~\cite{2014arXiv1410.8516D,2016arXiv160508803D,rezende2015variational} (see also \cite{kobyzev2020normalizing, papamakarios2021normalizing} for reviews), could be quite successful at fast and accurate emulation of Dataset~1 calorimeter showers. Generalizing the \cf{} method to Datasets~2~\cite{CaloChallenge_ds2} and 3~\cite{CaloChallenge_ds3}, which are a factor of $\sim 10$ and $\sim 100$ larger in dimensionality compared to Dataset~1, is the focus on this work. 

The primary obstacle to scaling the \cf{} approach up to Datasets 2 and 3 is memory consumption. While normalizing flows are a powerful method for density estimation and generative modeling in high dimensional spaces, they can be very memory intensive, since they attempt to  parametrize a bijective transformation between the data space and a latent space of the same dimension. Scaling the same architecture used in \cf{} up to Datasets 2 and 3 would easily outstrip the locally available GPU memory, since the total number of model parameters scales as $\mathcal{O}(d^2)$, where $d$ is the data dimension. Only diffusion models have been applied to these high-dimensional dataset so far~\cite{Mikuni:2022xry}.

Recently, \LtoLF\ \cite{Diefenbacher:2023vsw} was proposed in order to overcome this obstacle. \LtoLF\ is based on the physical intuition that --- since particles propagate through the calorimeter primarily in one direction --- the pattern of energy deposition in one layer should depend in large part on the pattern in the previous layers. Trained on a toy dataset (derived from the one used here~\cite{Buhmann:2020pmy, Buhmann:2021lxj}) for the International Large Detector (ILD)~\cite{ILD-TDR_V4,ILD-IDR} of comparable size to Dataset 2, \LtoLF\ introduced one flow per calorimeter layer, conditioned on up to five previous layers through an embedding network. By not training a single flow to learn the voxel energies of the entire calorimeter, as was the case in~\cite{Krause:2021ilc,Krause:2021wez,Krause:2022jna}, but instead breaking up the model into separate flows (each no larger than a single layer of the calorimeter), \LtoLF\ was able to keep the memory footprint of the model manageable while still scaling it up to a higher granularity calorimeter.

In this paper, we take the approach of \LtoLF\ one step further and replace the separate flows for each calorimeter layer with a {\it single} flow that generates the voxel energies in each layer, conditioned on the previous layer. This allows for the entire calorimeter shower to be learned inductively: training not on entire events over all layers, but rather on pairs of layers. Like in a mathematical induction proof, the initial layer is learned separately. Then, new events are simulated layer by layer, with the results for the $i^{\rm th}$ layer serving as conditional input for the same normalizing flow now generating the $(i+1)^{\rm th}$. 

In more detail, our new framework, which we dub {\it Inductive CaloFlow} (\icalo), uses three normalizing flows to learn and generate calorimeter events. First, \flowone\ learns the pattern of total energy deposition in each layer of the calorimeter, conditioned on the incident energy entering the detector. This is a relatively low-dimension dataset, with the dimension being the number of layers (45 for the examples used in this paper). Next, \flowtwo\ learns the pattern of normalized energy deposition within the first layer of the calorimeter, conditioned on the total energy deposited in the layer. Finally, \flowthree\ learns the pattern of normalized energy deposition in the $i^{\rm th}$ layer, conditioned on both the total energy deposited in the layer and the pattern of energy deposition in the previous $(i-1)^{\rm th}$ layer. This last flow is trained simultaneously over the deposition pattern of every layer beyond the first. 

To achieve both high fidelity and ultra-fast generation speed, we follow the teacher-student method of \cf{}. First we train 
Masked Autoregressive Flow (MAF) \cite{papamakarios2017masked} versions of  \flowone, \flowtwo\ and \flowthree\ on the \geant\ data. MAFs are fast in density estimation but slow (by a factor of ${\mathcal O}(d)$) in generation. Then we fit ``student" versions of \flowtwo\ and \flowthree\ to their MAF teacher counterparts, 
using the technique of Probability Density Distillation (PDD) \cite{pmlr-v80-oord18a}.\footnote{\flowone{} is not paired with an IAF as the output dimension is relatively small, and so generation time for the MAF is comparatively short.} 
The student models are Inverse Autoregressive Flows (IAFs) \cite{kingma2016improved}, which are fast in generation but slow in density estimation.

The result is a new deep learning architecture, capable of learning and quickly generating calorimeter events even for the largest detector layout of {\it CaloChallenge2022}. We quantify the fidelity of the events generated by both the \icalo\ teacher and student flows, using a combination of histograms of physical features, classifier-based metrics  (as in Ref.~\cite{Krause:2021ilc,Krause:2021wez,Krause:2022jna,Diefenbacher:2023vsw}), and other metrics. We also measure and report on the generation speed of \icalo\ for different batch sizes and hardware.

We reserve a detailed comparison with \LtoLF\ for future work: while such a comparison would be very interesting, it is non-trivial to perform. \LtoLF\ was trained on a completely different dataset, comparable in size to Dataset 2, so direct comparisons are not possible without significant additional computational effort.\footnote{\LtoLF\ would have to be retrained on Dataset 2 and also generalized to Dataset 3 which is a factor of $\sim 10$ larger.}
Here we just note that the main differences with \LtoLF\ --- using \flowthree\ instead of separate flows for each calorimeter layer, and conditioning on the previous layer instead of up to five previous layers --- make \icalo\ significantly more memory efficient, but likely at the cost of being
less expressive (and hence a worse fit to the data). The \icalo{} IAF student model also results in a speed advantage over \LtoLF's teacher-only MAF, again probably at the expense of fidelity.

In Section~\ref{sec:data}, we describe the datasets we train our algorithm on and generate new data to compare against. In Section~\ref{sec:icaloflow}, we describe the multiple flows that make up the \icalo\ algorithm in detail, along with our training procedure for both the teacher and student. In Section~\ref{sec:results}, we show the results of our event generation, and use a classifier to quantitatively compare with the original dataset. We conclude in Section~\ref{sec:conclusions}. 

\section{CaloChallenge Data}\label{sec:data}

\begin{figure*}
\includegraphics[width=.5\columnwidth]{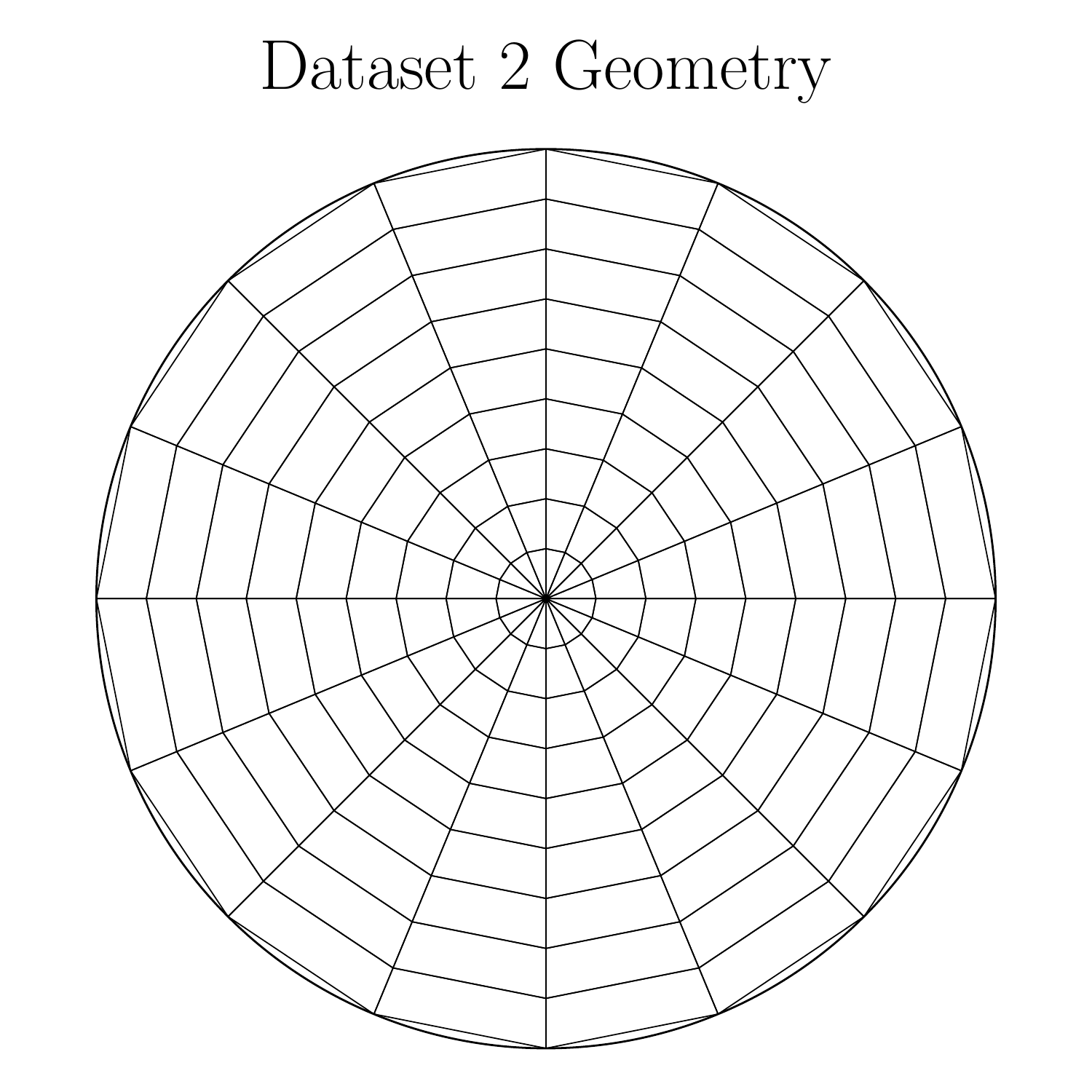}\includegraphics[width=.5\columnwidth]{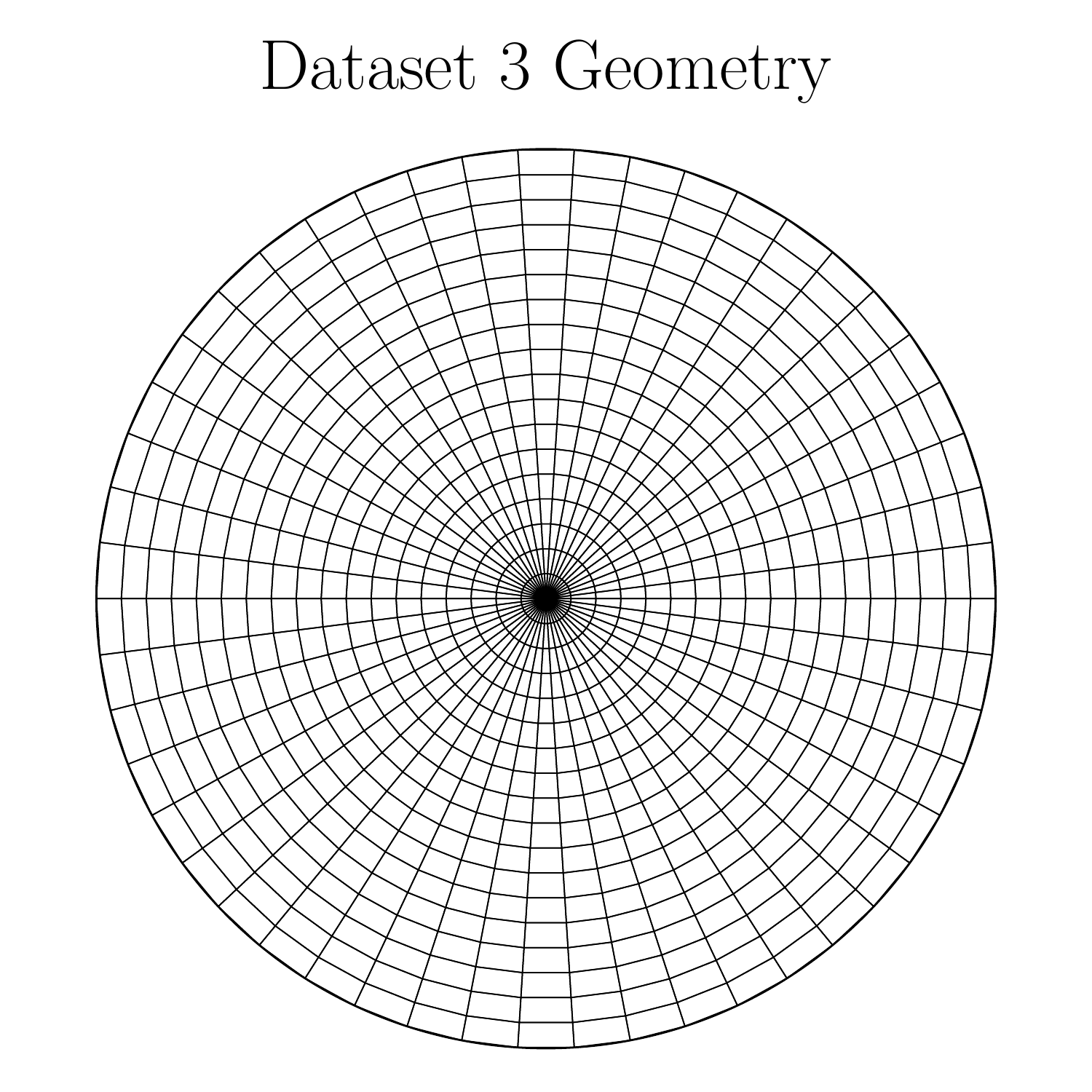}\includegraphics[width=.7\columnwidth]{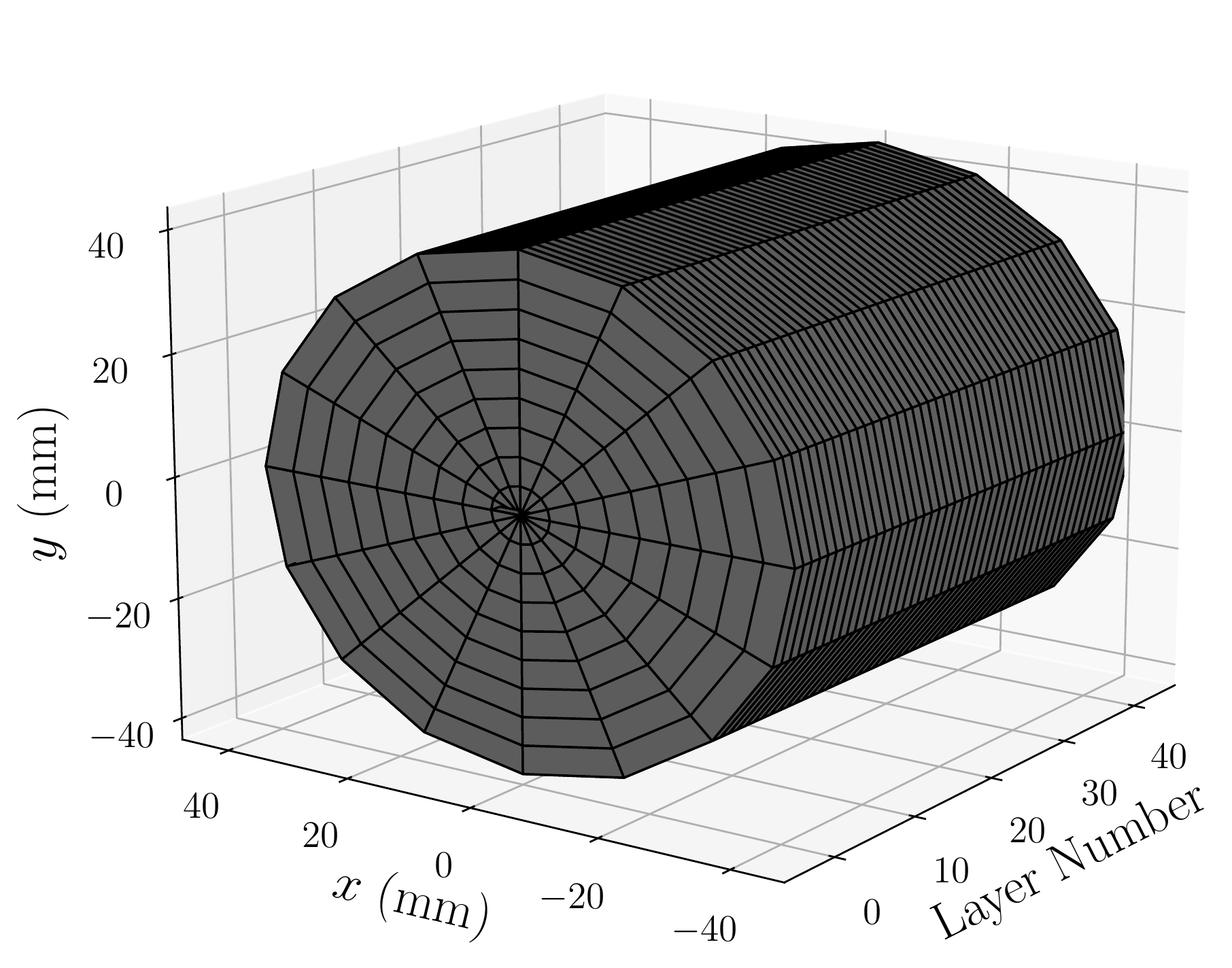}

\caption{Geometry of the detector voxels in each layer for Dataset 2 (left) and Dataset 3 (center) and the three-dimensional geometry of Dataset 2 (right). Dataset 2 has 9 concentric rings, each divided into 16 voxels in $\alpha$, while Dataset 3 has 18 rings, each divided into 50 segments. Both Dataset 2 and 3 contain 45 layers in depth (as shown for Dataset 2).}
\label{fig:geometry}
\end{figure*}

Datasets 2 and 3~\cite{CaloChallenge_ds2,CaloChallenge_ds3} of the {\it CaloChallenge2022}~\cite{calochallenge} consist of 100k \geant-simulated electron showers each, with incident energies sampled uniformly in log-space from 1~GeV to 1~TeV. The simulated detector volume has a cylindrical geometry of radius 80~cm, with 45 layers of  active silicon detector (thickness 0.3~mm), alternating with inactive tungsten absorber layers (thickness 1.4~mm). The length of the voxel along the $z$-axis is 3.4~mm, which corresponds to two physical layers (tungsten-silicon-tungsten-silicon). Taking into account only the absorber value of radiation length ($X_0 =3.504~\text{mm}$) it makes the $z$-cell size equal 0.8 $X_0$.

Each detector layer is segmented in read-out voxel cells in the azimuthal angle $\alpha$ and radial distance $r$ from the center of the cylindrical detector volume. The two datasets only differ in their voxelization. Dataset 2 has each layer divided into 9 concentric rings of voxels in the radial direction. Each ring is then divided into 16 voxels in $\alpha$, for 144 voxels in each of the 45 layers (6480 for the entire detector). Dataset 3 has 18 radial segments and 50 azimuthal, for 900 voxels in the 45 layers (40500 total). Figure~\ref{fig:geometry} shows the geometry of the voxels within a layer for both datasets, as well as the positioning of layers along the calorimeter axis.

Each event record consists of the total energy of the incident electron $E_{\rm inc}$, together with the energy depositions recorded in each voxel $\mathcal{I}_{ia}$, where $i$ is the layer index and $a$ is the voxel index within the layer. The minimum energy deposition in each voxel is 15~keV. The time required to generate a \geant\ shower depends strongly on the shower incident energy. It is approximately ${\cal O}(100\,{\rm s})$ when averaged over the incident energies of these datasets~\cite{AnnaZaborowska}, but the time required per shower is much longer for the showers with higher incident energies. Since the underlying detector geometry of Datasets 2 and 3 is the same and only the voxelization is different, the generation times with \geant\ are the same for both datasets.

\section{\icalo} \label{sec:icaloflow}

\subsection{Overview}
\label{sec:overview}

\begin{figure*}
\includegraphics[width=1.9\columnwidth]{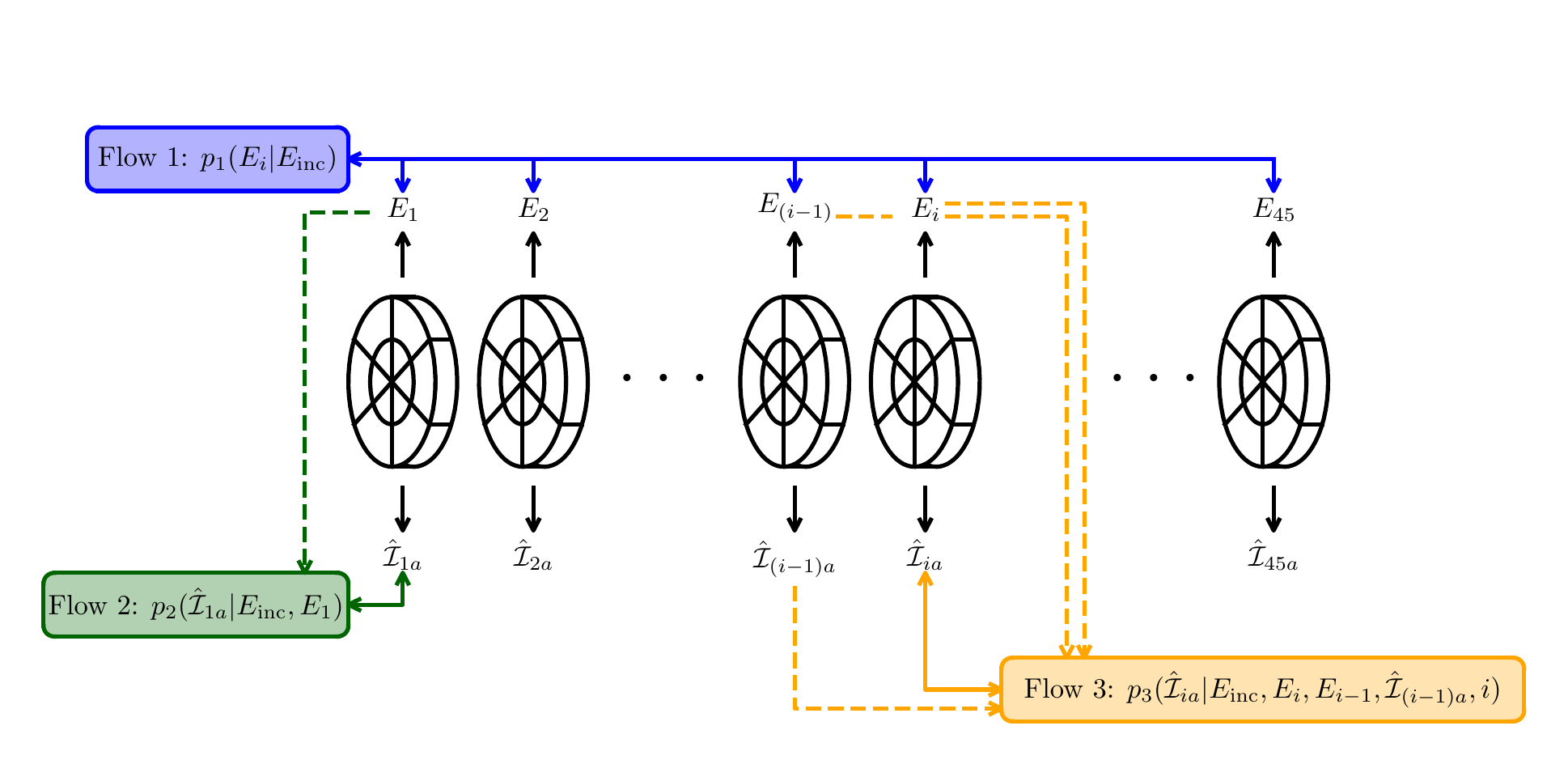}

\caption{Schematic of the three \icalo{} flows. Solid lines are bidirectional --- the direction into each flow denotes the density estimation step and the direction out of the flow denotes the sample generation step. Note that there are postprocessing steps (see main text) after each generation step, which are omitted in the schematic. Dashed lines indicate the conditional input to the respective flows. \flowthree{} is used iteratively on subsequent layers.}
\label{fig:icaloflow}
\end{figure*}

The purpose of \icalo\ (and of the {\it CaloChallenge2022} generative modeling problem more generally) is to learn and sample from the conditional probability density $p({\mathcal I}_{ia}|E_{\rm inc})$
that describes the \geant\ reference data. In Figure~\ref{fig:icaloflow} we show a schematic of \icalo, which consists of three flows:

\begin{itemize}

\item \flowone{} learns the joint probability distribution of total energy deposited in each layer $E_i$, conditioned on the incident energy of the event $E_{\rm inc}$: $p_1(E_i|E_{\rm inc})$. It is necessary to learn this probability distribution as $E_i$ is a conditional input for \flowtwo{} and \flowthree{} in the generation step. We found that removing $E_i$ as a conditional input in \flowtwo{} and \flowthree{} decreased the quality of our generated samples.

\item \flowtwo{} learns the probability distribution of the unit-normalized voxel energies in the first layer of the calorimeter, $\hat{\mathcal{I}}_{1a} \equiv \mathcal{I}_{1a}/\sum_b \mathcal{I}_{1b}$, conditioned on $E_{\rm inc}$ and the energy deposited in the first layer, $E_1$: $p_2\left(\hat{\mathcal{I}}_{1a}|E_{\rm inc},E_1\right)$. Here $a$ is the voxel index.  

\item Finally, \flowthree{} learns the probability distribution of unit-normalized voxel energies in every layer after the first, $\hat{\mathcal{I}}_{ia} \equiv \mathcal{I}_{ia}/\sum_b \mathcal{I}_{ib}$ for $i\in[2,45]$, where the $i^{\rm th}$ layer is conditioned on the energy deposited in the layers $i$ and $i-1$ ($E_i$ and $E_{i-1}$), $E_{\rm inc}$, the unit-normalized voxel energies in the $(i-1)^{\rm th}$ layer $\hat{\mathcal{I}}_{(i-1)a}$, and the one-hot\footnote{One-hot encoding is used for layer numbers instead of ordinal encoding using the layer number directly, because other than the location in the detector, there is no information in the layer number, i.e., layer 30 is not 15 times more important than layer 2.} encoded layer number $i$: $p_3\left(\hat{\mathcal{I}}_{ia}|E_{\rm inc},E_i,E_{i-1}, \hat{\mathcal{I}}_{(i-1)a}, i\right)$. 

\end{itemize}

The conditional inputs, dimension of conditionals, and the outputs for each of the three flows are summarized in Table \ref{tab:architecture}. When generating new showers with \icalo, first the nominal energies per layer $\tilde{E}_i$ are generated for a given incident $E_{\rm inc}$ with \flowone{}\footnote{This does not directly correspond to the energy deposited in the layer, $E_i$, see the discussion on postprocessing below.}. Then the unit-normalized voxel energies $\hat{{\mathcal I}}_{1a}$ of layer 1 is generated using \flowtwo{}, conditioned on $\tilde{E}_1$ and $E_{\rm inc}$. 
Finally, the normalized voxel energies $\hat{{\mathcal I}}_{ia}$ of layers $i=2,\dots,45$ are generated sequentially (inductively), using \flowthree{} with the conditional inputs of the previous layer's $\hat{\cal I}$ distribution provided by \flowtwo{} (for generating Layer 2) or \flowthree{} (for all subsequent layers). By generating all layers beyond the first from a single normalizing flow rather than each layer separately as in \LtoLF~\cite{Diefenbacher:2023vsw}, our model is far more efficient in memory usage. The potential downsides are that training cannot be trivially parallelized and our model may be less expressive.

One important subtlety with the construction of \icalo{} is that \flowtwo{}\ and \flowthree{} cannot learn the unit-normalization constraint of the $\hat{\cal I}_{ia}$ training data perfectly. In practice, the sum of the generated showers will be distributed around unity, and there will always be some mismatch between the nominal layer energy generated by \flowone{} and the actual layer energy produced by the combination of all three flows. We will address this mismatch here\footnote{This problem is also present in the previous iterations of \cf{}~\cite{Krause:2021ilc,Krause:2021wez,Krause:2022jna} and in \LtoLF{}~\cite{Diefenbacher:2023vsw}, but differences in the reference datasets led to different treatments of the problem. For more details, see Appendix~\ref{sec:preprocessing}.}  by taking the latter as the actual layer energy: after multiplying $\hat{\cal I}_{ia}$ with the output of \flowone{} and applying a minimum energy threshold of $E_{\rm min}= 15$~keV to obtain ${\cal I}_{ia}$, we take the sum over voxels $a$ as the total energy in the layer:
\begin{equation}
    E_i \equiv \sum_a \tilde{E}_i \hat{\cal I}_{ia} \Theta(\tilde{E}_i \hat{\cal I}_{ia}-E_{\rm min}). 
\end{equation}
We will think of the output of \flowone\ as just an intermediate step needed for the subsequent conditioning, and refer to  output of \flowone{} when generating new events as the {\it proxy} $\tilde{E}_i$. We note that referring to the output of a flow as a proxy for the distributions it was trained on is non-standard, but necessary in this case due to the differences in normalization and the multiple ways of defining the energy deposited in a layer. In a sense, also the the normalized shower shapes $\hat{\cal I}_{ia}$ that \flowtwo{} and \flowthree{} learn can be thought of as proxies, since the resulting samples need to be unnormalized and thresholded as a postprocessing step,
\begin{equation}
    {\cal I}_{ia} = \tilde{E}_i \hat{\cal I}_{ia} \Theta(\tilde{E}_i \hat{\cal I}_{ia}-E_{\rm min}).
\end{equation}

\begin{table}
    \centering
    \begin{tabular}{|c|c|c|c|} \hline
         & Conditionals &\begin{tabular}{@{}c@{}}Dim of \\ conditional\end{tabular} & Output  \\ \hline\hline
\flowone & $E_{\rm inc}$ & 1  & $\tilde{E}_i$ \\ \hline 
\flowtwo & $E_{\rm inc}, E_1$  & 2& $\hat{\mathcal{I}}_{1a}$ \\ \hline
\flowthree & $E_{\rm inc}, E_i, E_{i-1},\hat{\mathcal{I}}_{(i-1)a},i $  & 191 or 947&$\hat{\mathcal{I}}_{ia}$ \\ \hline
    \end{tabular}
    \caption{The conditional inputs for each flow, and the features whose probability distributions are the output of each flow (for both teachers and their paired student). $E_{\rm inc}$ is the incident energy of the particle, $E_i$ $(i=1,\ldots,45)$ is the total energy deposited in layer $i$, $\tilde{E}_i$ is a proxy for $E_i$ (see text), and $\hat{{\cal I}}_{ia}$ is the pattern of normalized energy deposition in the voxels $a$ in layer $i$. For \flowthree{}, Dataset 2 has smaller conditional dimension (191) compared to Dataset 3 (947).}
    \label{tab:architecture}
\end{table}

\subsection{Architectures} 

\label{sec:architecture}

\subsubsection{Teacher MAFs} \label{sec:teachers}

The \icalo\ teacher flows (including \flowone) are MAFs \cite{papamakarios2017masked}, which learn a bijective transformation $f$ between a latent space $z$, with a simple $N$-dimensional probability distribution,\footnote{In this work, the latent space follows an $N$-dimensional Gaussian distribution.} and the target space $x$. For consistency with the ``forward'' function of the code, we define $z = f(x)$.

Following \cite{Krause:2021ilc,Krause:2021wez,Krause:2022jna}, all three teacher flows use compositions of rational quadratic spline (RQS) transformations~\cite{durkan2019neural} as their transformation function $f$. The neural networks defining the parameters $\vec{\kappa}$ of the RQS consist of MADE blocks \cite{germain2015made}. The MADE blocks allow for tractable training given the large dimensionality of the training data for \flowtwo{} and \flowthree, at the cost of longer evaluation time for generating new events. For details of the architectures used for the teacher flows, see  Table~\ref{tab:flow_architecture}.

\subsubsection{Student IAFs}\label{sec:students}

Though the three MAFs are capable of generating complete events simulating the \geant{} output, the sampling time is quite slow.  
While this is not a particular concern for \flowone{}, the sampling time is an issue especially for \flowthree{}, which has a large output dimension and must be sampled 44 times sequentially (i.e., cannot be parallelized) in order to generate a single event.

The solution proposed in~\cite{Krause:2021wez}, which we carry over here to Datasets 2 and 3 of {\it CaloChallenge2022}, is to pair every slow-sampling MAF (i.e., \flowtwo\ and \flowthree) with a fast-sampling Inverse Autoregressive Flow (IAF). Since training the IAF with a negative log-likelihood of the probability of the data is prohibitively slow, the IAF is trained by fitting it to a pre-trained MAF using the Probability Density Distillation (PDD) method  developed in \cite{Krause:2021wez}. The goal of this training is  for the IAF to learn $f_{\text{IAF}} = f_{\text{MAF}}$, or equivalently --- since only the fast passes through the flows can be used meaningfully for optimization --- to have $f_{\text{MAF}}$ and $f^{-1}_{\text{IAF}}$ be each other's inverse. For more details of the training and loss terms, see Sec.~\ref{sec:students_training} and \cite{Krause:2021wez}.
 
In order to carry out the PDD method, we require $f_{\text{IAF}} = f_{\text{MAF}}$ at every individual step of the normalizing flow. Hence, the IAF must be composed of the same number of MADE-RQS blocks as its corresponding pre-trained MAF. However, the hidden layer sizes in the blocks can be different. We use a larger hidden layer, with 384 nodes, for \flowthree{} student which led to better performance in Dataset 2.
For Dataset 3, \flowthree{} student has the same hidden layer size (256 nodes) as the teacher due to memory constraints. For a detailed listing of the architecture hyperparameters for the student IAFs, see
Table~\ref{tab:flow_architecture}.

\begin{table*}
\begin{center}
\begin{tabular}{|c|c|c|c|c|c|c|c|}
\hline
\multicolumn{2}{|c|}{\multirow{2}{*}{}}&dim of & number of & \multicolumn{3}{c|}{layer sizes} & number of \\[-0.2ex]
\multicolumn{2}{|c|}{} &base distribution & MADE blocks & input & hidden & output & RQS bins \\
\hline
\hline
\multirow{5}{*}{DS2}&\flowone{} & 45 & 8 & 256  & $1\times 256$ & 1035 & 8\\
\cline{2-8}&\flowtwo{} teacher & 144 &8 & 256  & $2\times 256$ & 3312 & 8\\
\cline{2-8}&\flowtwo{} student & 144 &8 & 256  & $2\times 256$ & 3312 & 8\\
\cline{2-8}&\flowthree{} teacher & 144 &8 & 256  & $2\times 256$ & 3312 & 8\\
\cline{2-8}&\flowthree{} student & 144 &8 & 384  & $2\times 384$ & 3312 & 8\\
\hline
\multirow{5}{*}{DS3} &\flowone{}& 45 & 8 &  256   &  $1\times 256$ & 1035 & 8\\
\cline{2-8}&\flowtwo{} teacher & 900 &8 & 256  & $2\times 256$ & 20700 & 8\\
\cline{2-8}&\flowtwo{} student & 900 &8 & 256  & $2\times 256$ & 20700 & 8\\
\cline{2-8}&\flowthree{} teacher & 900 &8 & 256  & $1\times 256$ & 20700 & 8\\
\cline{2-8}&\flowthree{} student & 900 &8 & 256  & $1\times 256$ & 20700 & 8\\
\hline
\end{tabular}
\caption{Summary of architecture of the various MAF teacher and IAF student models used in \icalo{}. For the hidden layer sizes, the first number is the number of hidden layers in each MADE block and the second number is the number of nodes in each hidden layer (e.g., $2\times 256$ refers to 2 hidden layers per MADE block with 256 nodes per hidden layer).}
\label{tab:flow_architecture}
\end{center}
\end{table*}

\subsection{Training} \label{sec:training}

Prior to training the teachers and students, we must standardize and preprocess the datasets. New events generated from the trained flows will be in the standardized space, and thus the transformations are inverted to produce events in the physical units. We detail this process for all three flows in Appendix \ref{sec:preprocessing}. The same preprocessing was used when training both the teachers and students.

\subsubsection{Teachers}

The teacher MAFs are trained using the mean negative log-likelihood of the data evaluated on the output of the MAF as the loss function,
\begin{align}
    \label{eq:teacher.loss}
    L_{\text{teacher, }1} &= -\langle \log p_1(E_i|E_{\rm inc}) \rangle, \\
    L_{\text{teacher, }2} &= -\langle \log p_2\left(\hat{\mathcal{I}}_{1a}|E_{\rm inc},E_1\right) \rangle,\\
    L_{\text{teacher, }3} &= -\langle \log p_3\left(\hat{\mathcal{I}}_{ia}|E_{\rm inc},E_i,E_{i-1}, \hat{\mathcal{I}}_{(i-1)a}, i\right) \rangle.
\end{align} 
All teacher MAFs in this work are trained with independent \textsc{Adam} optimizers~\cite{kingma2014adam}.

\begin{figure}
     \centering
     \includegraphics[width=0.4\textwidth]{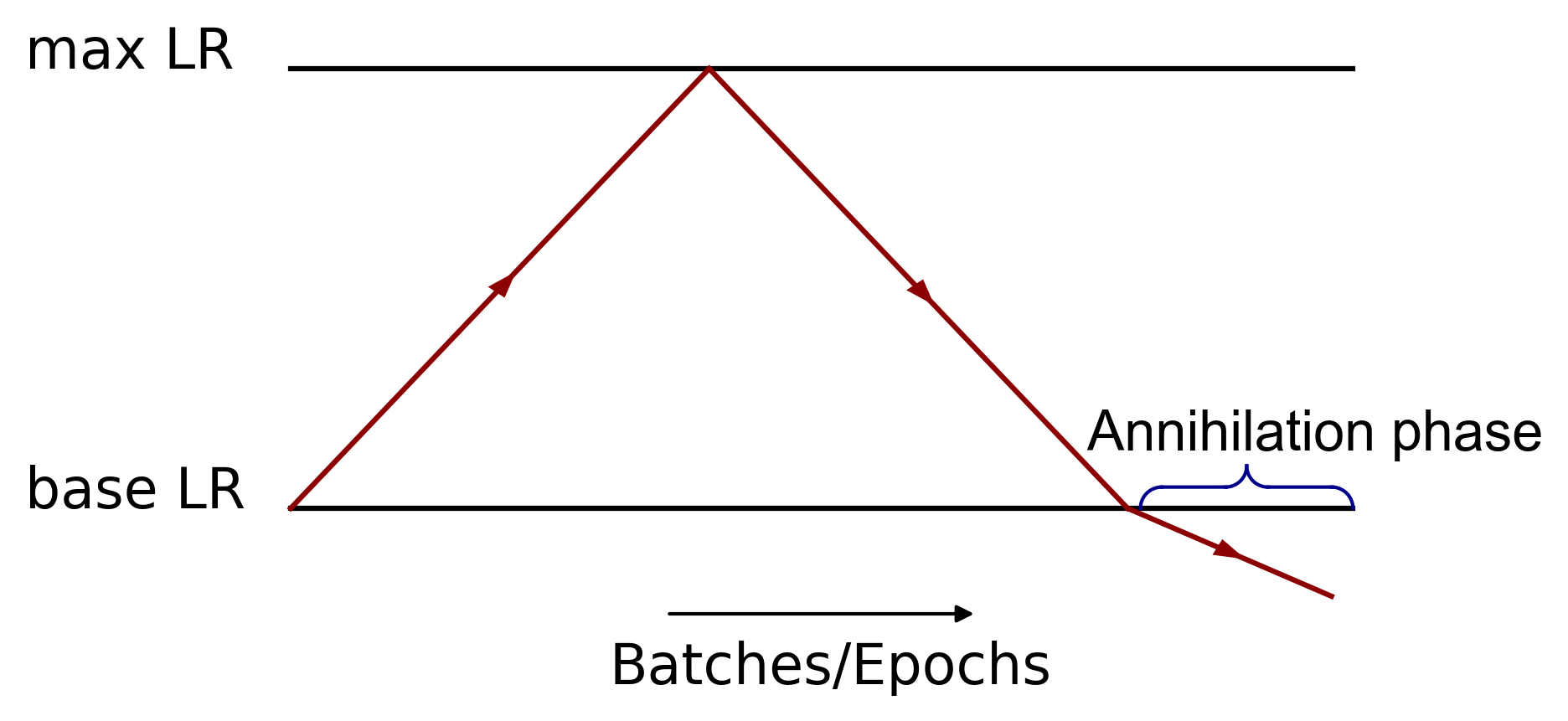}
     \caption{Illustration of OneCycle LR schedule with annihilation phase \cite{Krause:2022jna}.}
     \label{fig:onecycle}
 \end{figure}

Given the different sizes of Datasets 2 and 3, we use a slightly different training strategy for the two datasets. We use 70k samples of Dataset 2 for training the flows and the remaining 30k samples for model selection. The OneCycle learning rate (LR) schedule \cite{smith2019super} was implemented for all three flows. With this LR schedule (see
Figure~\ref{fig:onecycle}), the LR begins at a chosen base LR and is updated after each batch such that it increases up to a maximum
LR. After this, the LR is made to decrease from the maximum
LR to the base LR. The schedule finishes with an annihilation phase where the base LR is further decreased up to a factor of 10. As in \cite{Krause:2022jna}, we find that using the OneCycle LR schedule for Dataset 2 enabled us to obtain a lower training loss within a shorter number of epochs. We show a summary of the training hyperparameters in Table \ref{tab:learning_rates}. The reason for the smaller number of epochs in \flowthree{} compared to the other ones is that the dataset is now 44 times larger (consisting of data from layer $i$ and $i-1$ for $i\in [2, 45]$).

Training \flowone{} and \flowtwo{} for Dataset 3 uses a similar strategy as used for Dataset 2. However, a different learning rate schedule was used when training the flows for Dataset 3. Note that the {\it CaloChallenge2022} provides Dataset 3 as two files of 50k events each (Files 1 and 2). For the first stages of training, we combine the showers of these two files into one dataset with 100k showers --- 70k of these are used for training and the remaining 30k are used for model selection. 

We train \flowthree{} for Dataset 3 in two stages. First, we train it for 20 epochs using 40k samples from File 1 for training and 10k for model selection. Then, using the same optimizer, we train \flowthree{} for another 20 epochs using 40k samples of File 2. We again use the remaining 10k samples for model selection. Simultaneous handling of all 100k showers was not possible due to prohibitive memory requirements. A multistep LR schedule was used when training the Dataset 3 teacher flows, where we halve a chosen initial LR after epochs 400 and 500 (see Table \ref{tab:learning_rates} for summary of training hyperparameters). We found that training the flows with OneCycle LR schedule resulted in a slightly poorer performance for Dataset 3. 

For \flowone{} and \flowtwo{} of both datasets, the epoch with the lowest test loss is selected for subsequent sample generation. For \flowthree{} of both datasets, due to the large training data, the test loss is evaluated after every 250 batches and also at the end of each epoch. The model checkpoint with the lowest test loss is selected for subsequent sample generation.

\subsubsection{Students}
\label{sec:students_training}

The main idea behind teacher-student training of the IAF via PDD is to enforce that they are each other's inverses using only their fast passes:
\begin{equation}
\label{eq:xpass}
 x \equiv f^{-1}_{\text{IAF}}(f_{\text{MAF}}(x)),
\end{equation}
and
\begin{equation}
\label{eq:zpass}
    z \equiv f_{\text{MAF}}(f^{-1}_{\text{IAF}}(z)).
\end{equation}
We refer to these conditions as $x$-pass and $z$-pass, respectively. For each of the passes, we can construct a set of mean-squared error (MSE) losses that force the IAF to converge to the MAF:
\begin{align}
\label{eq:x-loss}
    L_x &= \text{MSE}\left(x, f^{-1}_{\text{IAF}}(f_{\text{MAF}}(x))\right) \\ \label{eq:z-loss}
    L_z &= \text{MSE}\left( z, f_{\text{MAF}}(f^{-1}_{\text{IAF}}(z))\right).
\end{align}

In addition, \cite{Krause:2021wez} proposed two more MSE loss terms (which we will refer to as $L_{x-{\rm MADE}}$ and $L_{z-{\rm MADE}}$ here) that enforce the agreement between MAF and IAF at the level of the outputs and parameters of each individual MADE block (see \cite{Krause:2021wez} for details). These were found to improve the performance of the teacher-student training and we also include them here.

When training the \flowtwo{} students, we were able to achieve good agreement between the teachers and students by using the same loss function as in \cite{Krause:2021wez}:
\begin{equation}
    L = \left(L_x + L_{x-{\rm MADE}} \right) +\left(L_z + L_{z-{\rm MADE}}\right).
\end{equation}

However, for \flowthree{} students, using the same loss function resulted in a significant disagreement between the teachers and students for some distributions. Instead, we found that training with $x$-loss only,
\begin{equation}
\tilde L = L_x +L_{x-{\rm MADE}}
\end{equation}
 resulted in better agreement. We note that a similar behavior was found in \cite{pmlr-v115-huang20c} where the loss constructed only using the $x$-pass was more robust to the large dimensionality of the training data.

When training each student using PDD, the ratio of the data used for training and model selection was chosen to match that of the corresponding teacher. Like the teacher MAFs, all the student IAFs are trained with independent \textsc{Adam} optimizers~\cite{kingma2014adam}. The OneCycle LR schedule was implemented for all the student IAFs and the details of the training hyperparameters are shown in Table~\ref{tab:learning_rates}. Note that a smaller batch size was used when training Dataset 3 student flows due to memory constraints.

For \flowone{} and \flowtwo{} of both datasets, the epoch with the lowest mean Kullback–Leibler (KL) divergence\footnote{The KL divergence between the student $s(x)$ and teacher $t(x)$ distributions is defined as \\ $KL(s,t) = \int~dx~s(x)\log{\frac{s(x)}{t(x)}} \sim \sum_{x\sim s}\log{\frac{s(x)}{t(x)}}$. } is selected for subsequent sample generation. For \flowthree{} of both datasets, due to the large training data, the intermediate mean KL divergence of each epoch is evaluated after every 250 batches, while the final mean KL divergence is evaluated at the end of each epoch. The model checkpoint with the lowest evaluated mean KL divergence is selected for subsequent sample generation.

\begin{table}
\begin{center}
\resizebox{\columnwidth}{!}{%
\begin{tabular}{|c|c|c|c|c|c|}
\hline
\multicolumn{2}{|c|}{Trained with multistep LR} & \multicolumn{2}{|c|}{Initial LR} & Epochs & Batch size\\
\hline
\hline
\multirow{3}{*}{DS3 teacher}&\flowone{} & \multicolumn{2}{|c|}{$1\times10^{-4}$} & 750 & 1000\\
\cline{2-6}&\flowtwo{} & \multicolumn{2}{|c|}{$1\times10^{-4}$} & 750 & 1000\\
\cline{2-6} &  \flowthree{} & \multicolumn{2}{|c|}{$1\times10^{-4}$}& 20+20 & 500\\ 
\hline
\hline
\multicolumn{2}{|c|}{Trained with OneCycle LR} & base LR& max LR& Epochs & Batch size\\
\hline
\hline
\multirow{3}{*}{DS2 teacher}&\flowone{} & $1\times10^{-5}$  & $1\times10^{-4}$ & 500 & 1000\\
\cline{2-6}&\flowtwo{} & $2\times10^{-5}$  & $1\times10^{-3}$ & 200 & 1000\\
\cline{2-6} &  \flowthree{} & $2\times10^{-5}$ & $1\times10^{-3}$ & 60& 1000 \\ 
\hline
\multirow{2}{*}{DS2 student}&\flowtwo{} & $2\times10^{-5}$  & $1\times10^{-3}$ & 400 & 1000 \\
\cline{2-6} &  \flowthree{} & $2\times10^{-5}$ & $5\times10^{-4}$ & 100 & 1000 \\ 
\hline
\multirow{2}{*}{DS3 student}&\flowtwo{} &  $2\times10^{-5}$  & $1\times10^{-3}$ & 400 & 100 \\
\cline{2-6} &  \flowthree{} &  $4\times 10^{-6}$ &  $1\times10^{-4}$ & 15+15 & 100 \\ 
\hline
\end{tabular}
}
\caption{Hyperparameters used when training Dataset 2 and 3 \icalo{} models. For Dataset 3 \flowthree{}, the number of epochs is written as $N+N$ to indicate that we trained the flow for $N$ epochs on samples of File 1, followed by another $N$ epochs on samples of File 2.}
\label{tab:learning_rates}
\end{center}
\end{table}

\section{Results} \label{sec:results}

After training teacher and student flows on Datasets 2 and 3, respectively, we generate 100k calorimeter showers from each model for each dataset.
Incident energies are uniformly  sampled from log-space between 1~GeV and 1~TeV --- the same range and distribution of energies as the training data. 

\subsection{Shower images}

\begin{figure*}[ht!]
\includegraphics[width=0.88\columnwidth]{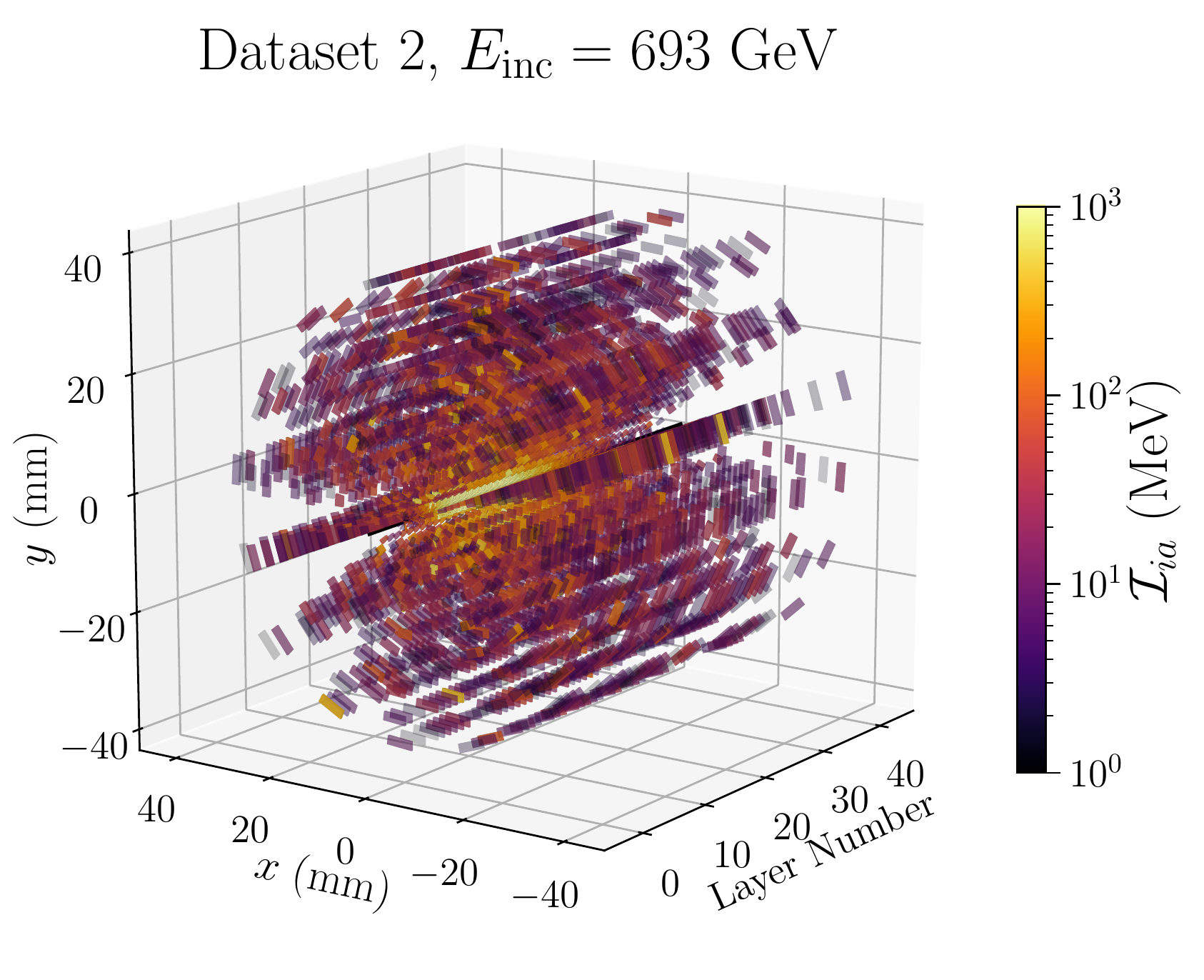}\includegraphics[width=0.88\columnwidth]{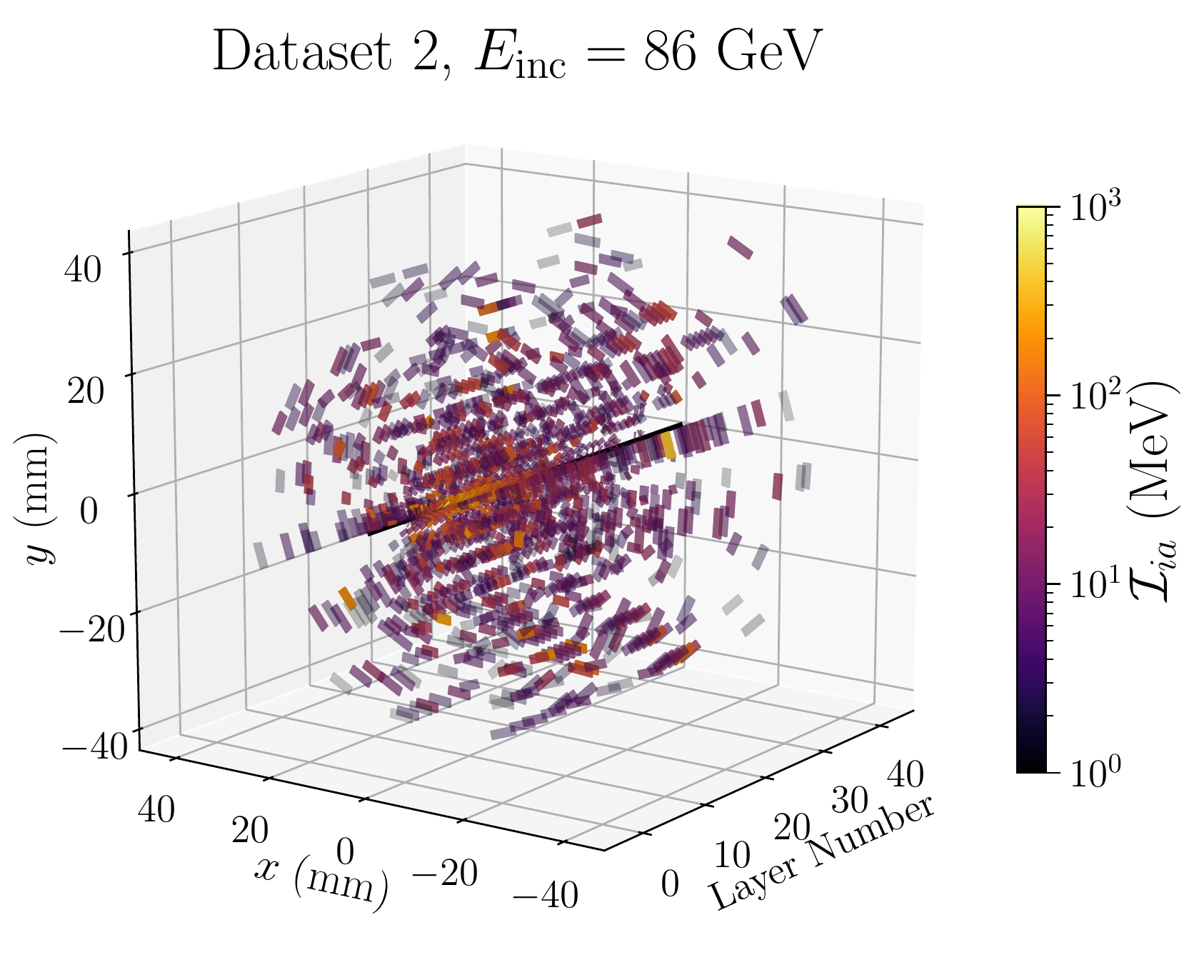}
\includegraphics[width=0.88\columnwidth]{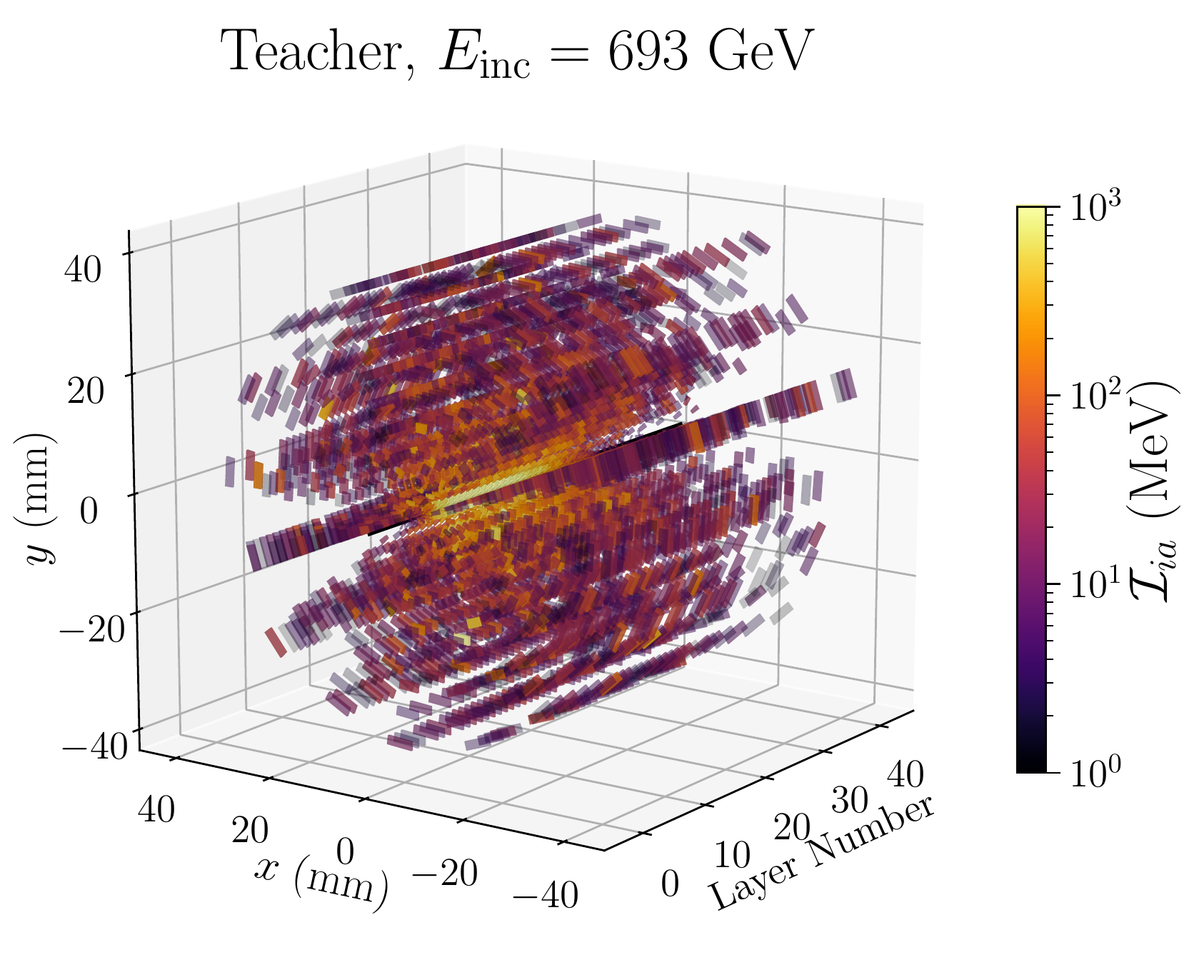}\includegraphics[width=0.88\columnwidth]{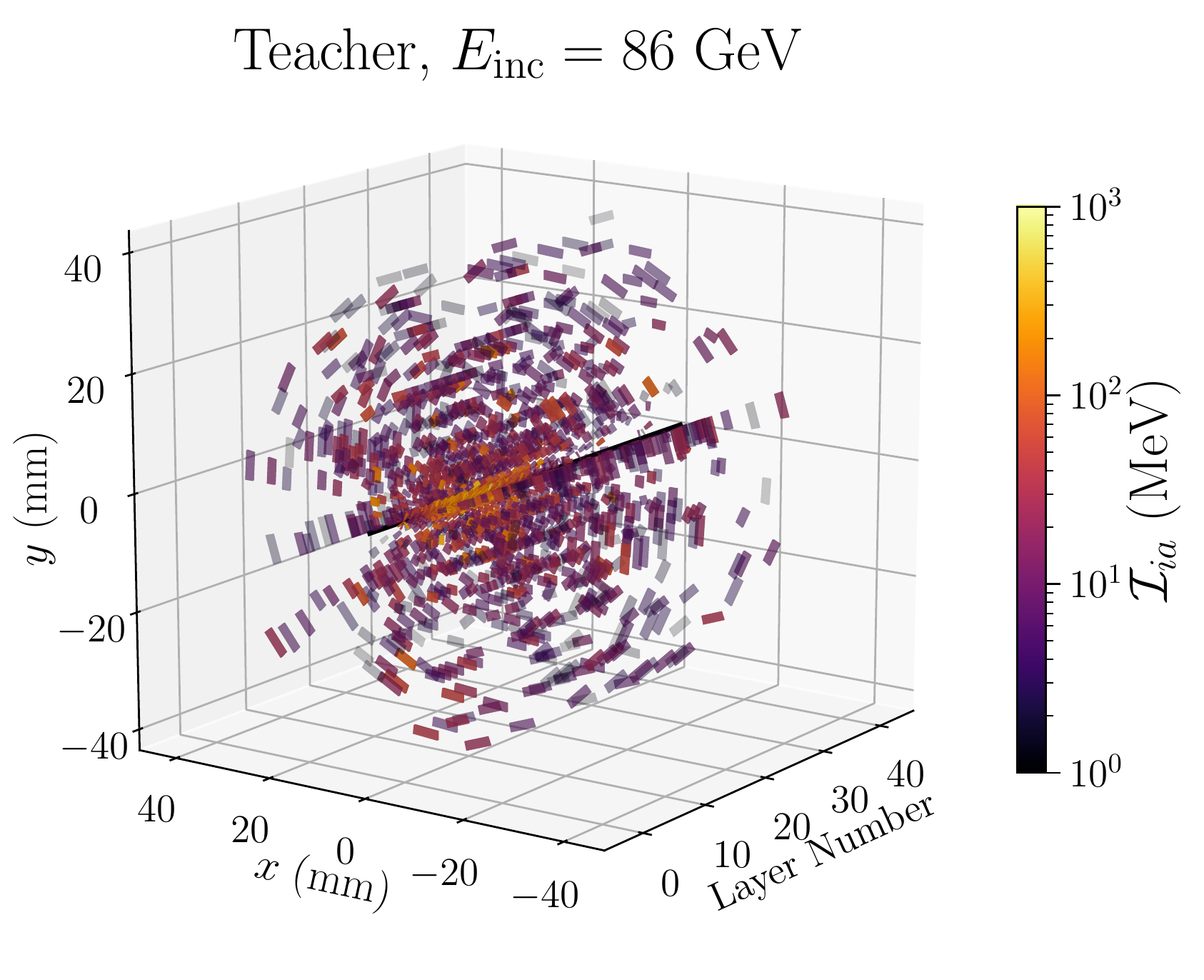}
\includegraphics[width=0.88\columnwidth]{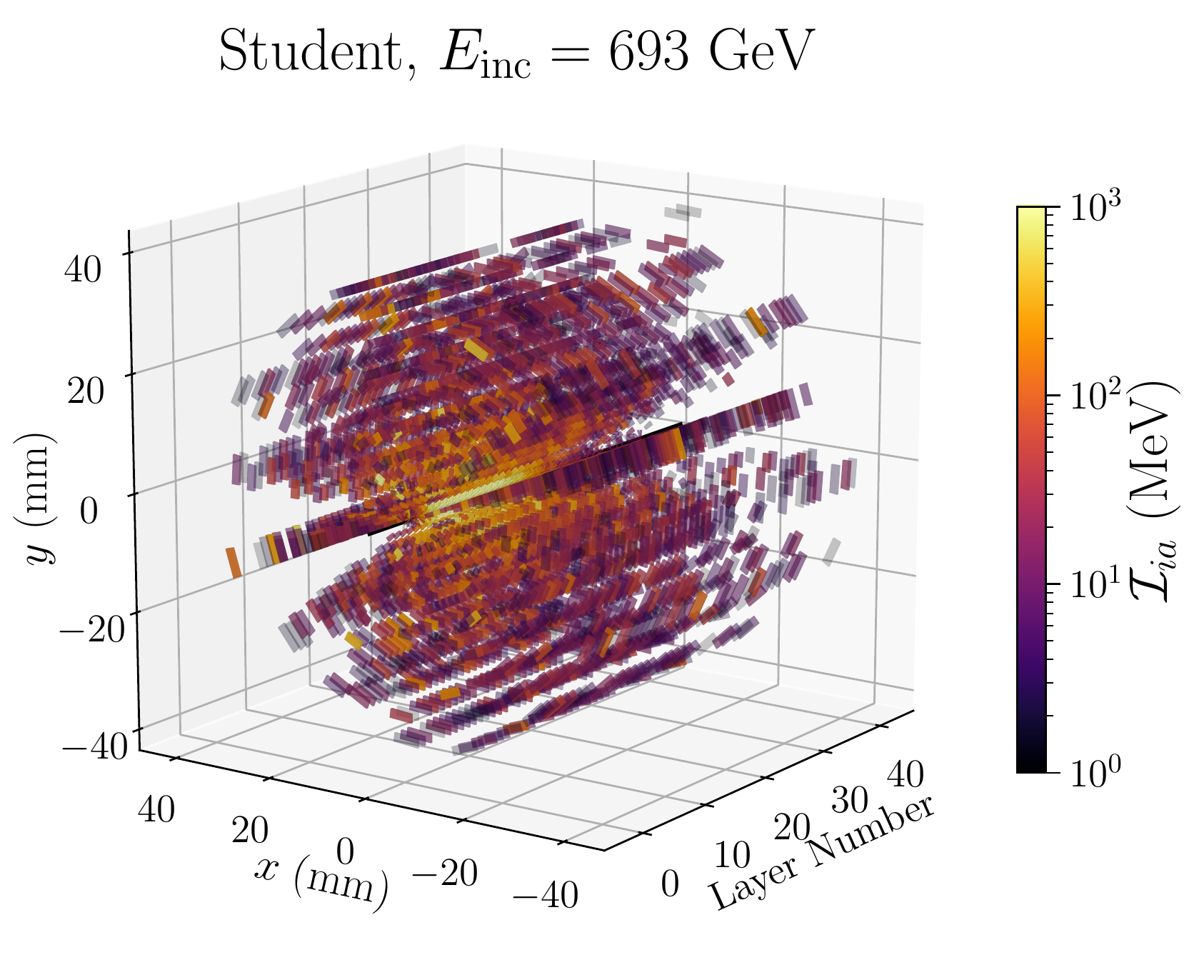}\includegraphics[width=0.88\columnwidth]{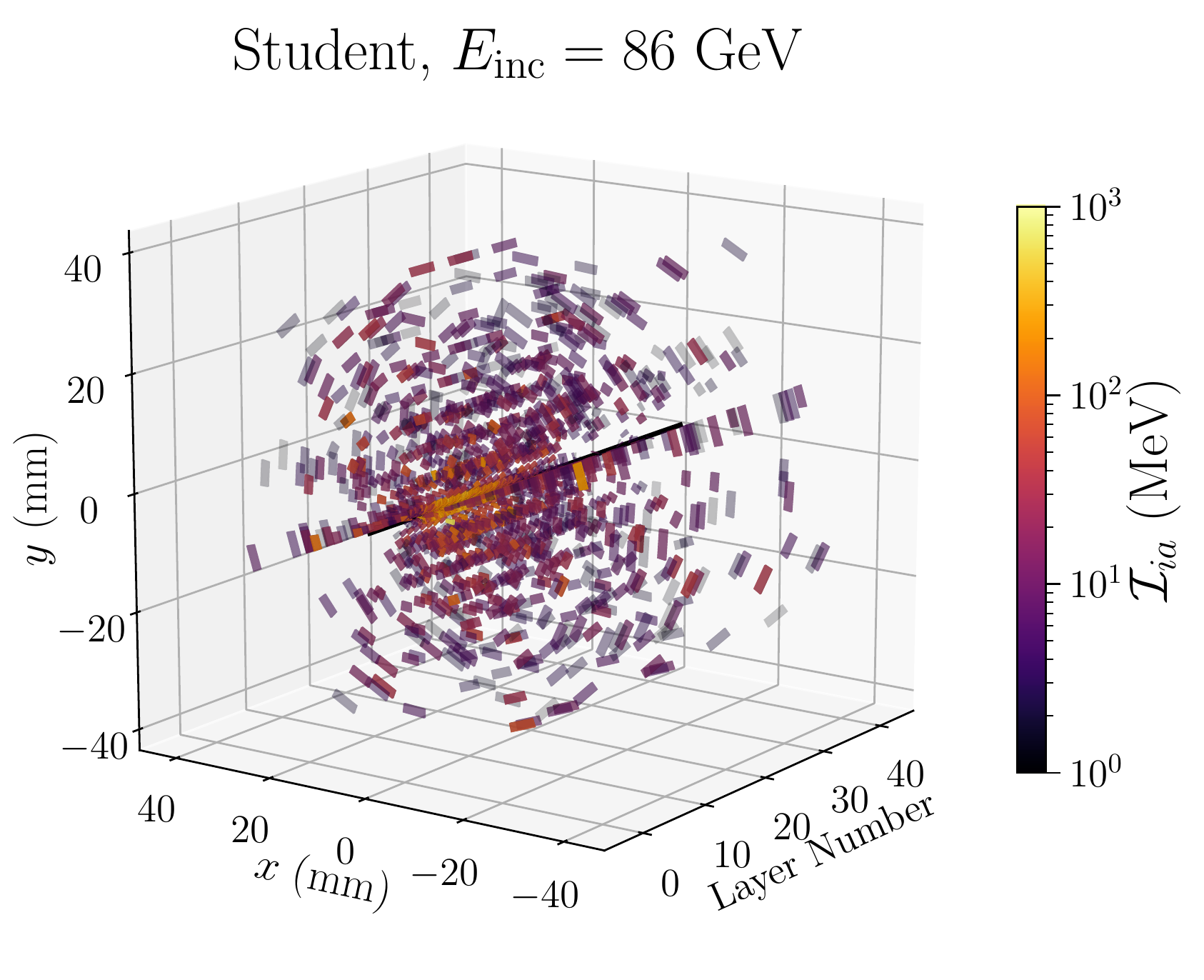}

\caption{Pattern of energy deposition from two example events generated by \geant{} in Dataset 2 (top row), \icalo{} teacher (middle row), and \icalo{} student (bottom row). Events have $E_{\rm inc} = 693$~GeV (left column) and $E_{\rm inc} = 86$~GeV (right column). For visual clarity, voxels with less than 1~MeV of energy have been suppressed. The beam axis is shown with a black line. For display purposes, the separation between layers and voxels within a layer have both been artificially increased from the real detector geometry. \label{fig:example_events}}
\end{figure*}

\begin{figure*}[ht]
\includegraphics[width=0.5\columnwidth]{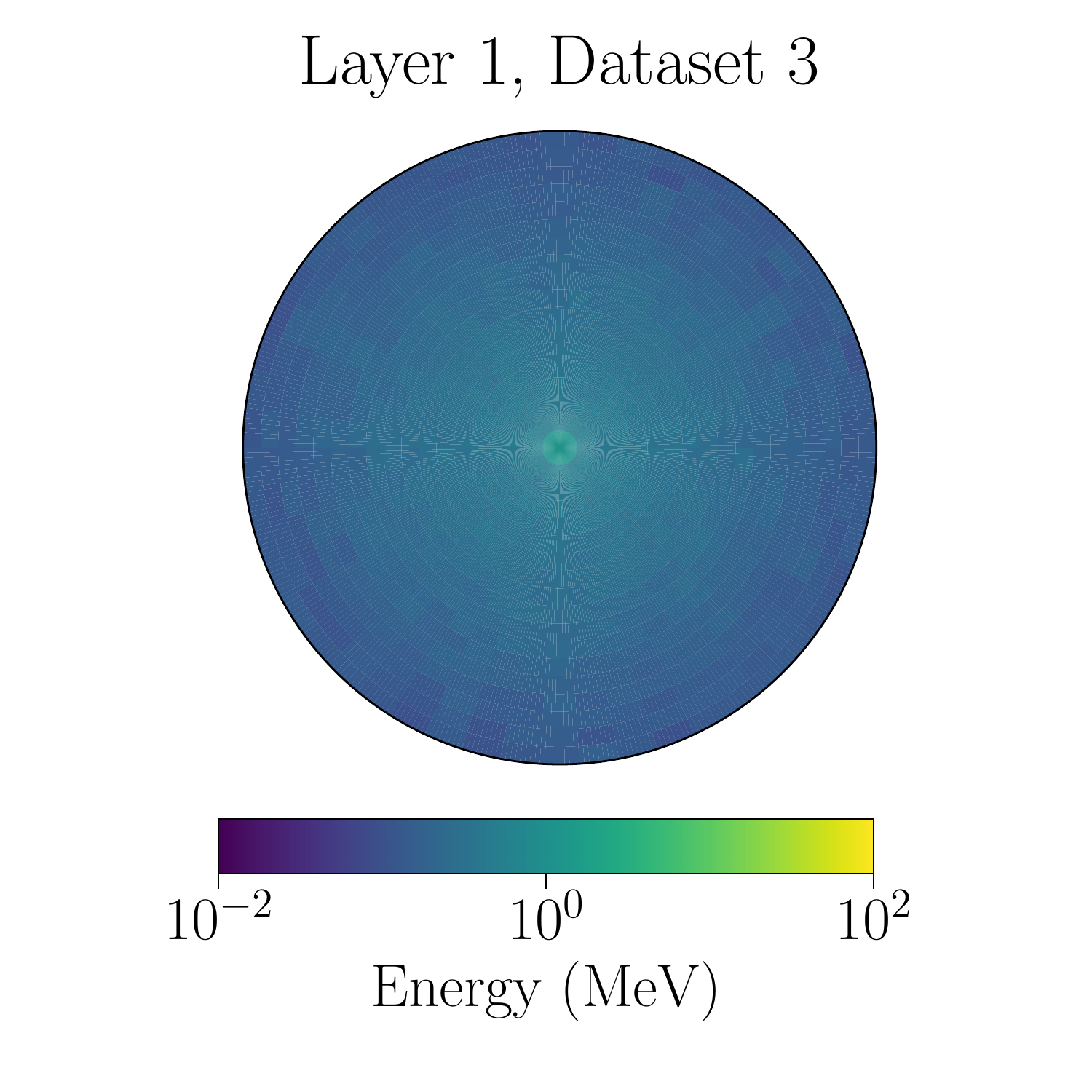}\includegraphics[width=0.5\columnwidth]{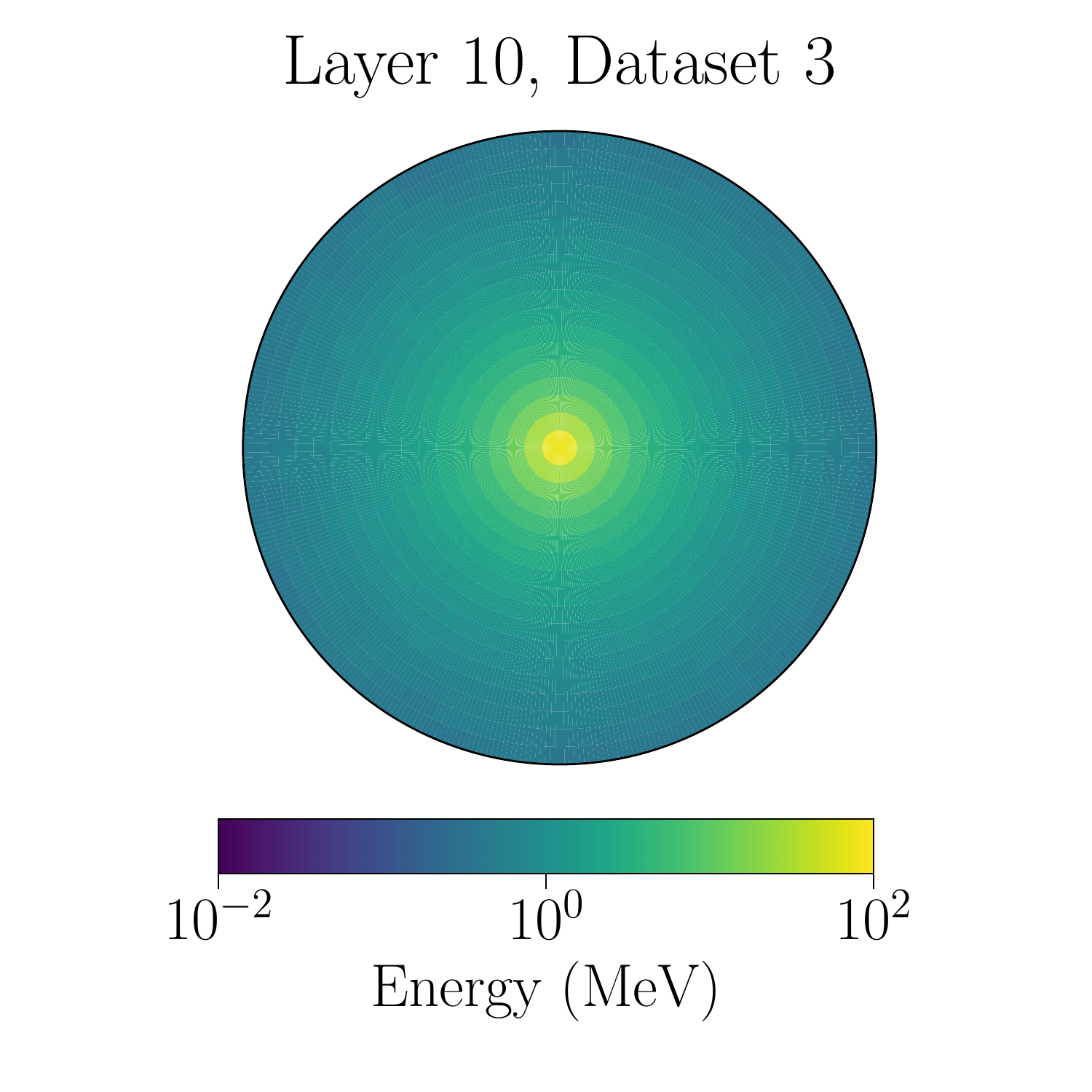}\includegraphics[width=0.5\columnwidth]{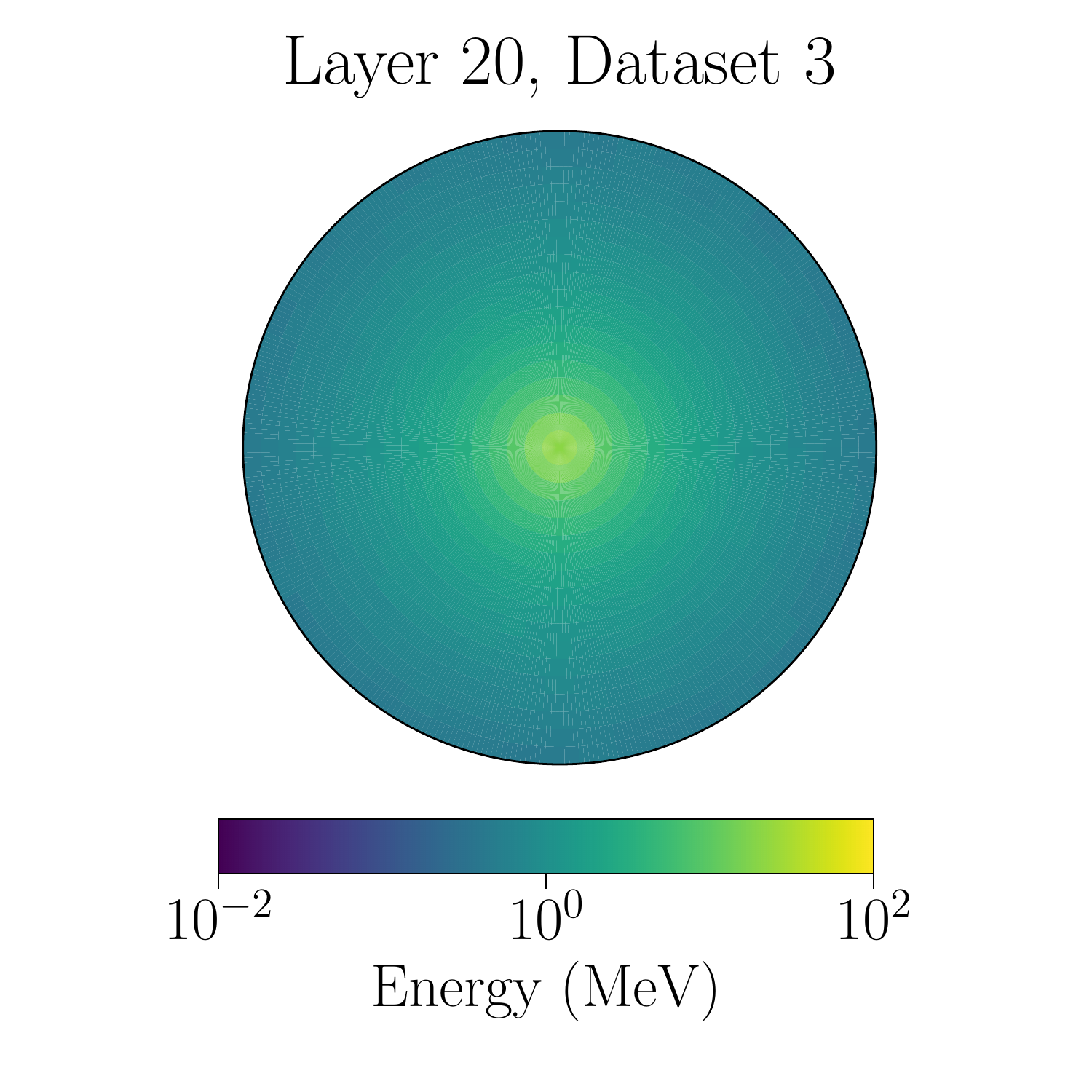}\includegraphics[width=0.5\columnwidth]{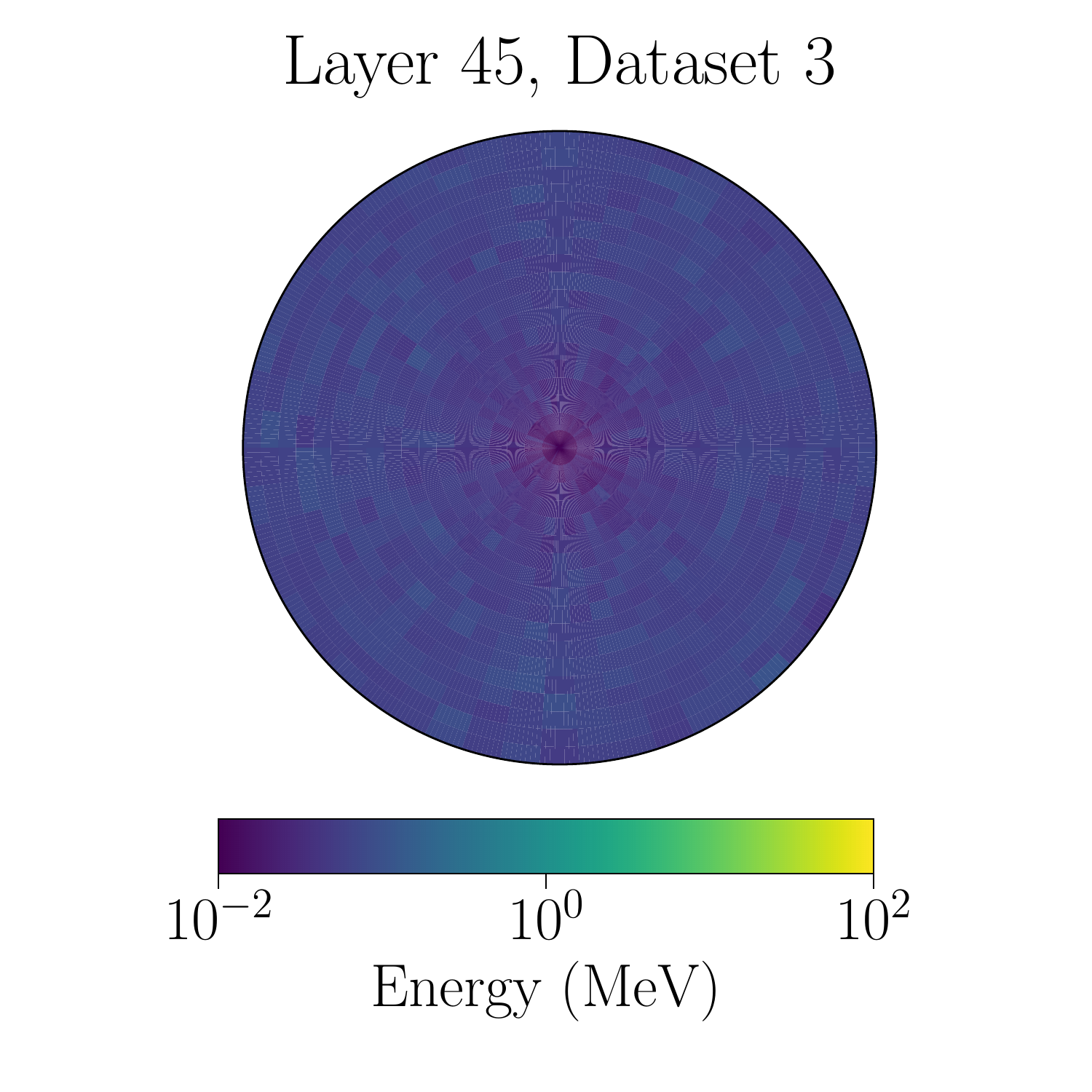}

\includegraphics[width=0.5\columnwidth]{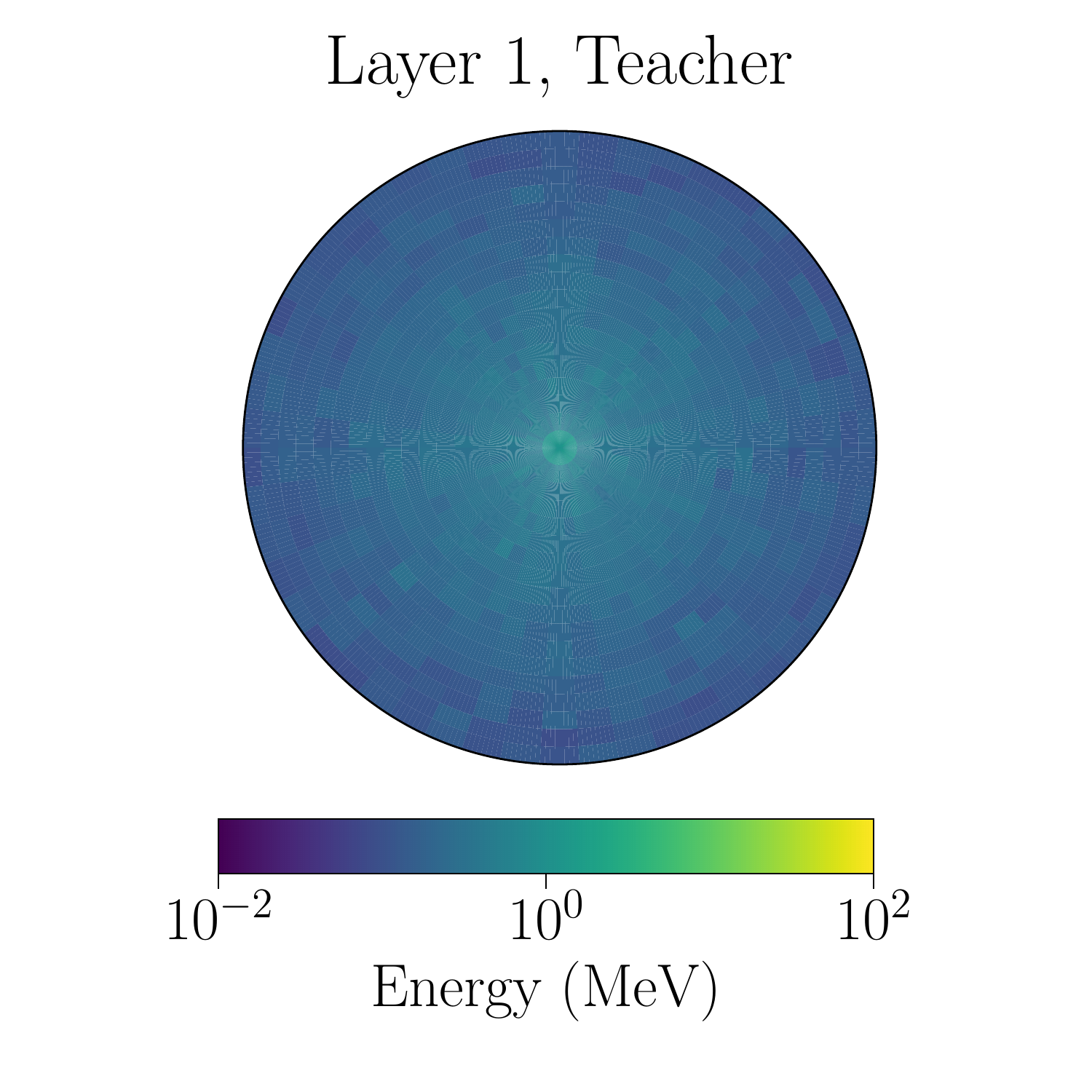}\includegraphics[width=0.5\columnwidth]{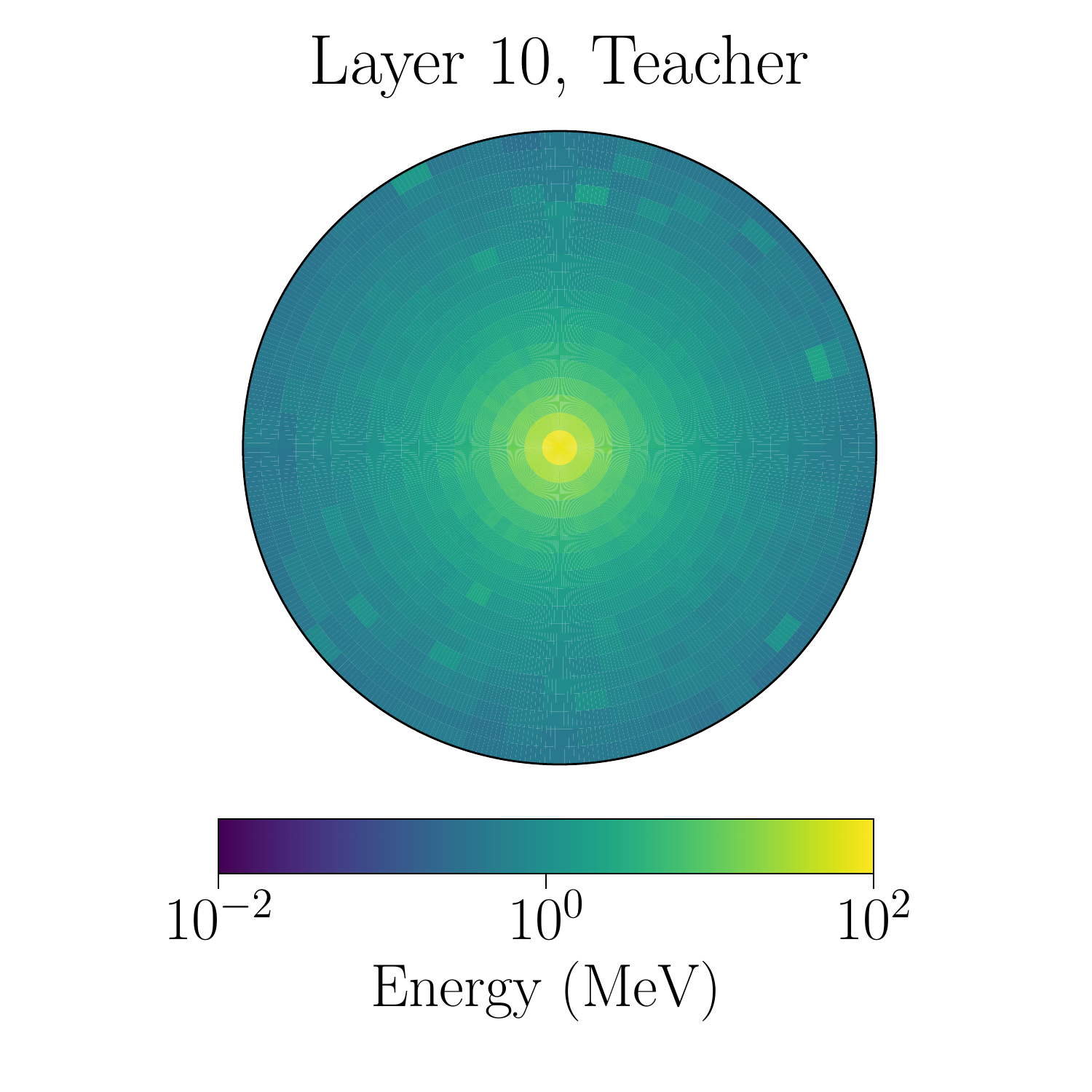}\includegraphics[width=0.5\columnwidth]{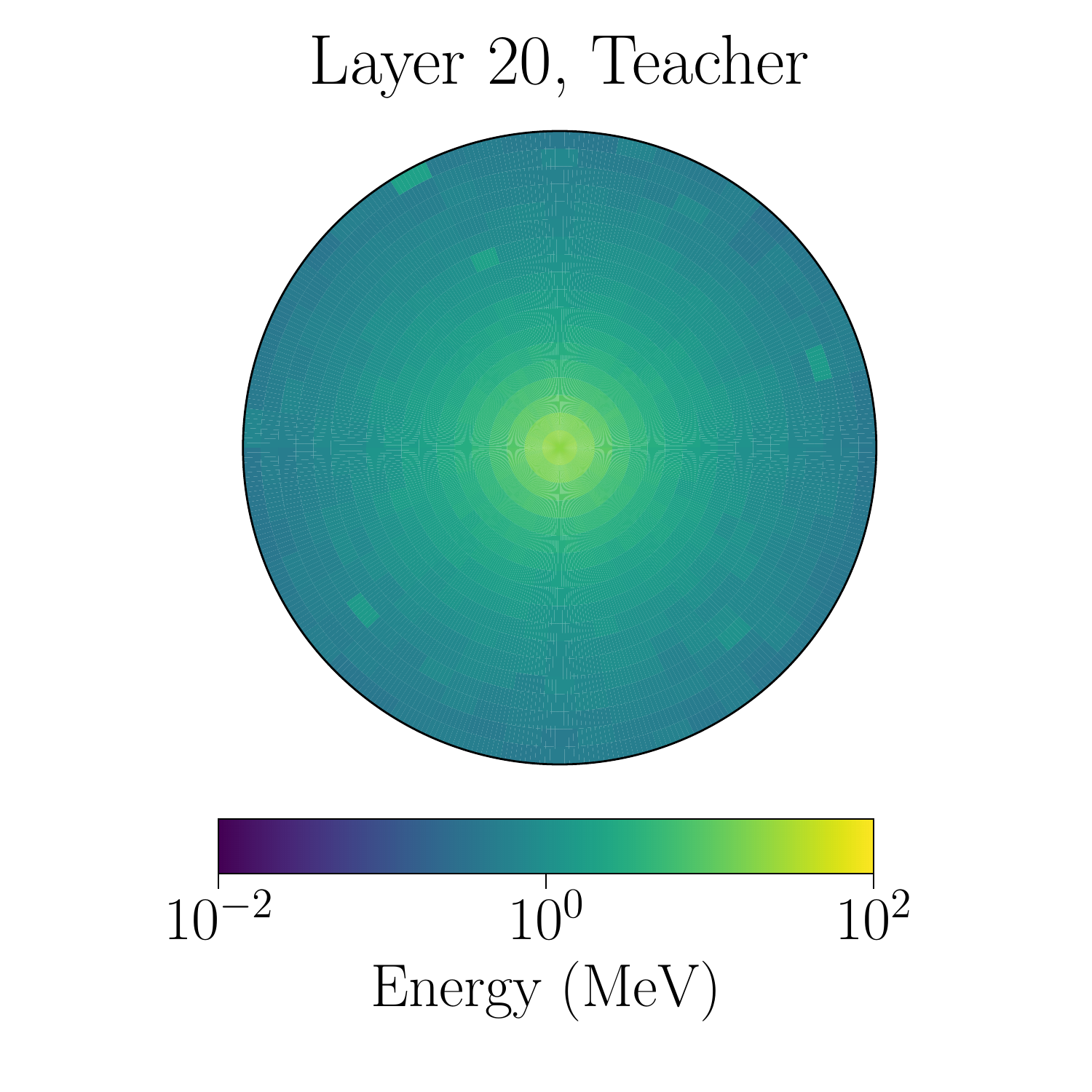}\includegraphics[width=0.5\columnwidth]{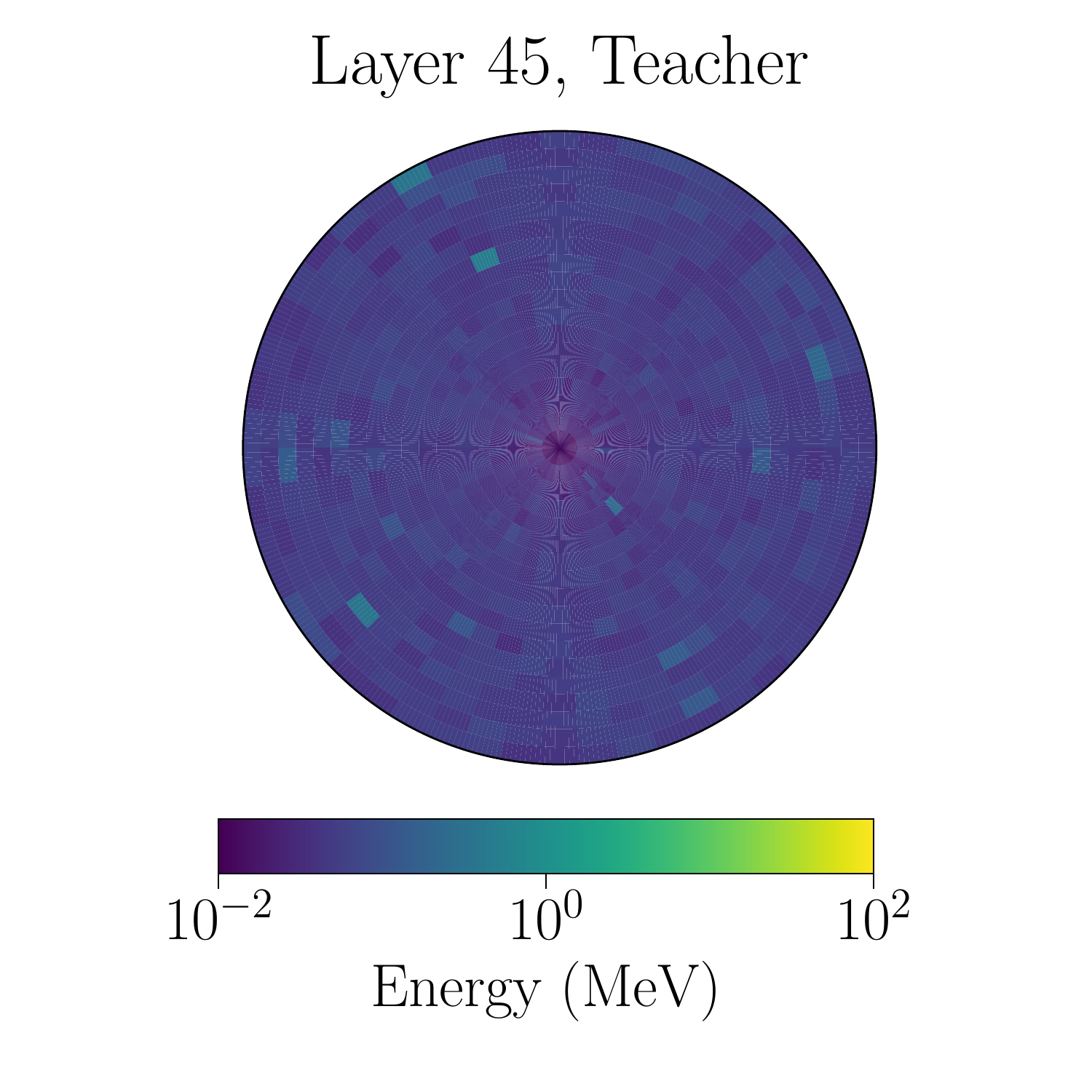}

\includegraphics[width=0.5\columnwidth]{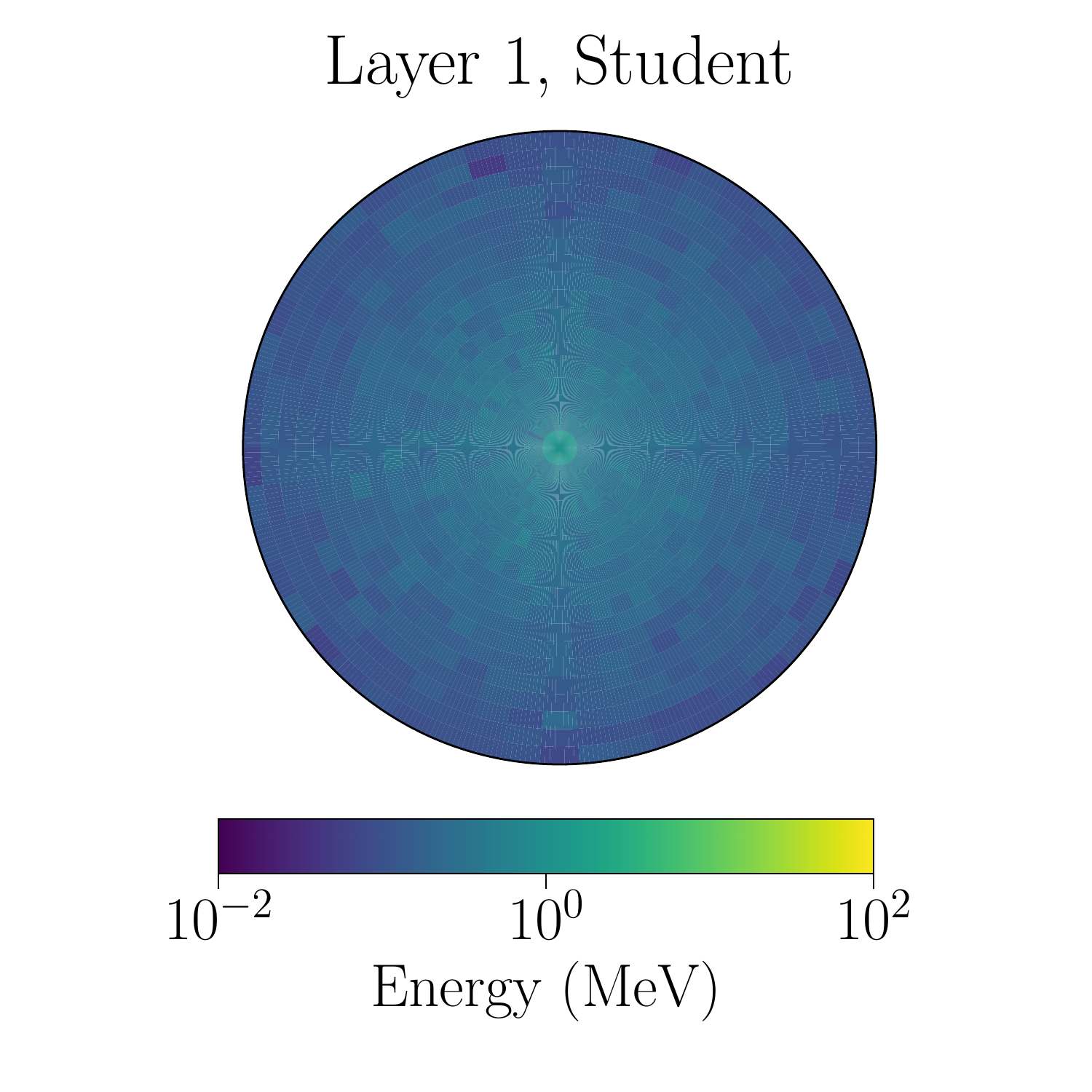}\includegraphics[width=0.5\columnwidth]{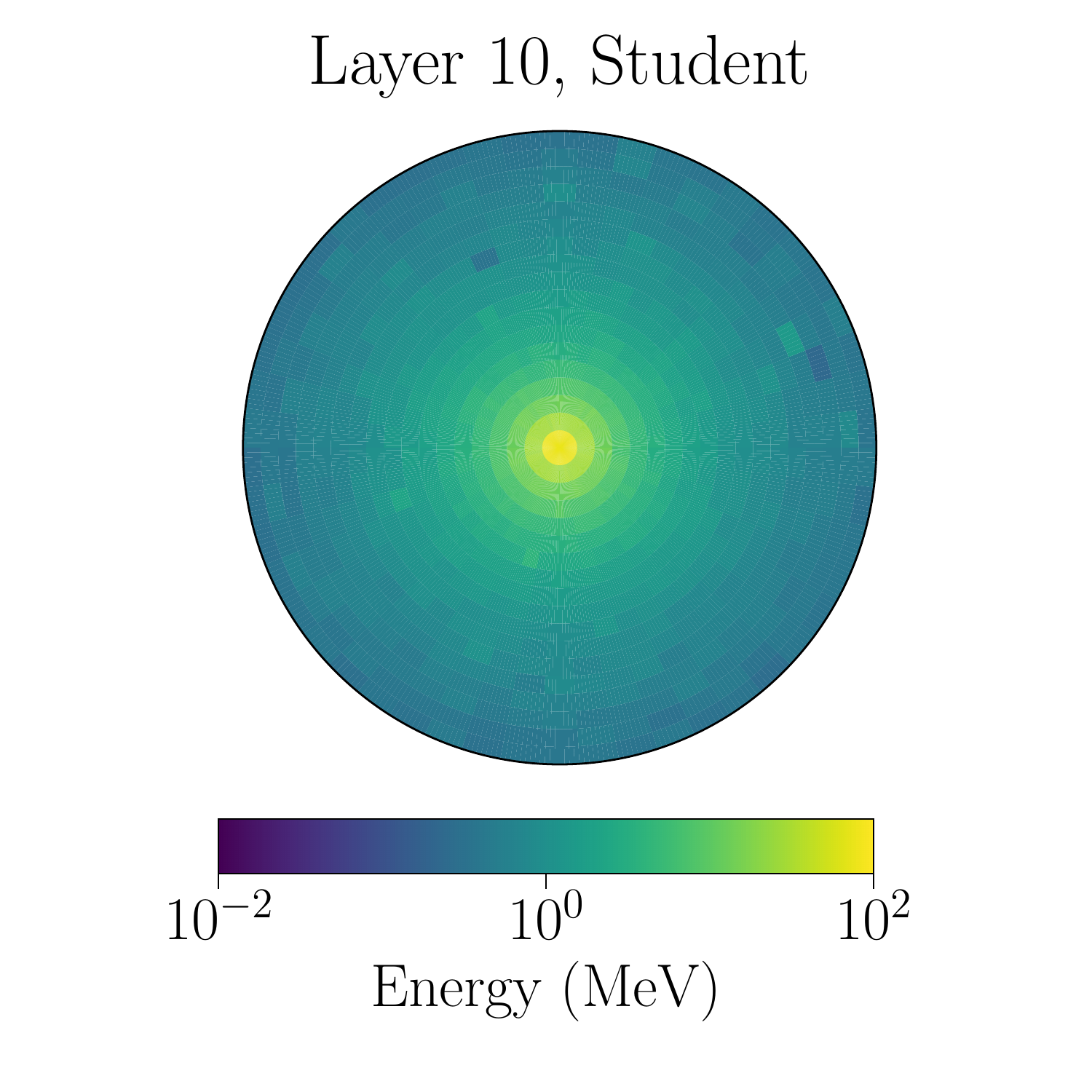}\includegraphics[width=0.5\columnwidth]{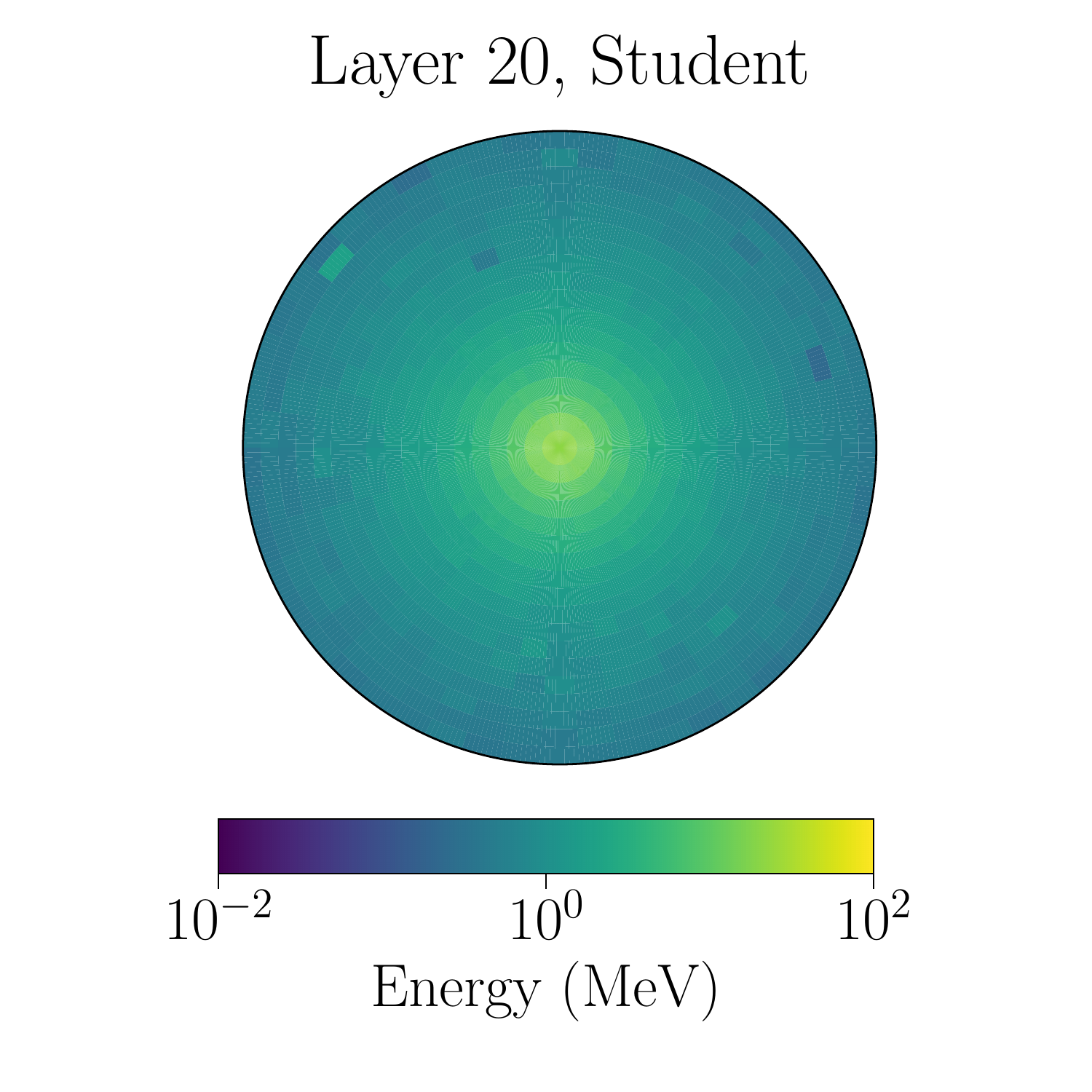}\includegraphics[width=0.5\columnwidth]{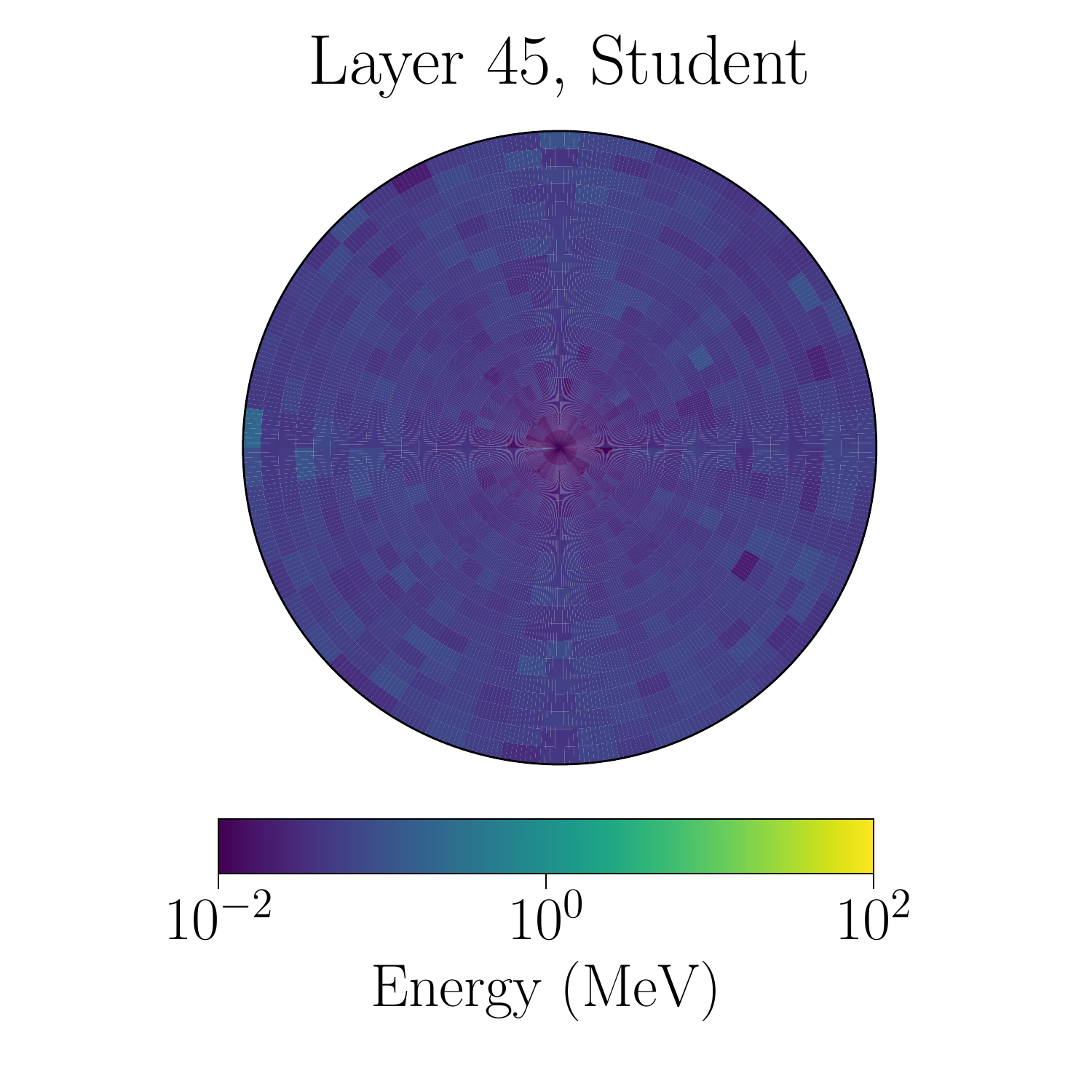}
\caption{Averaged energy deposition pattern of events in layers 1, 10, 20, and 45 (from left to right) for the \geant\ data in Dataset 3 (top row), events sampled from \icalo\ teacher trained on Dataset 3 (middle row), and events sampled from \icalo\ student trained on Dataset 3 (bottom row)\label{fig:example_averages}}.
\end{figure*}

In Figure~\ref{fig:example_events}, we show the pattern of energy deposition in all layers for two example events from Dataset 2 (one with $E_{\rm inc} = 693$~GeV and one with 86~GeV), compared with two events generated by \icalo{} with equal incident energies. (Similar plots for Dataset 3 events are difficult to visualize clearly due to the density of voxels.)  In Figure~\ref{fig:example_averages}, we show the pattern of energy deposition in layers 1, 10, 20, and 45, averaged over all the events in the dataset for both Dataset 3 and the sampled events from \icalo.

\subsection{Distributions}

\begin{figure*}[h!]
\includegraphics[width=1.9\columnwidth]{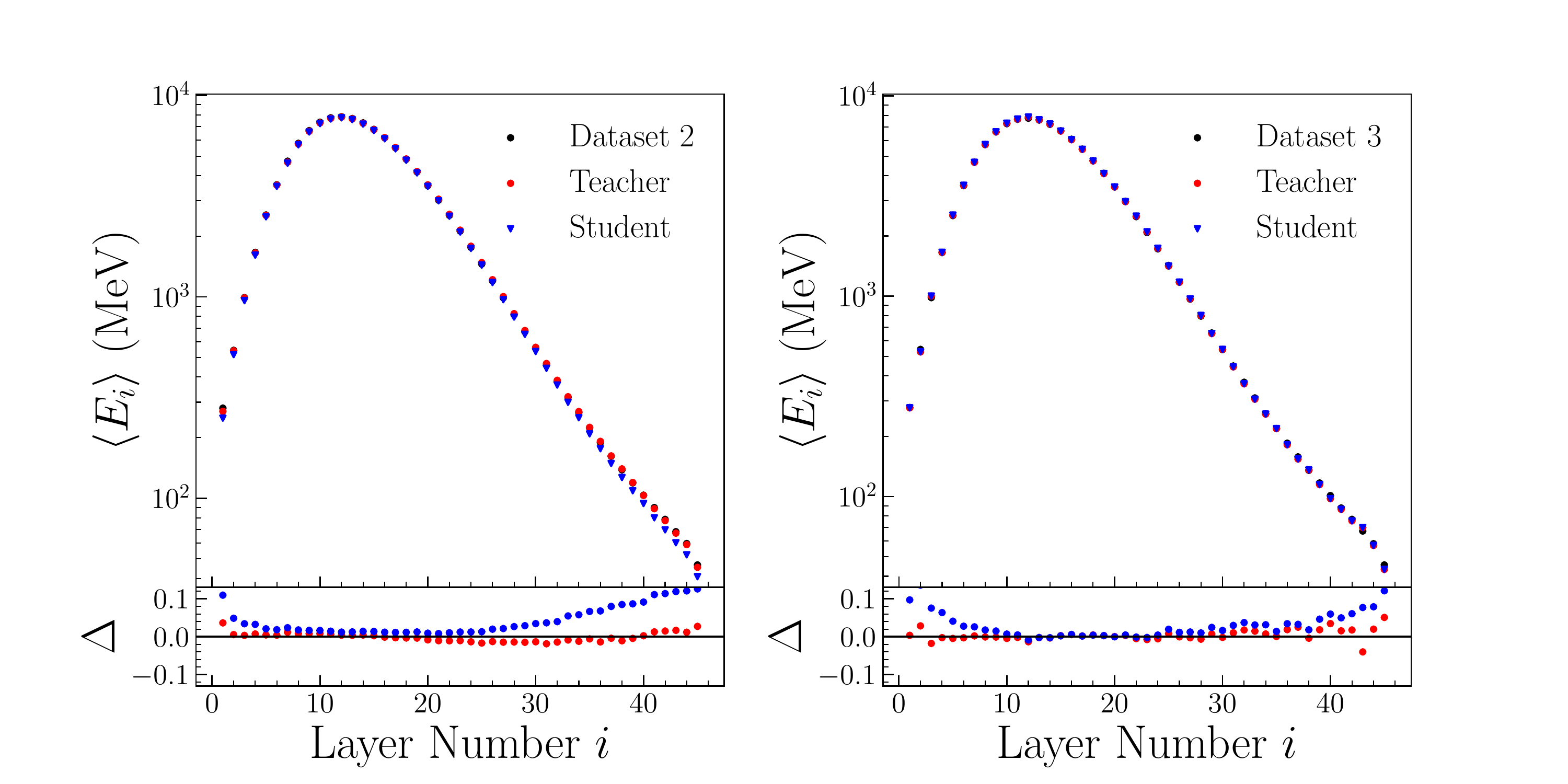}

\caption{Averaged total energy deposition $\langle E_i\rangle$ for each layer $i$ for Datasets 2 (left) and 3 (right). In each plot the averaged energy of the \geant{} data is shown in black, and the distribution generated by \icalo{} teacher (student)  in red (blue). The fractional difference $\Delta$ between the truth-level and generated distributions is shown below the main figures.}
\label{fig:gen_edep}
\end{figure*}

We next consider more detailed diagnostic plots, comparing the distribution of various high-level features between the \geant{}, teacher, and student events.
First, we examine the energy deposition in each layer (again noting this is obtained by the sum of the voxel energies output from \flowtwo{} and \flowthree{}). In Figure~\ref{fig:gen_edep}, we show the energies deposited in each layer, averaged over all the generated showers, which we denote by $\langle E_i \rangle$.
As expected, the generated distributions are similar for both Datasets 2 and 3, with small variations due to the different training regimes and normalization differences in the output of the flows. The output of the student networks largely follows that of the teachers, with the most significant deviations at both low and high layer numbers. 

\begin{figure*}[ht]
\includegraphics[width=0.5\columnwidth]{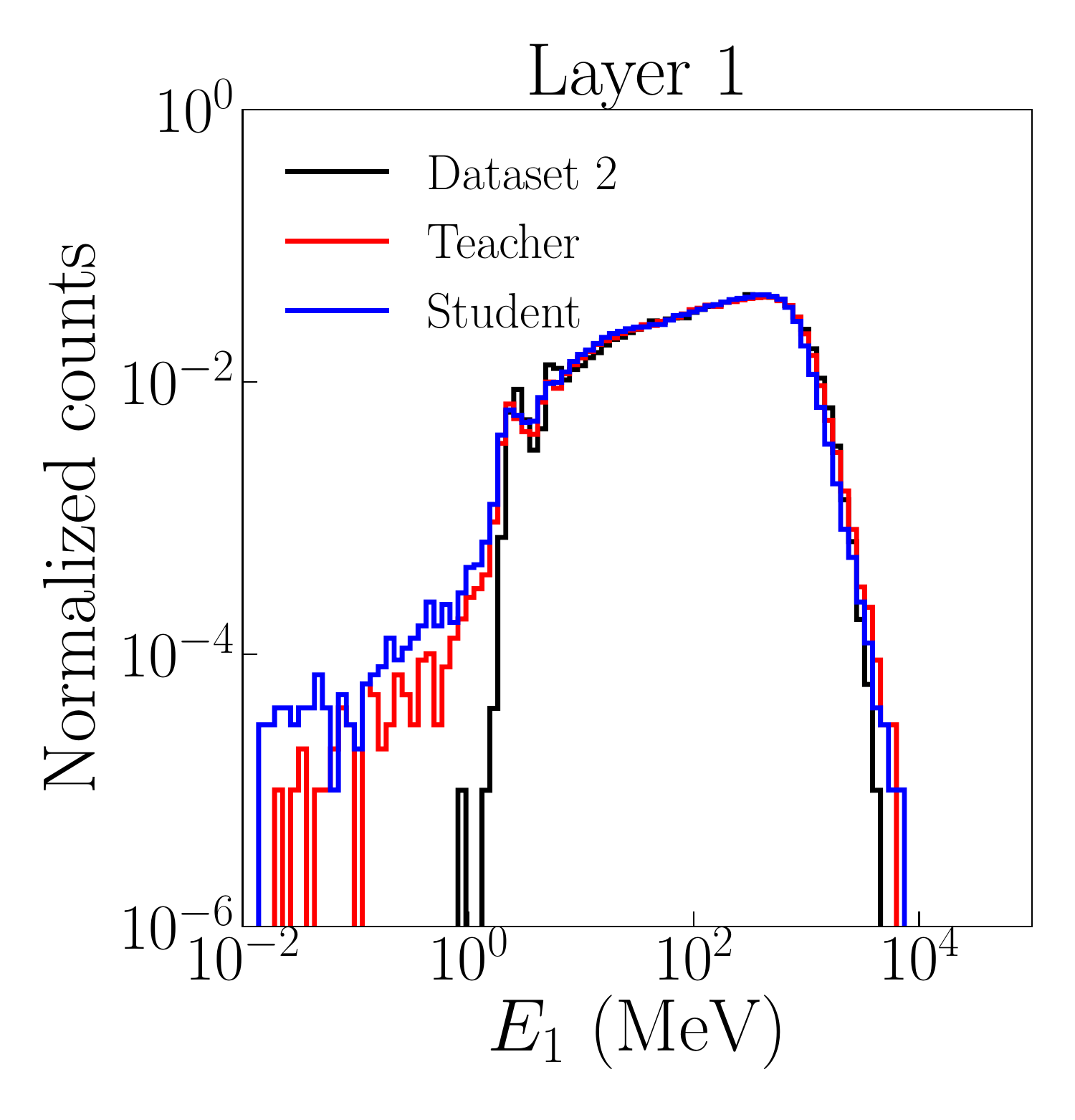}\includegraphics[width=0.5\columnwidth]{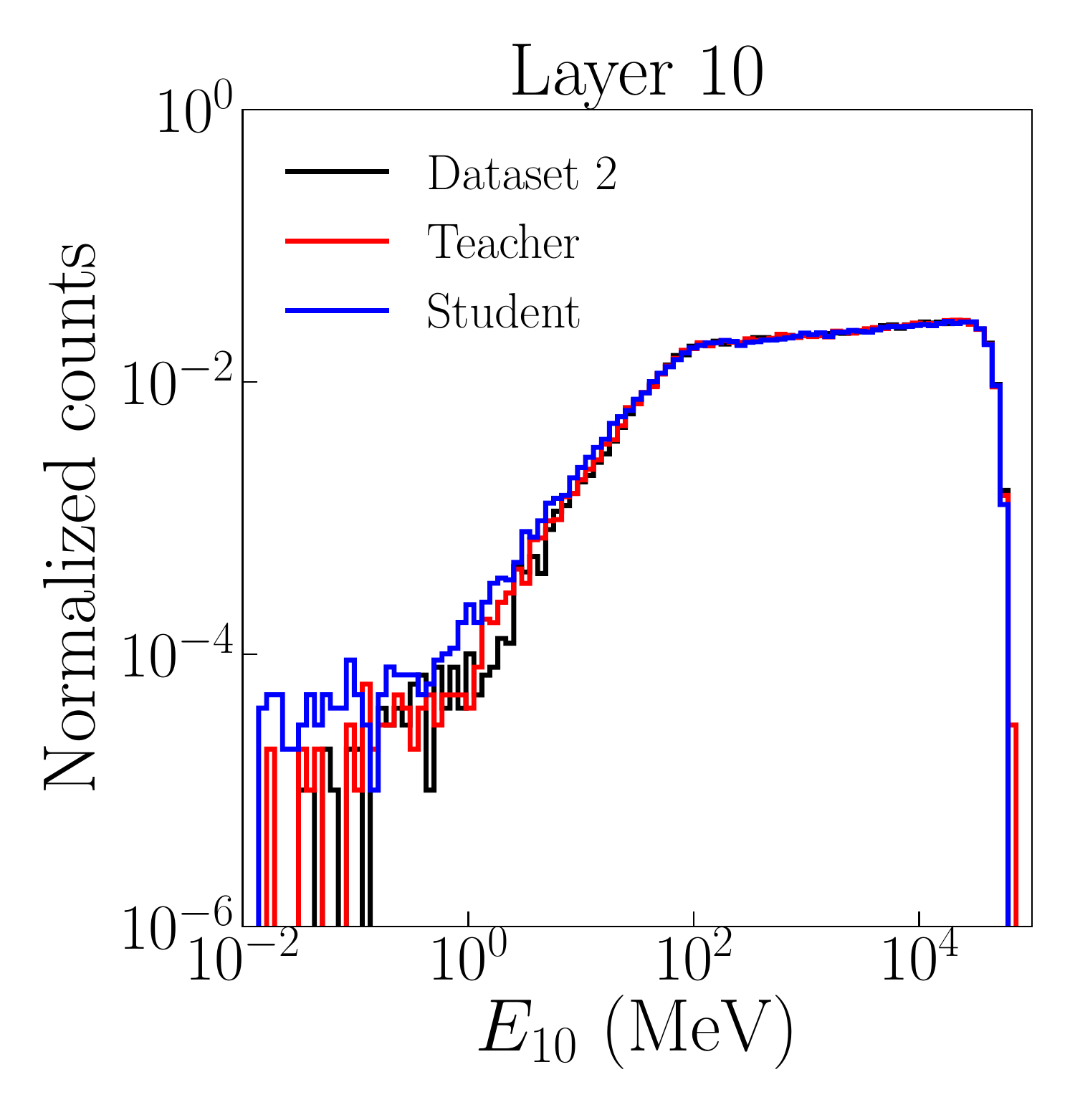}\includegraphics[width=0.5\columnwidth]{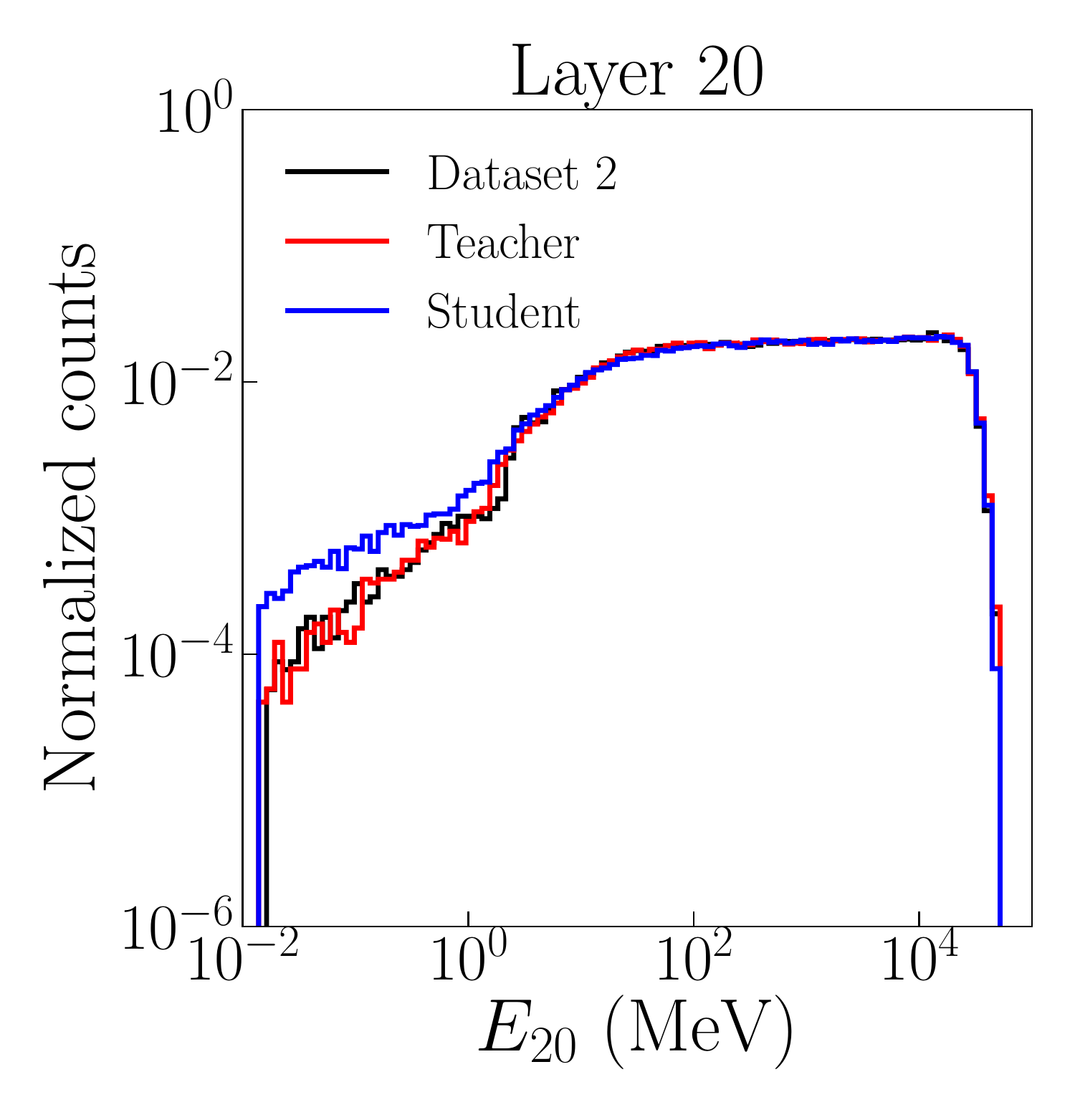}\includegraphics[width=0.5\columnwidth]{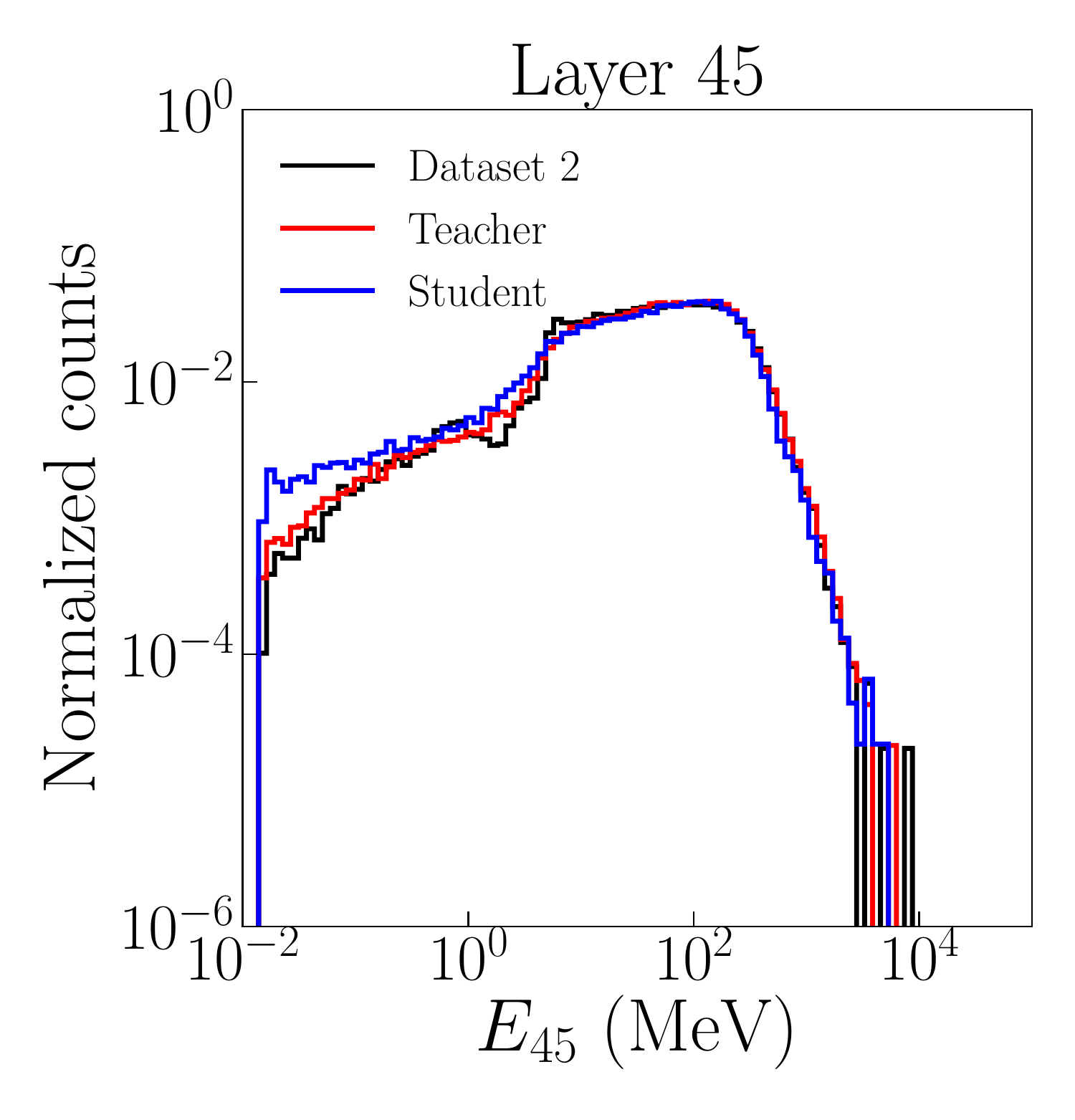}
\includegraphics[width=0.5\columnwidth]{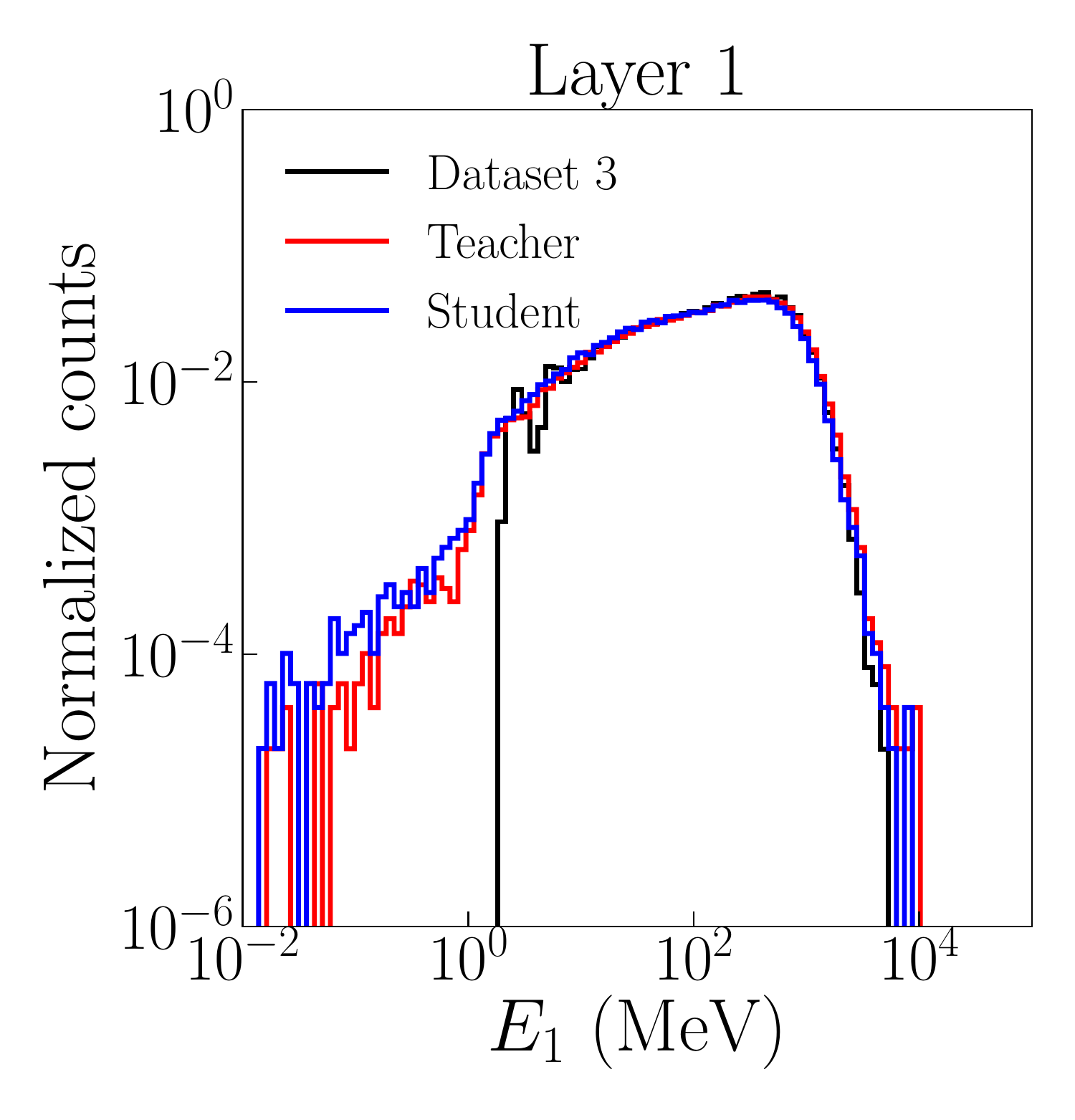}\includegraphics[width=0.5\columnwidth]{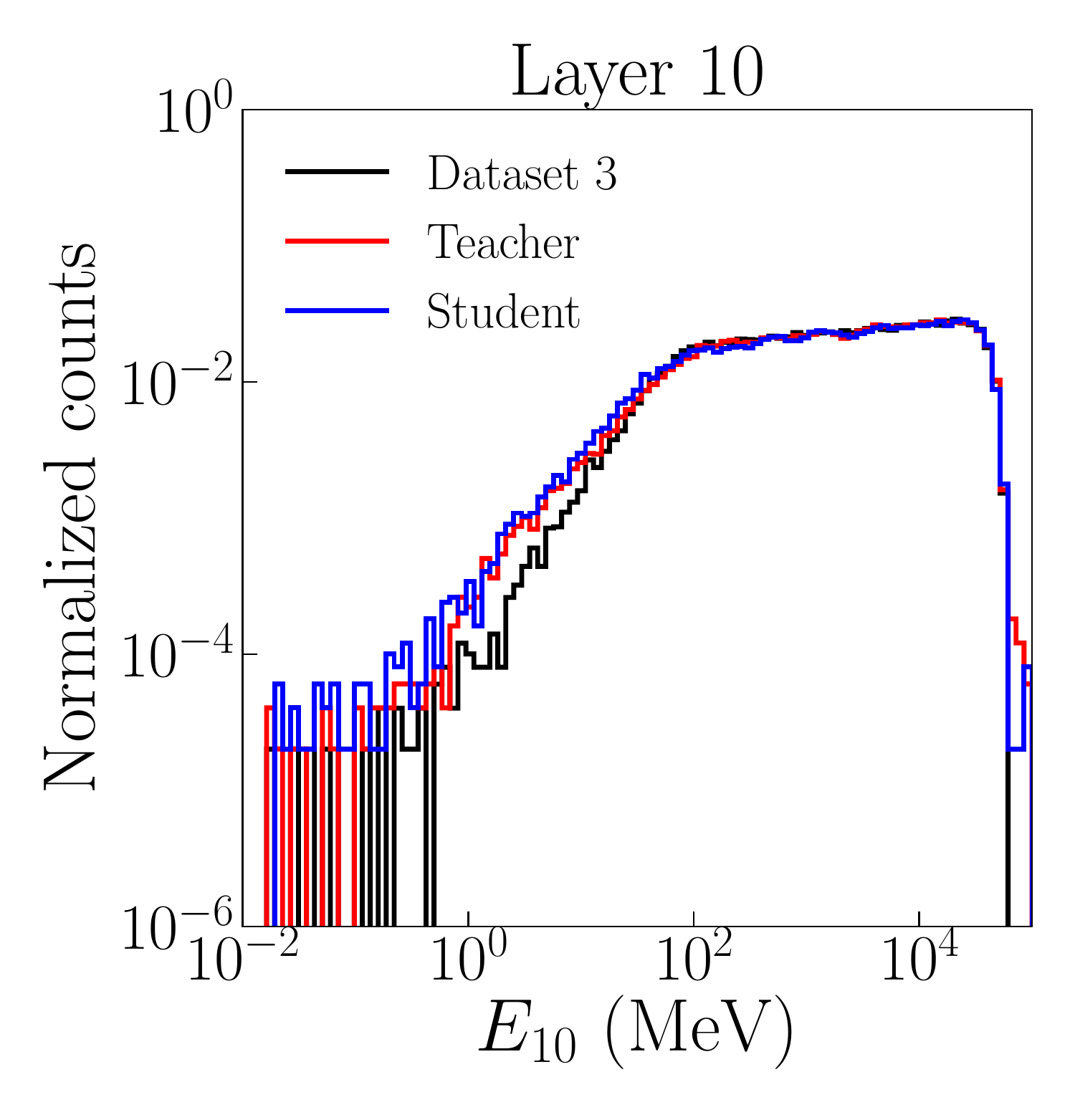}\includegraphics[width=0.5\columnwidth]{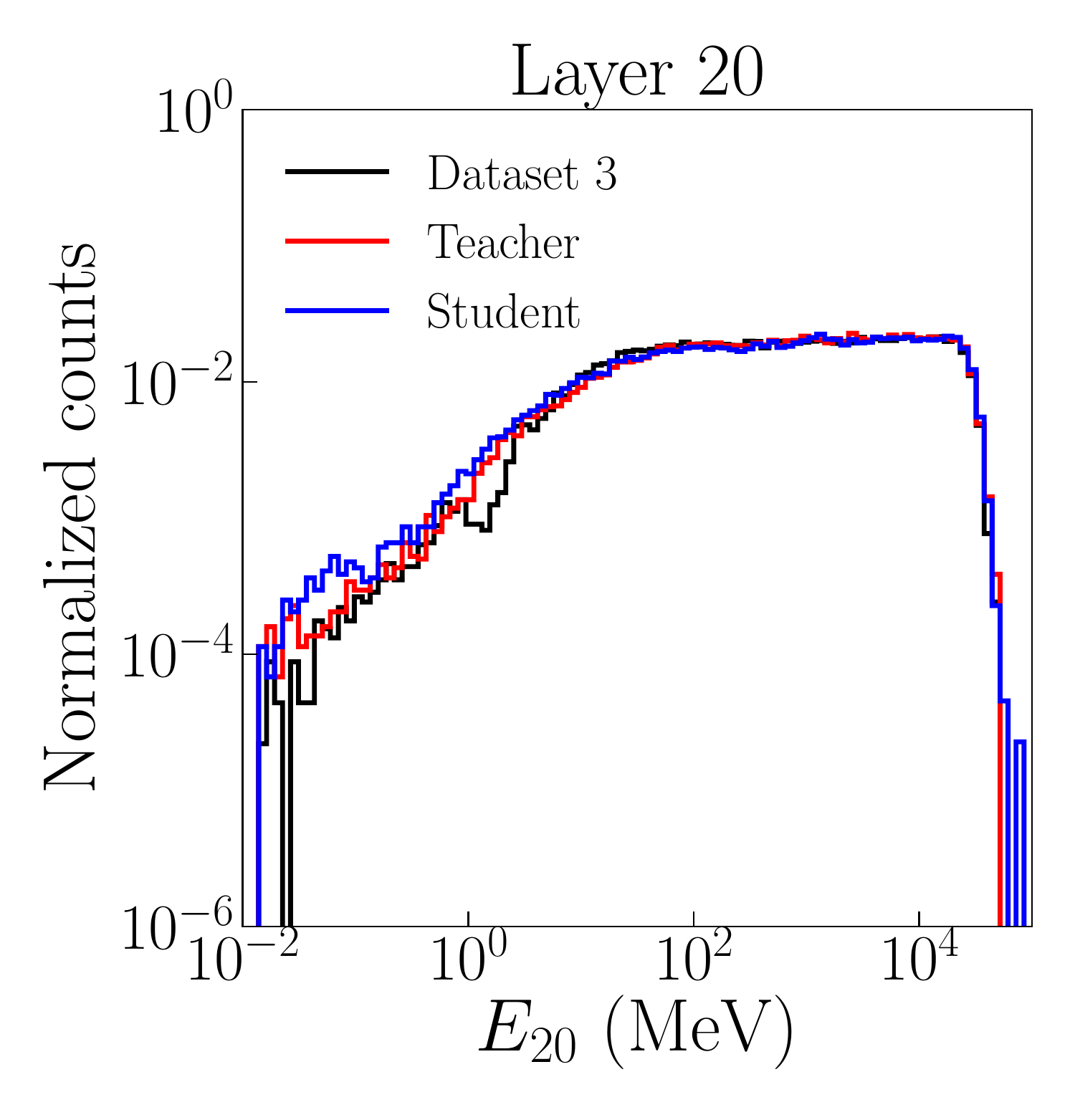}\includegraphics[width=0.5\columnwidth]{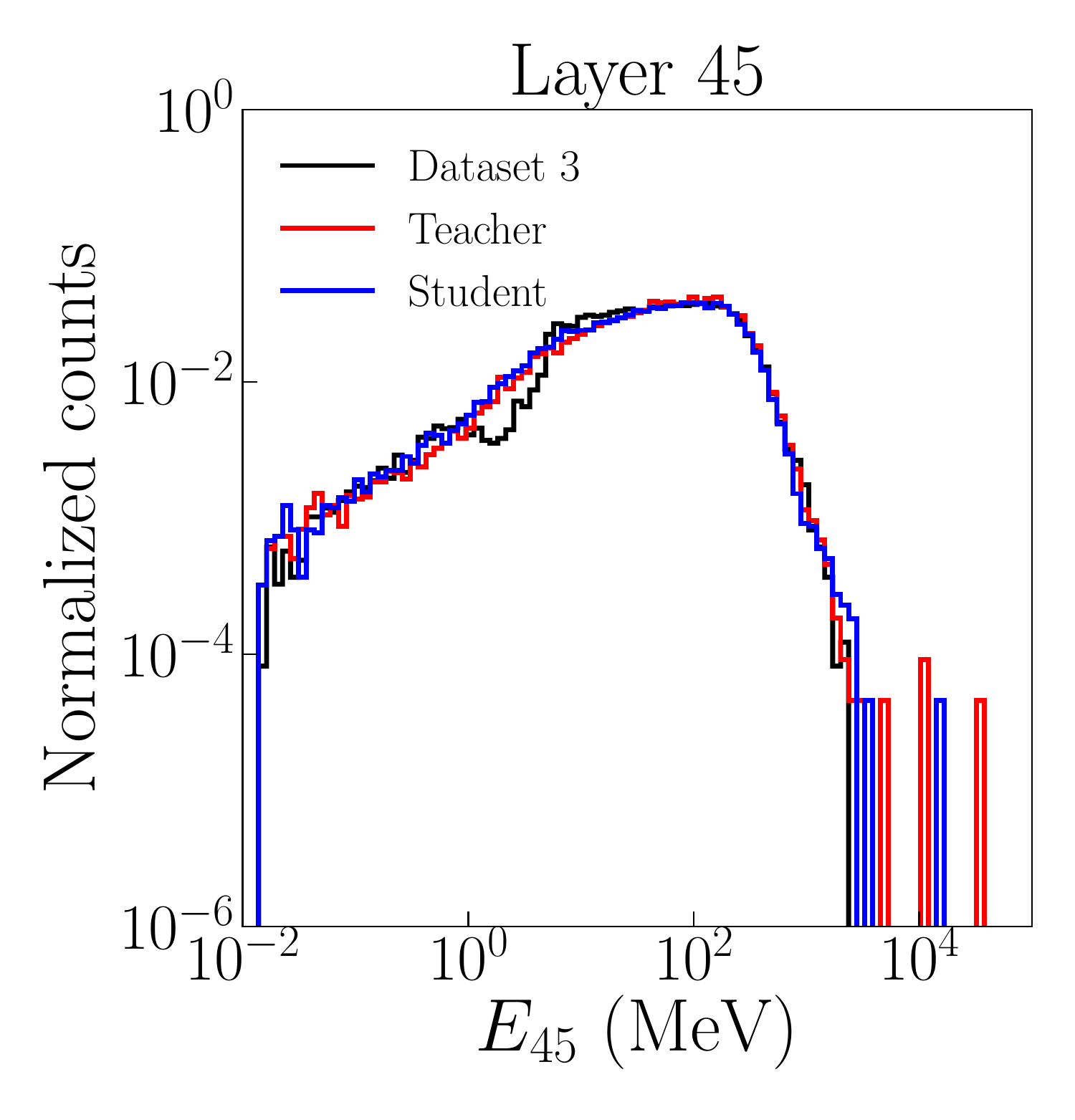}

\caption{Histograms of total energy deposited in a layer $i$ ($E_i$), for $i=1$, 10, 20, and 45 (from left to right), for Dataset 2 (top row) and Dataset 3 (bottom row). Distribution of \geant\ data is shown as black lines, and that of the \icalo\ teacher (student) trained on Dataset 2 or 3 (as appropriate) in red (blue).}
\label{fig:flowone_generated}
\end{figure*}

\begin{figure*}[ht]
\includegraphics[width=0.5\columnwidth]{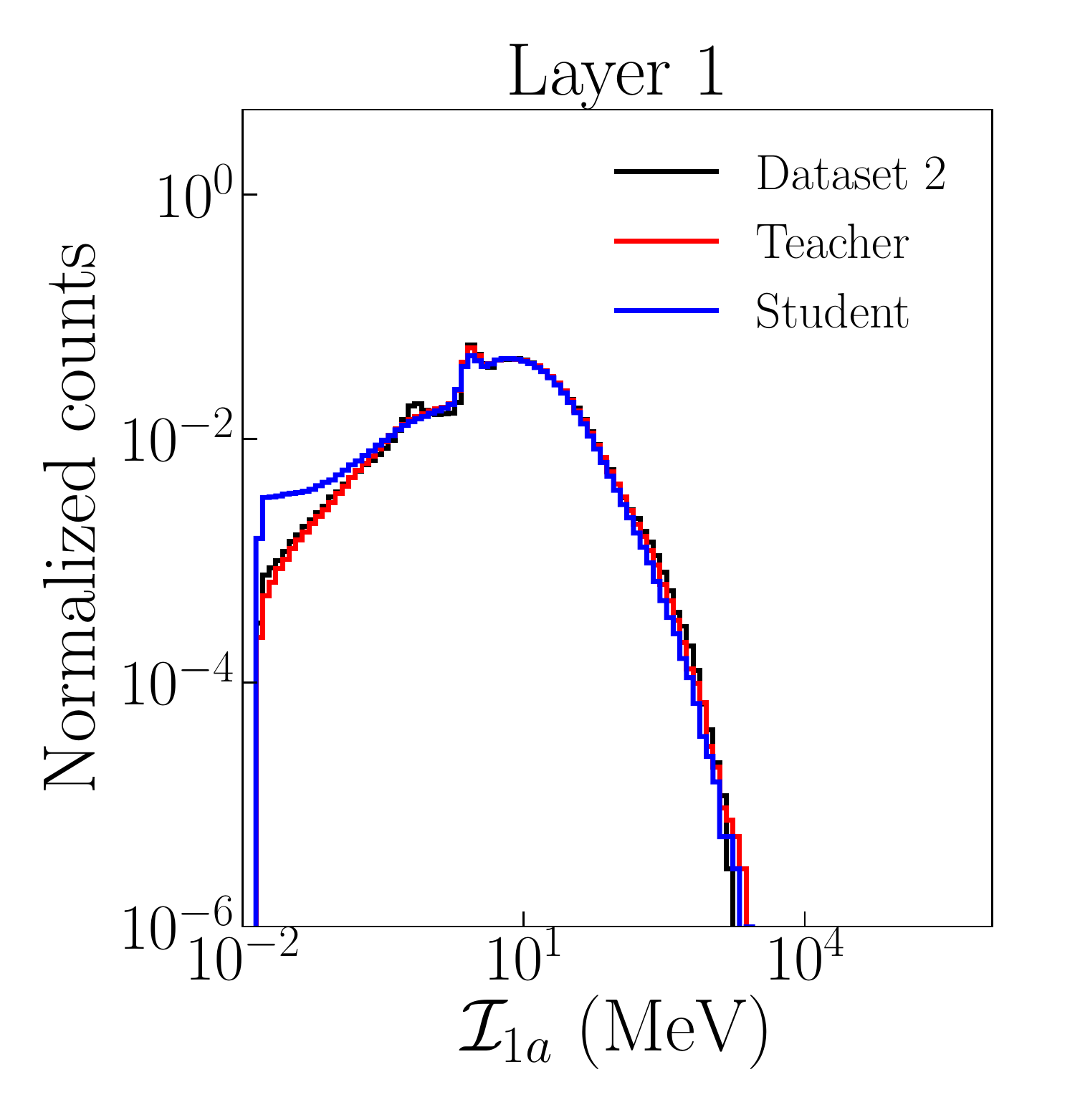}\includegraphics[width=0.5\columnwidth]{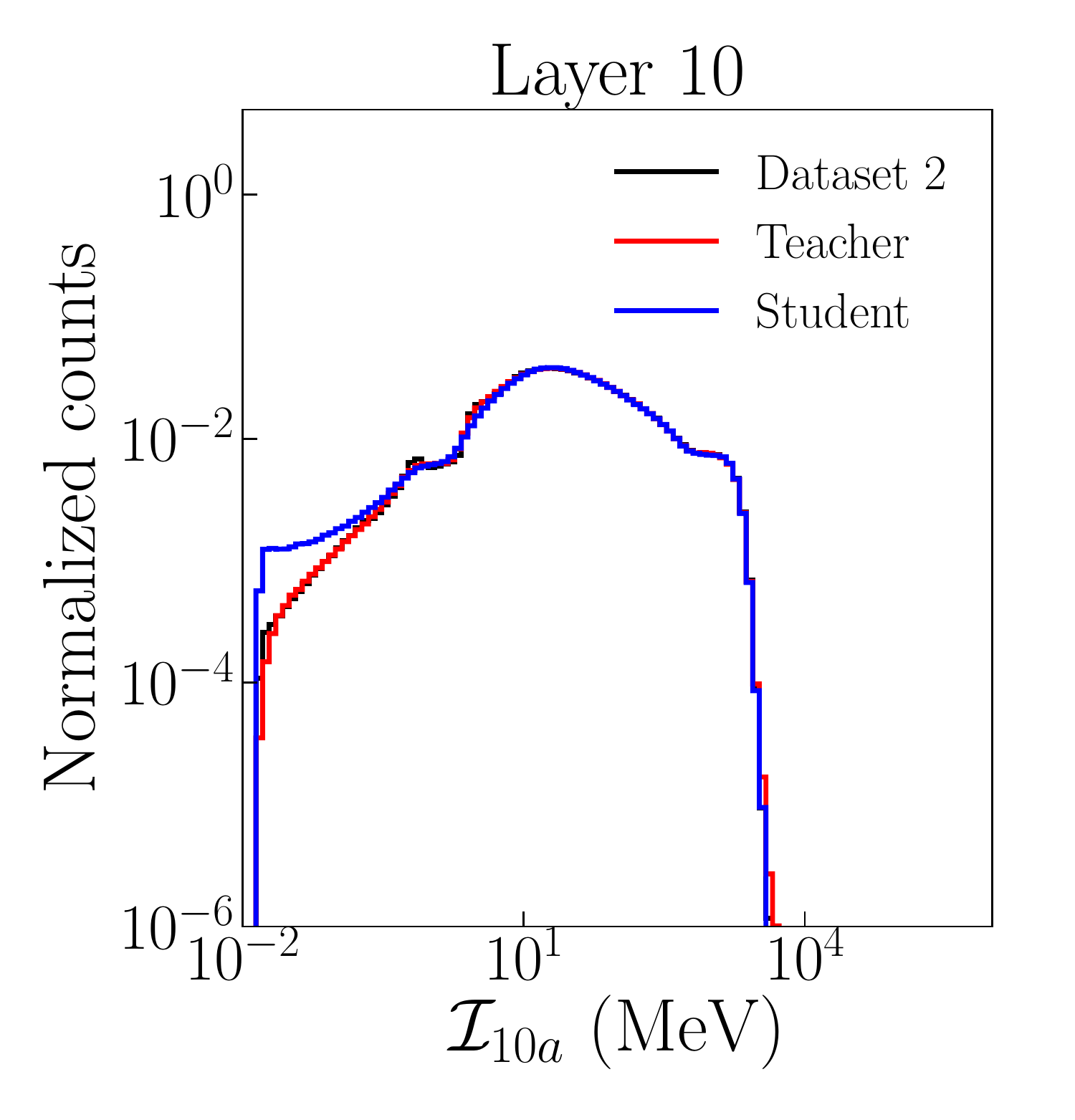}\includegraphics[width=0.5\columnwidth]{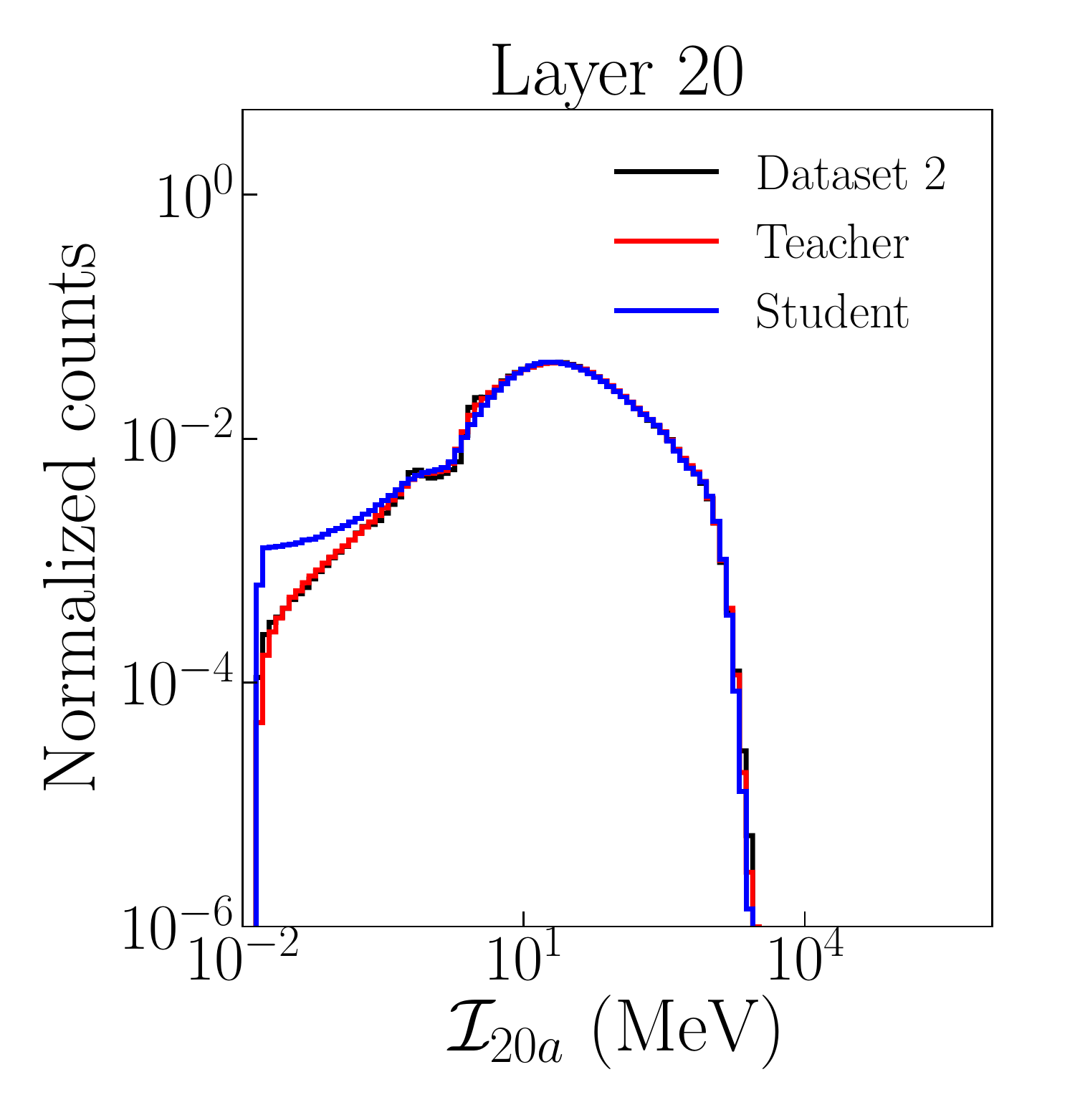}\includegraphics[width=0.5\columnwidth]{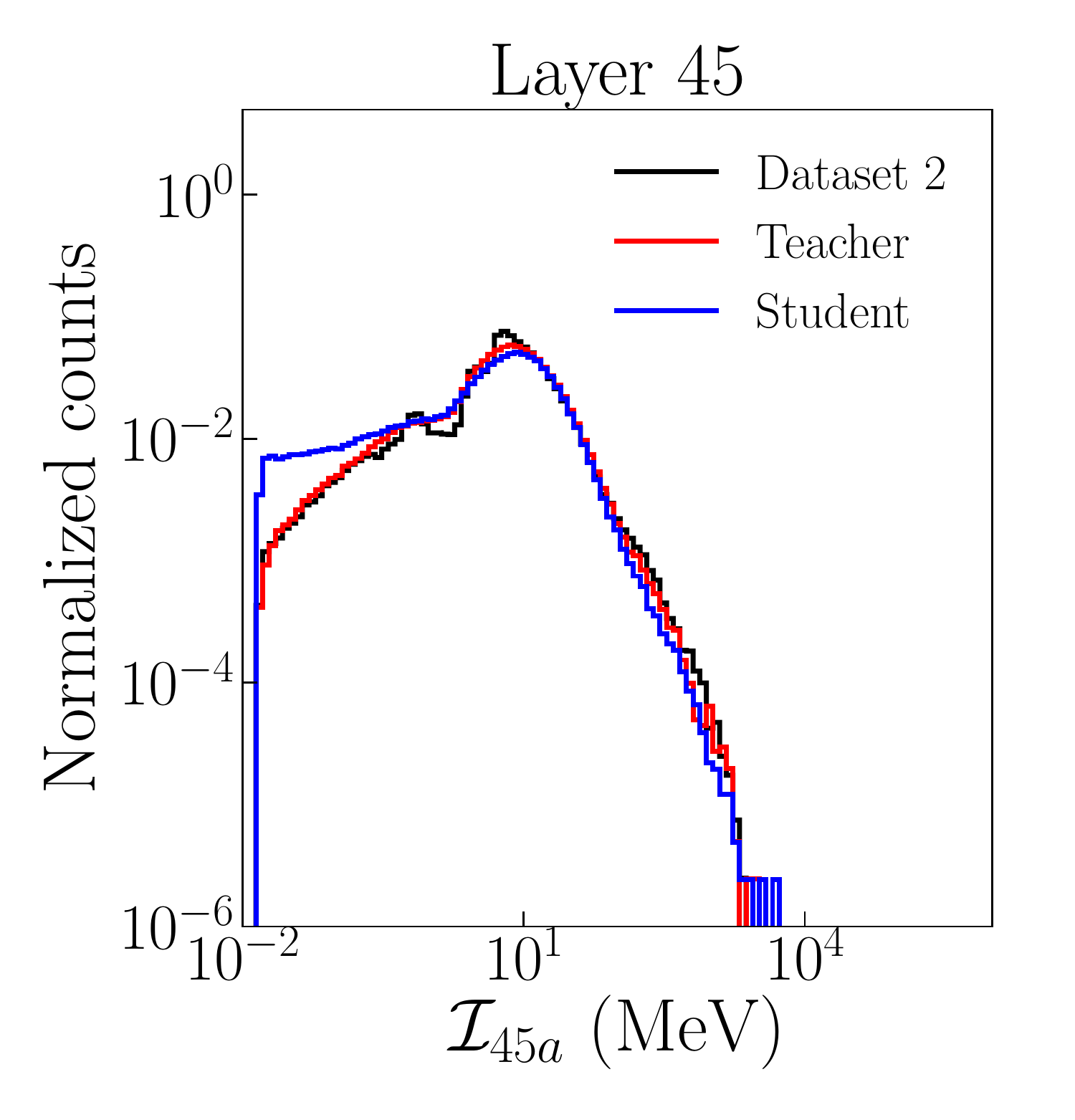}

\includegraphics[width=0.5\columnwidth]{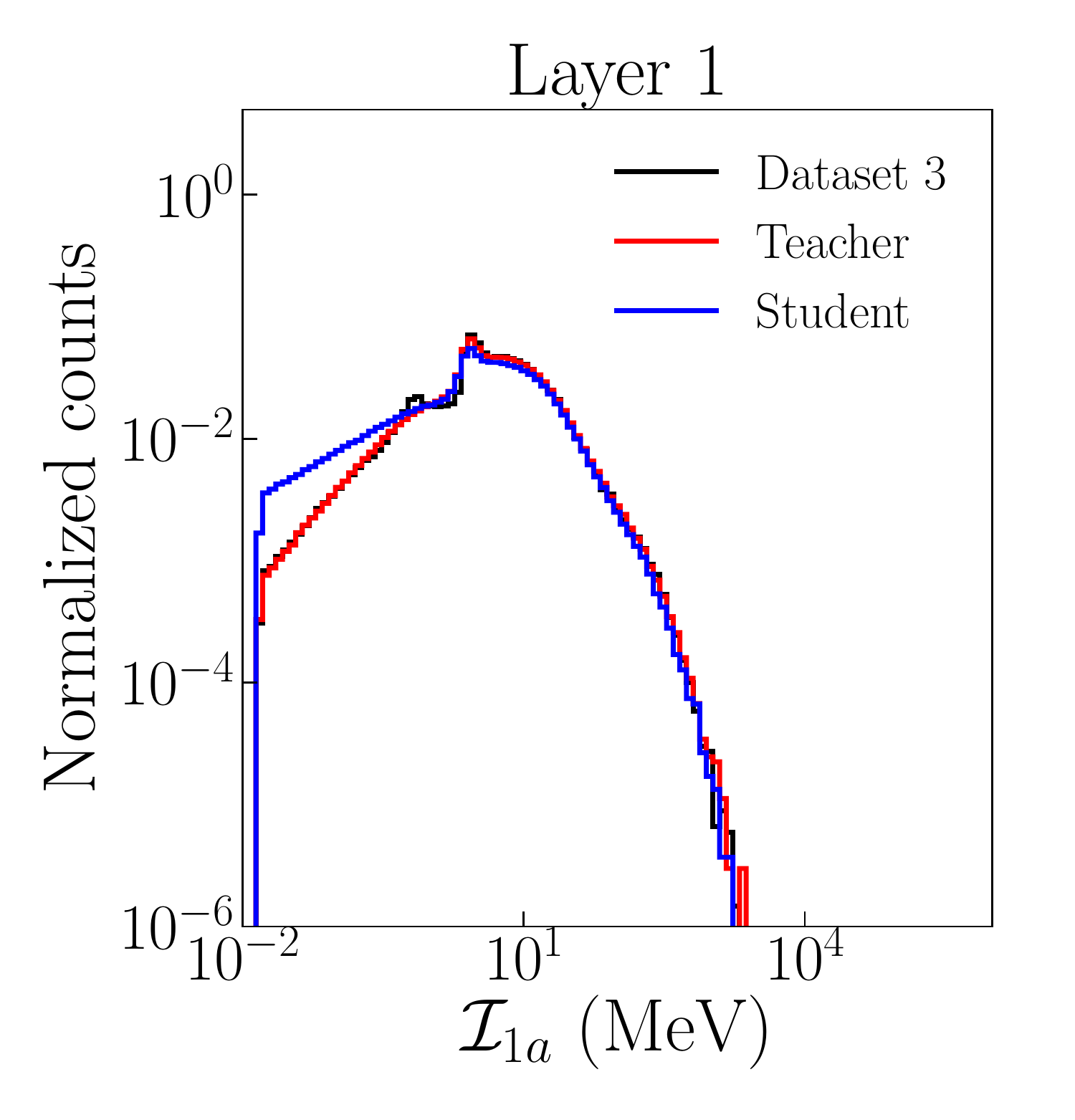}\includegraphics[width=0.5\columnwidth]{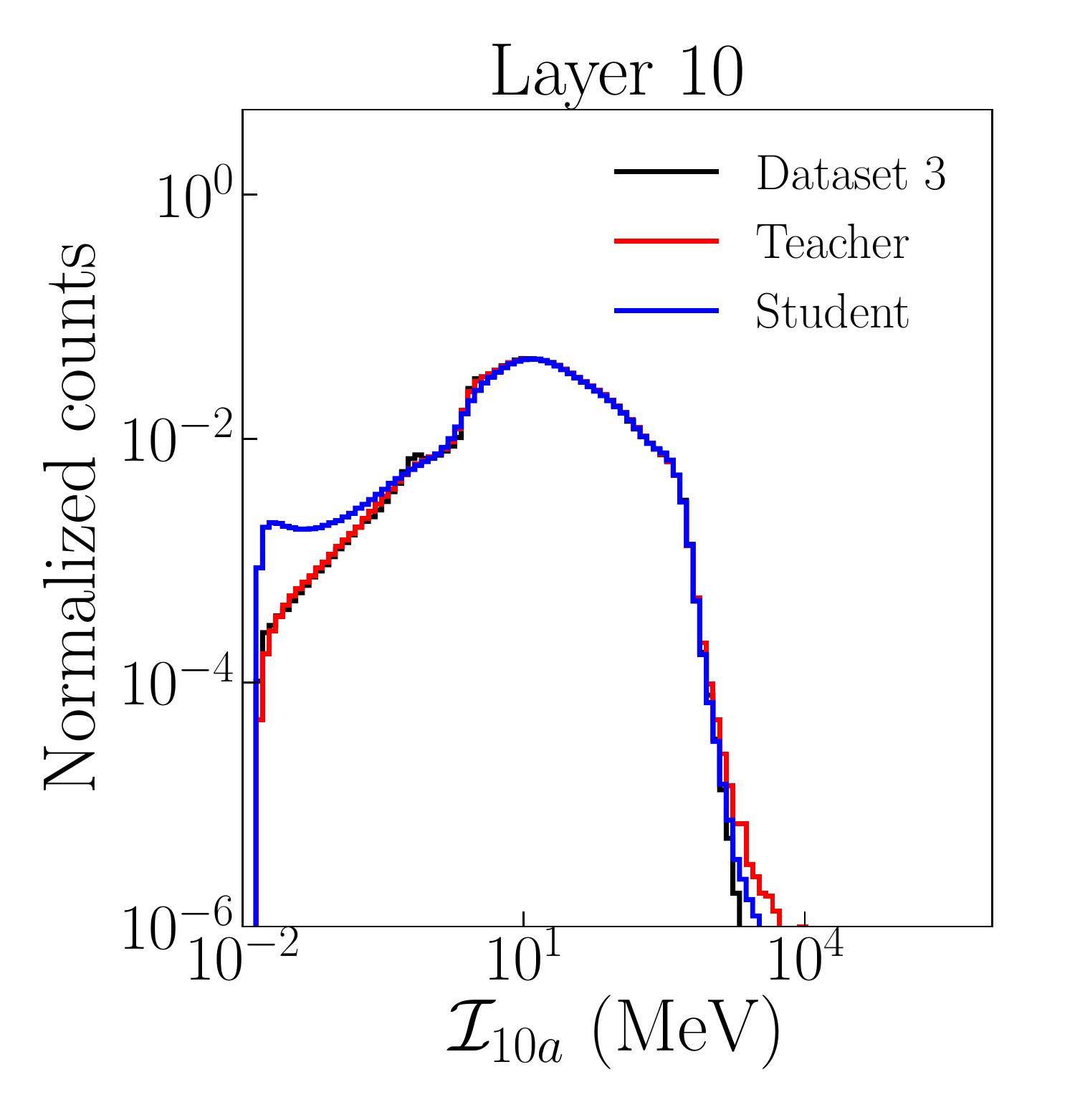}\includegraphics[width=0.5\columnwidth]{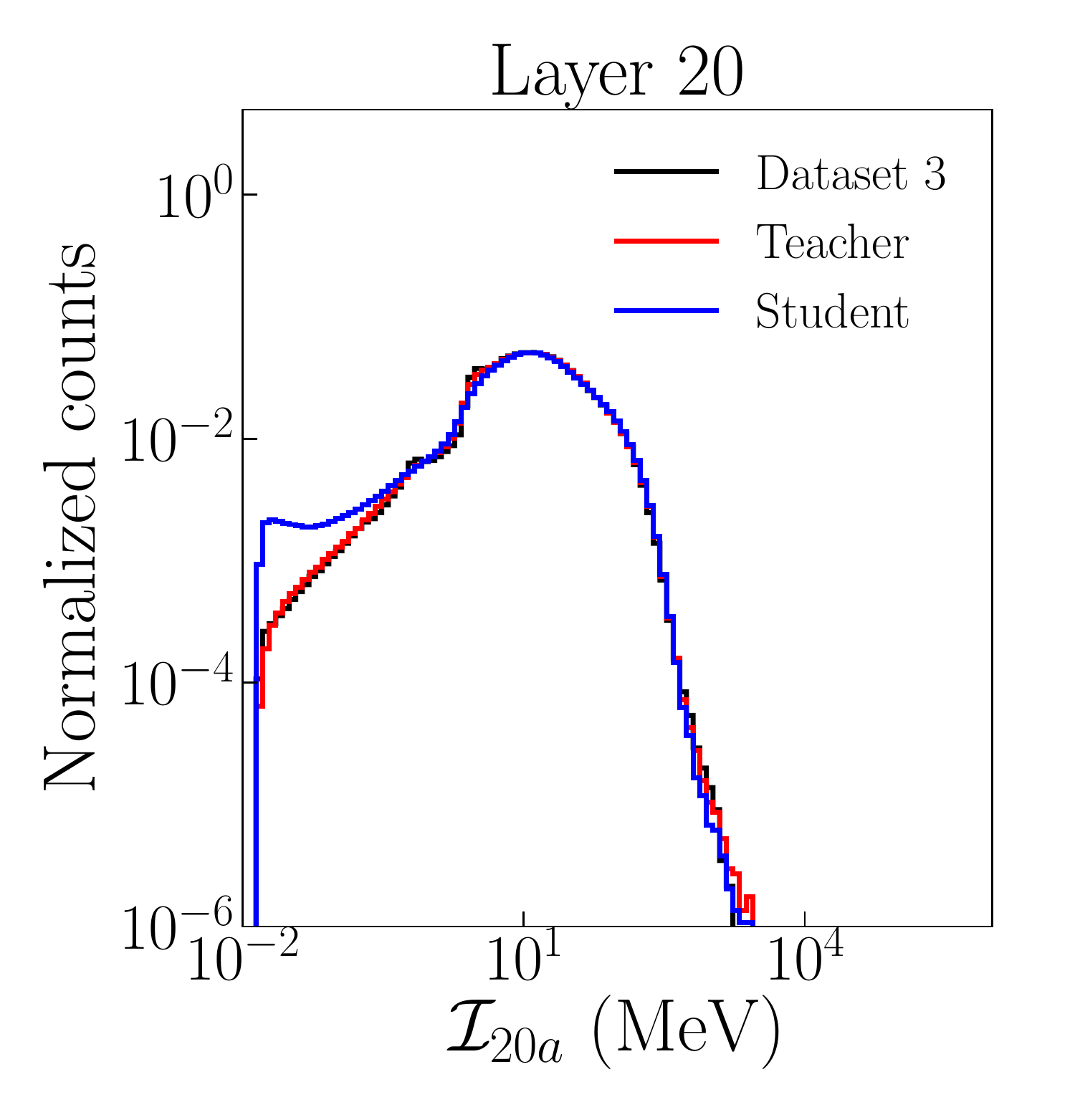}\includegraphics[width=0.5\columnwidth]{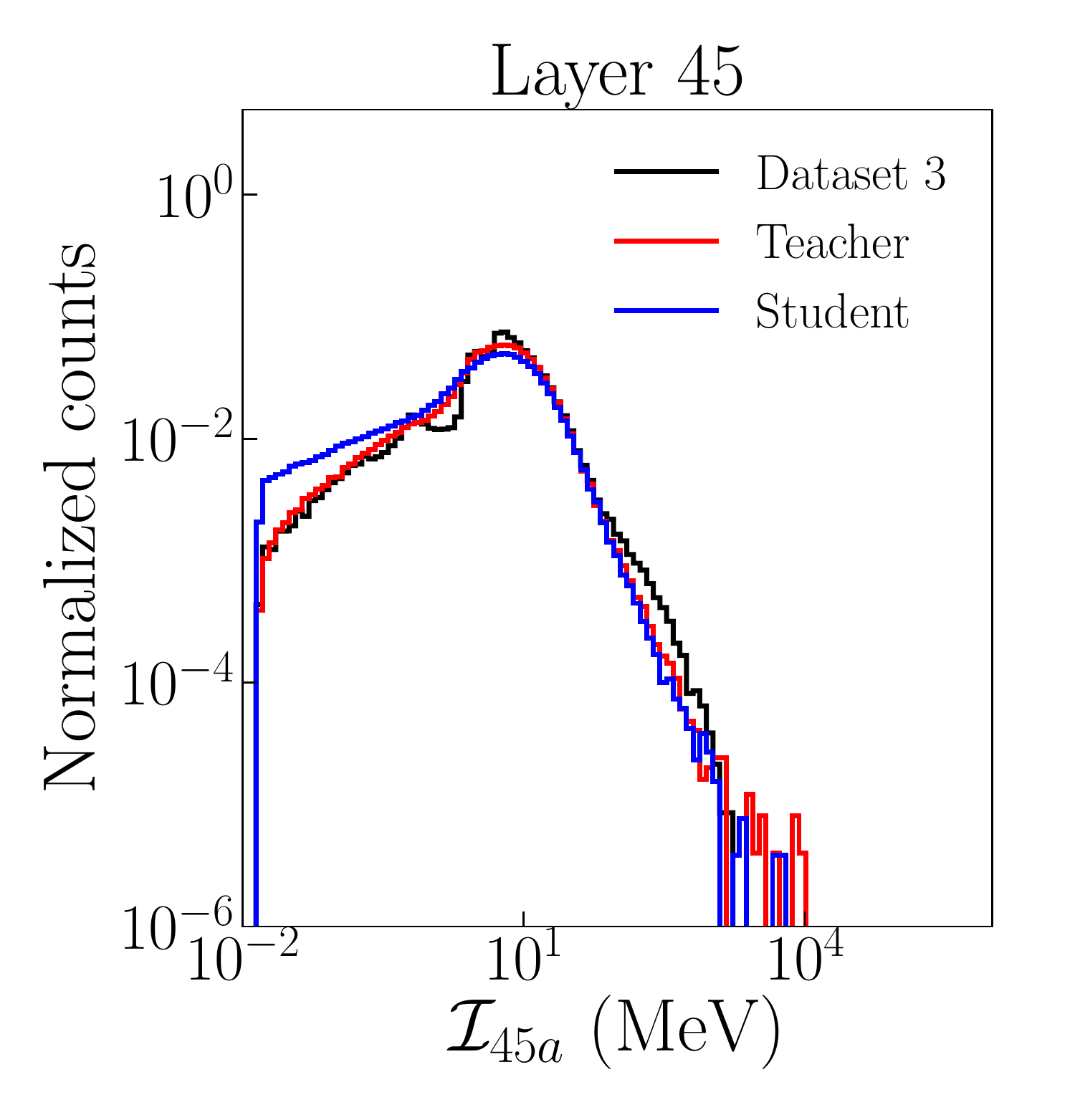}

\caption{Histograms of energy deposition per voxel in the layers 1, 10, 20, and 45 (from left to right) for Dataset 2 (upper row) and Dataset 3 (lower row). 
Distributions of \geant\ data are shown as black lines, and those of \icalo\ teacher (student) trained on Dataset 2 or 3 (as appropriate) in red (blue). \label{fig:layer1_energies}}
\end{figure*}

\begin{figure*}[ht]
\includegraphics[width=0.7\columnwidth]{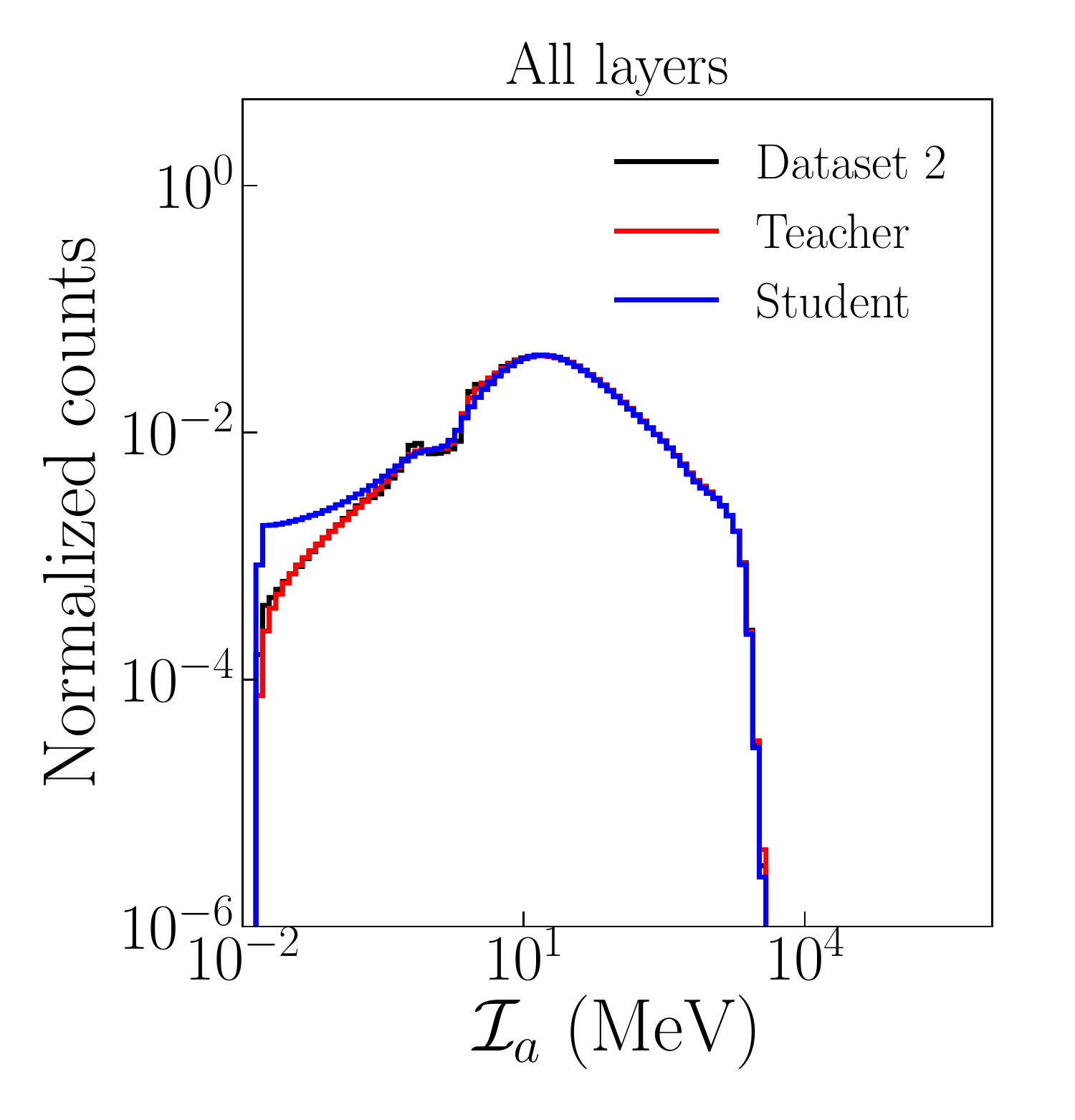}\includegraphics[width=0.7\columnwidth]{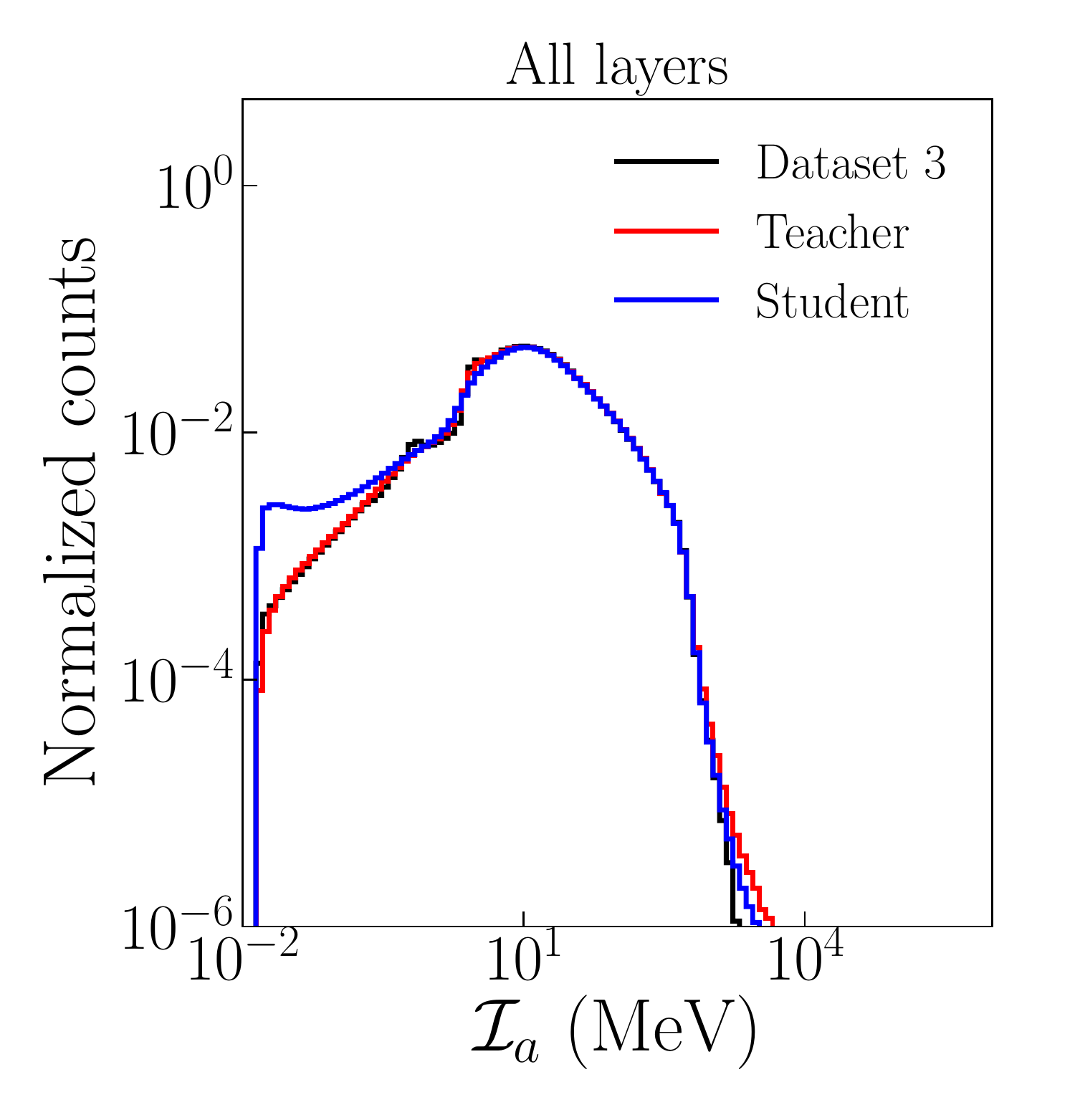}
\caption{Histograms of energy deposition per voxel in all layers for Dataset 2 (left) and Dataset 3 (right). 
Distributions of \geant\ data are shown as black lines, and those of \icalo\ teacher (student) trained on Dataset 2 or 3 (as appropriate) in red (blue). \label{fig:voxel_energies}}
\end{figure*}

In Figure~\ref{fig:flowone_generated}, we show the total energy deposition $E_i$ within a layer for our selected set of layers ($i = 1,10,20,45$). Here we see good overall agreement between \geant{} and \icalo{} distribution, with the exception of Layer 1 due to our choice of postprocessing. In particular, our decision not to normalize the outputs of \flowtwo{} and \flowthree{} to unity results in a difference between the energy deposited in a layer $E_i$ and the proxy output of \flowone{}, $\tilde{E}_i$. On the other hand, we find that the teacher voxel energy distributions in Figures \ref{fig:layer1_energies} and \ref{fig:voxel_energies} are generally in good agreement with the \geant{} distributions. (As discussed further in Appendix~\ref{sec:preprocessing}, enforcing $E_i=\tilde{E}_i$ would ``fix" the distribution in Fig.~\ref{fig:flowone_generated}, but at the cost of creating an excess of low-energy voxels in Figs.~\ref{fig:layer1_energies} and \ref{fig:voxel_energies}.) However, the student distributions suffer from an excess at low voxel energies. We found that this discrepant behavior is largely due to our choice of noise that is added to voxel energies during preprocessing. However, the addition of noise is necessary in our setup to ensure that the flow does not only learn zero energy voxels~\cite{Krause:2021ilc}. Note that many voxels have zero energy deposition, which are not captured in this logarithmic plot.

\begin{figure*}[ht]
\includegraphics[width=0.5\columnwidth]{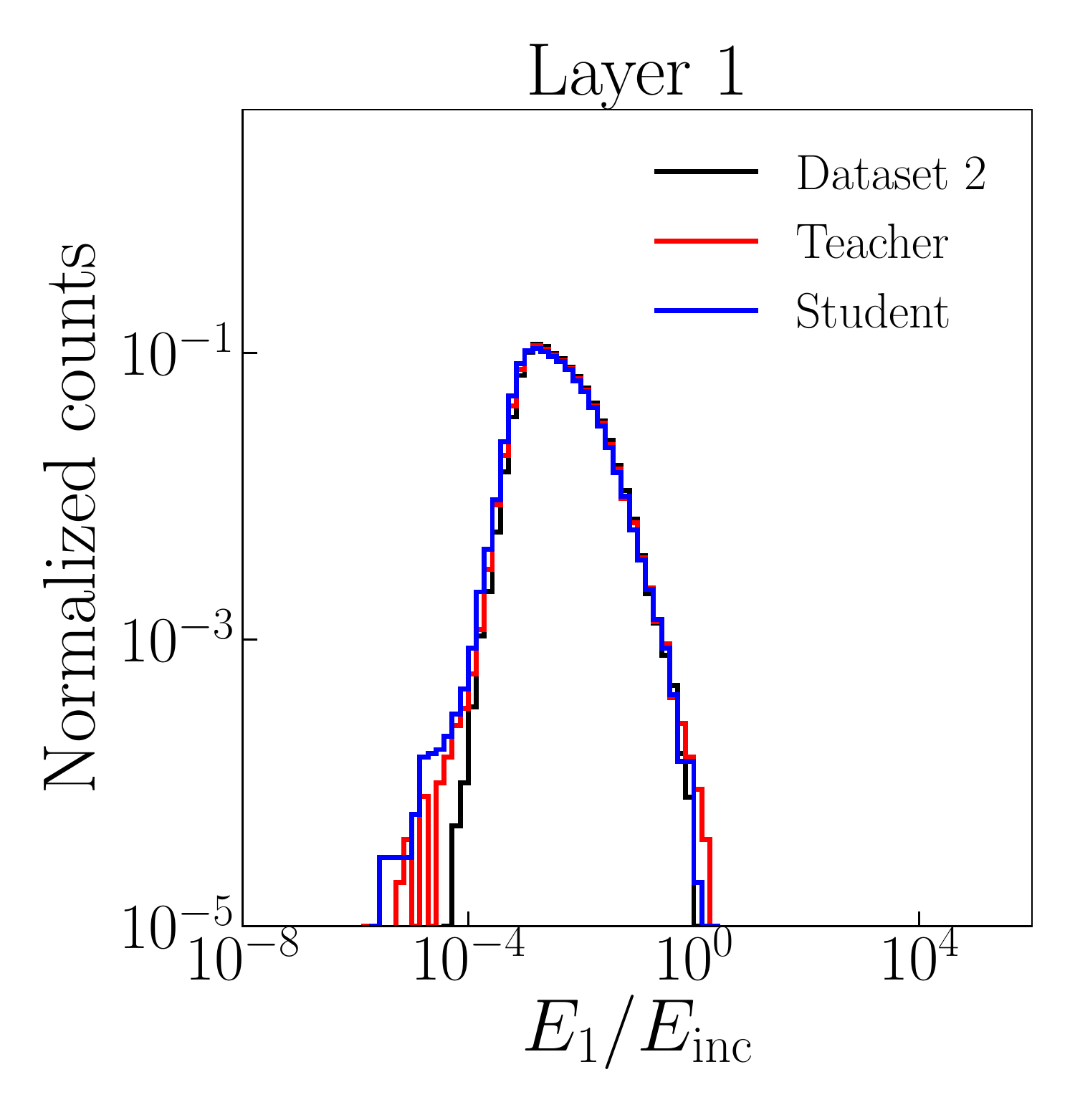}\includegraphics[width=0.5\columnwidth]{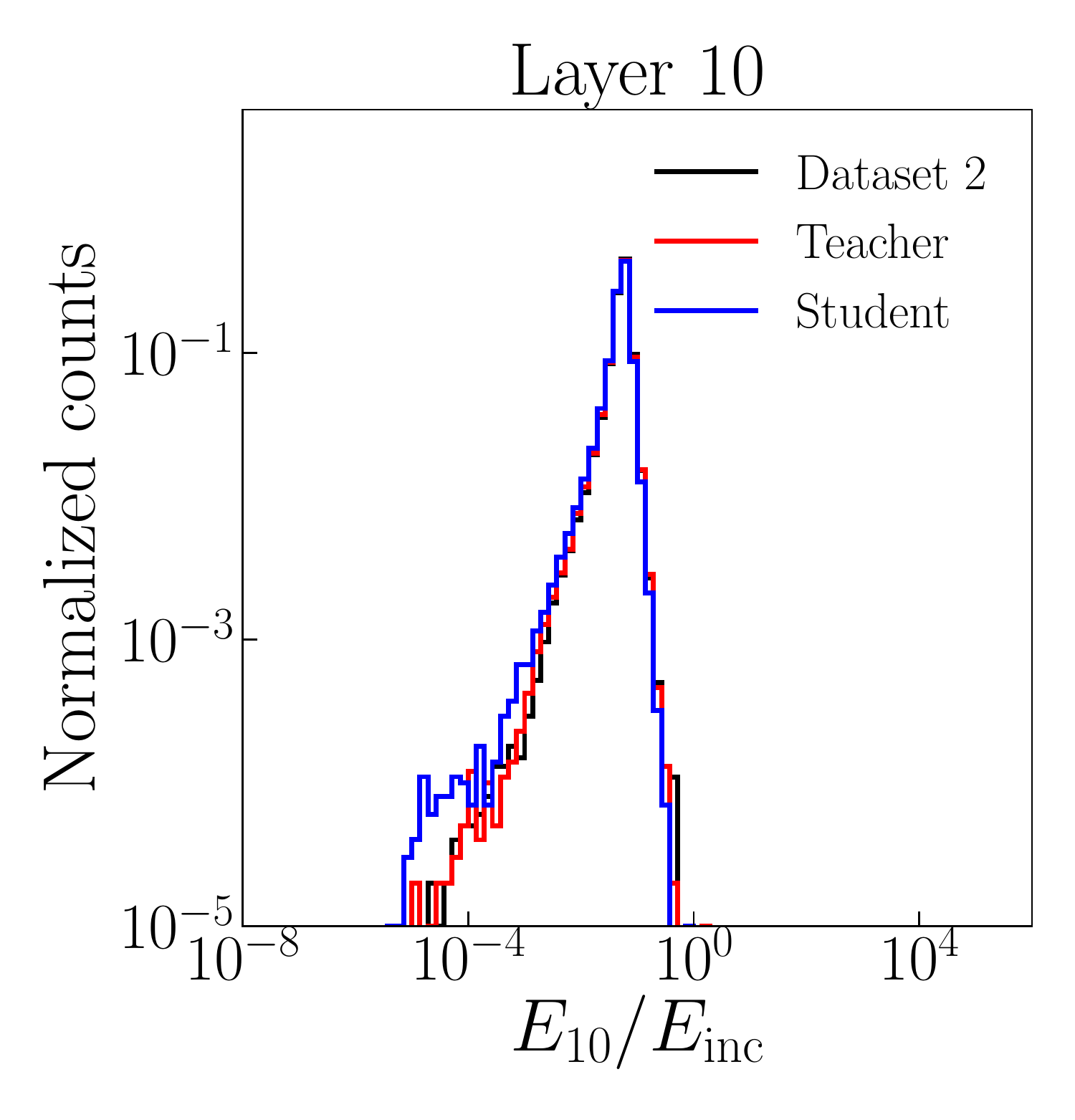}\includegraphics[width=0.5\columnwidth]{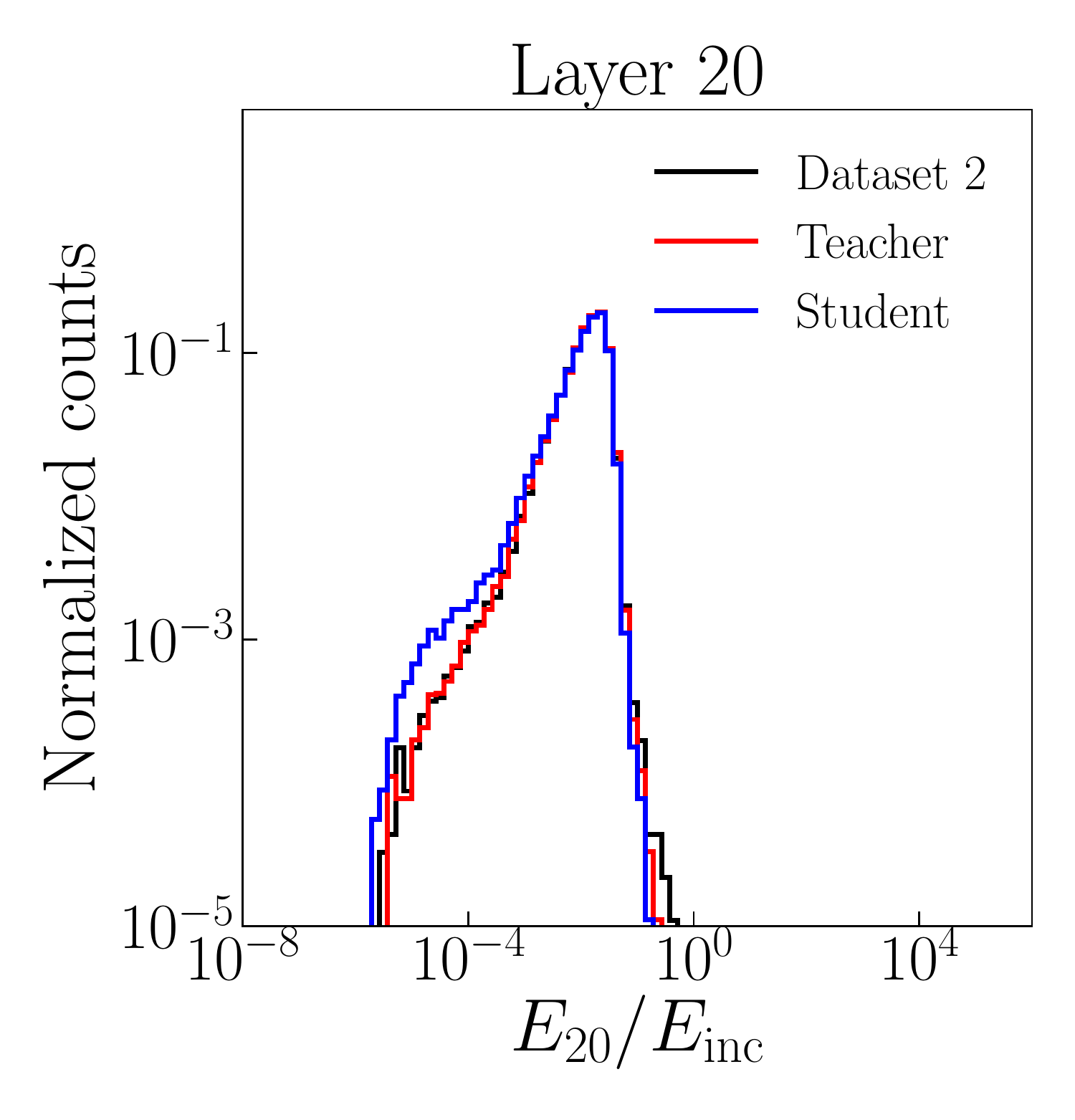}\includegraphics[width=0.5\columnwidth]{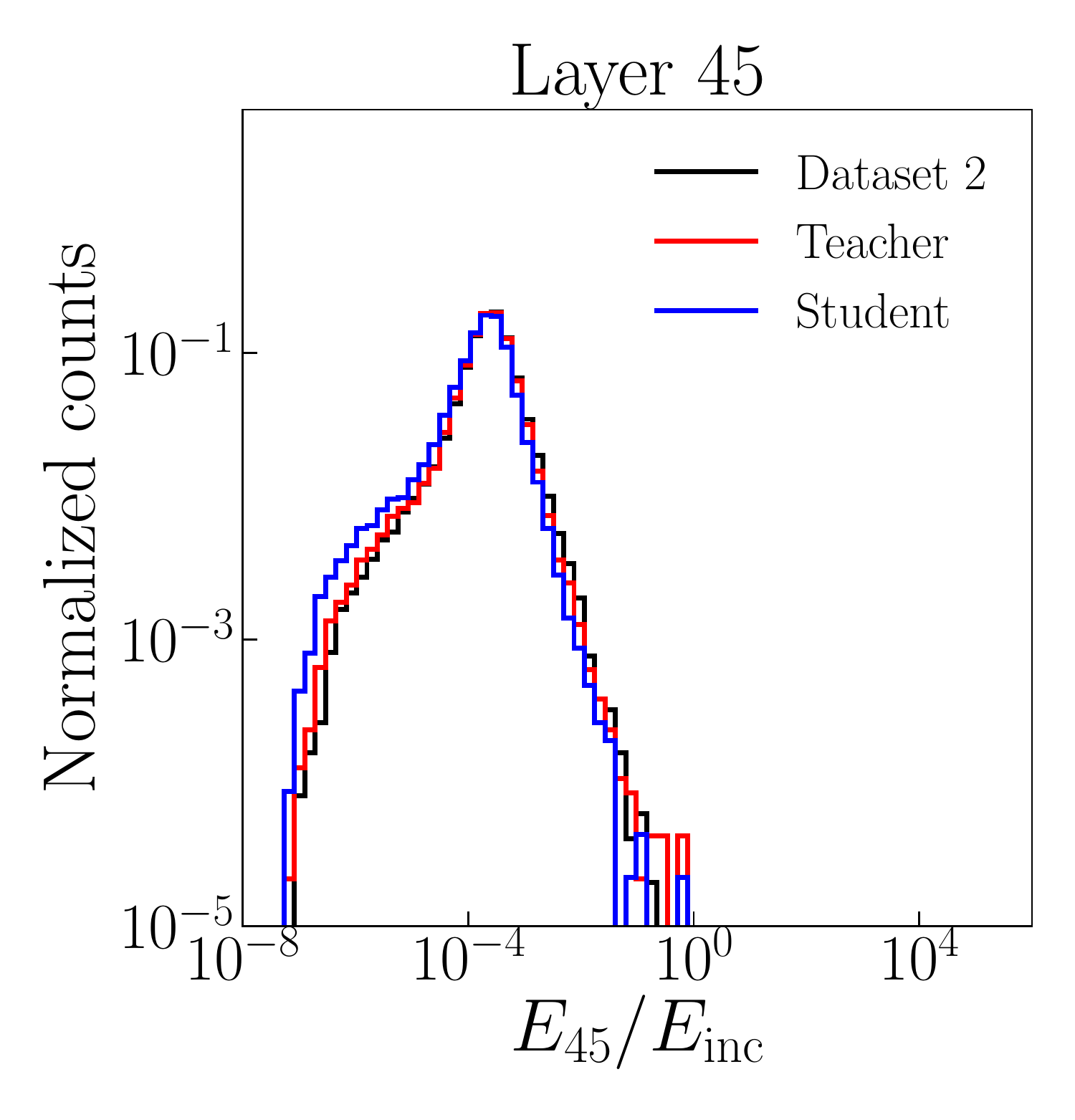}
\includegraphics[width=0.5\columnwidth]{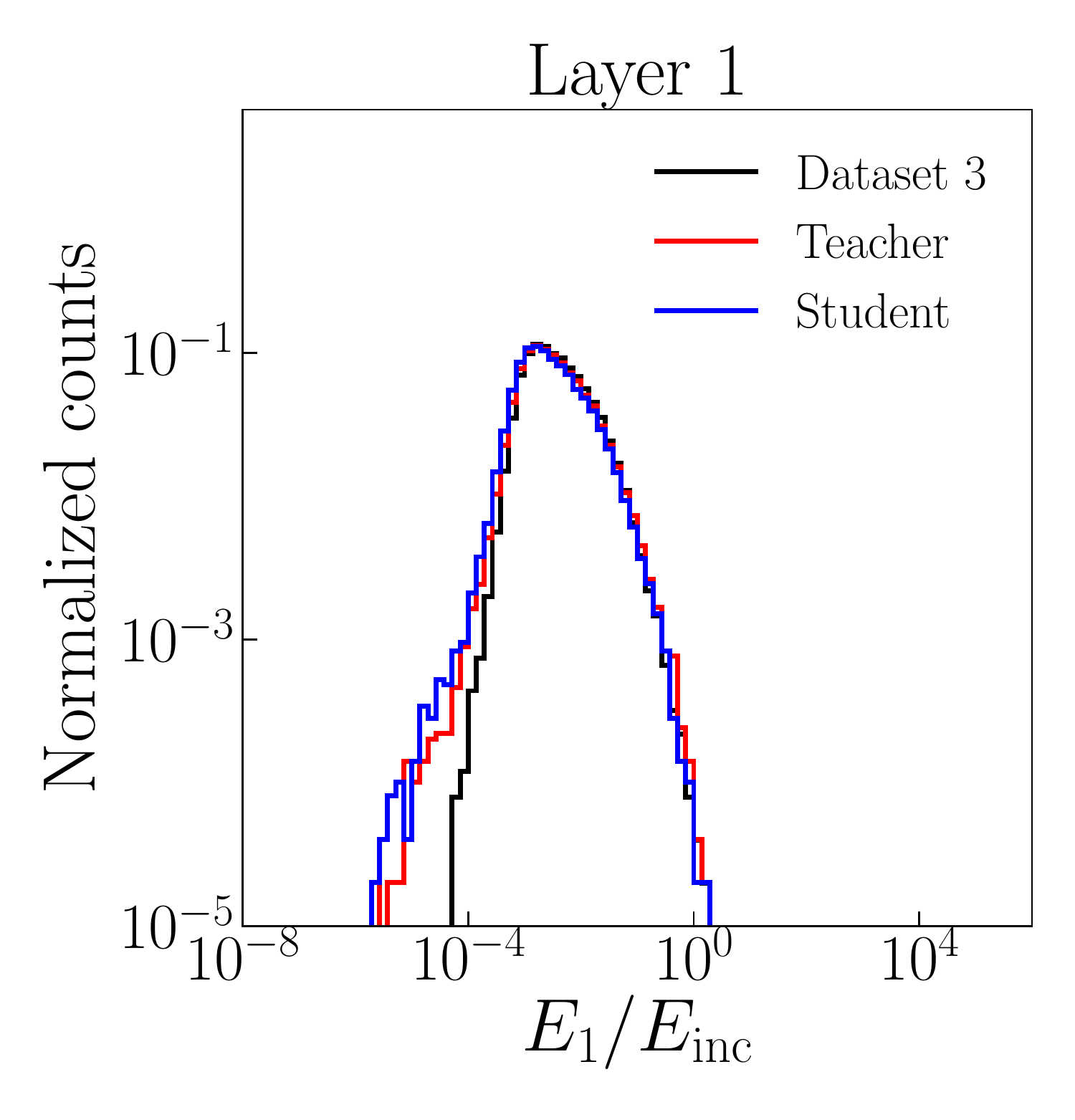}\includegraphics[width=0.5\columnwidth]{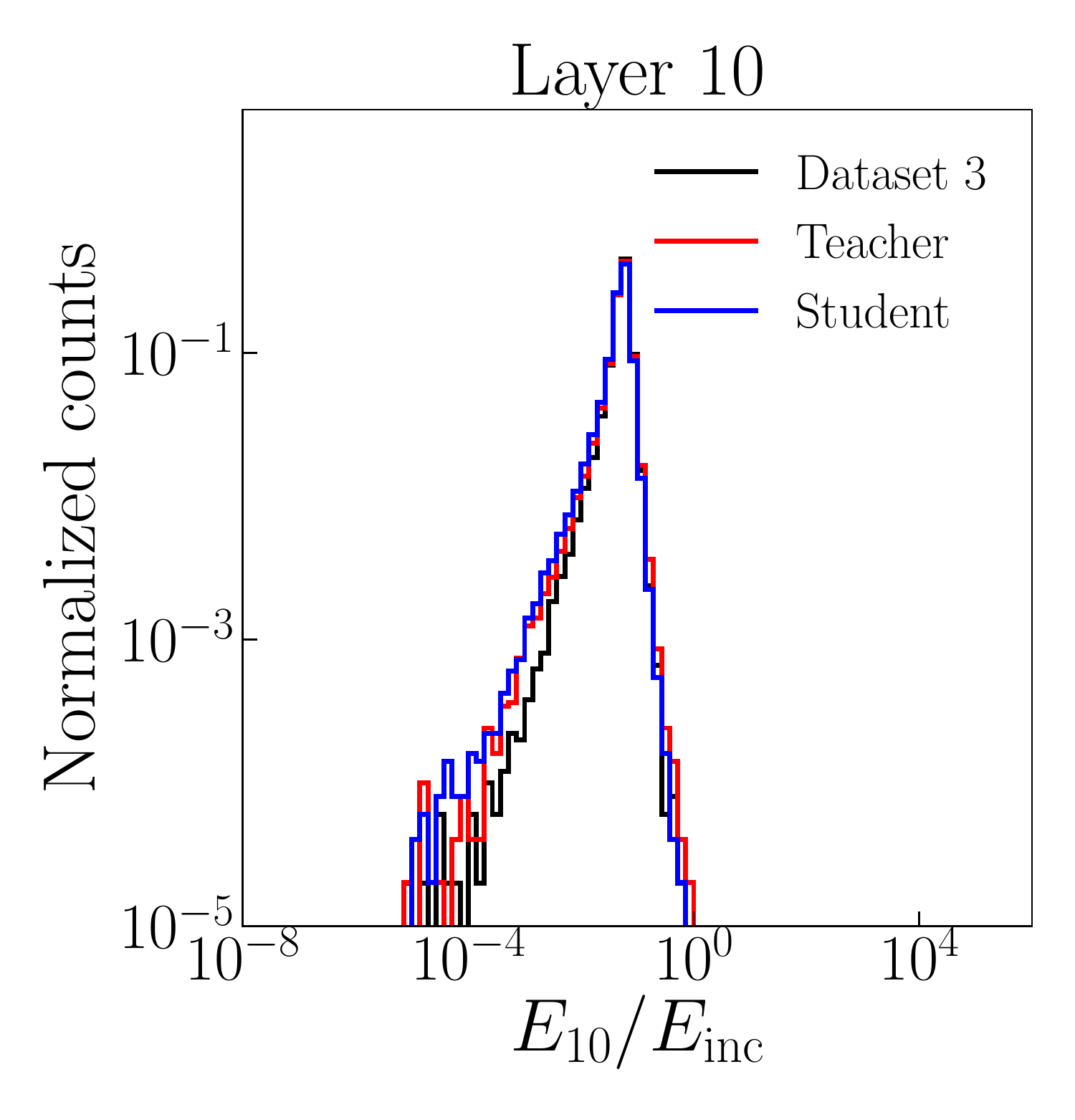}\includegraphics[width=0.5\columnwidth]{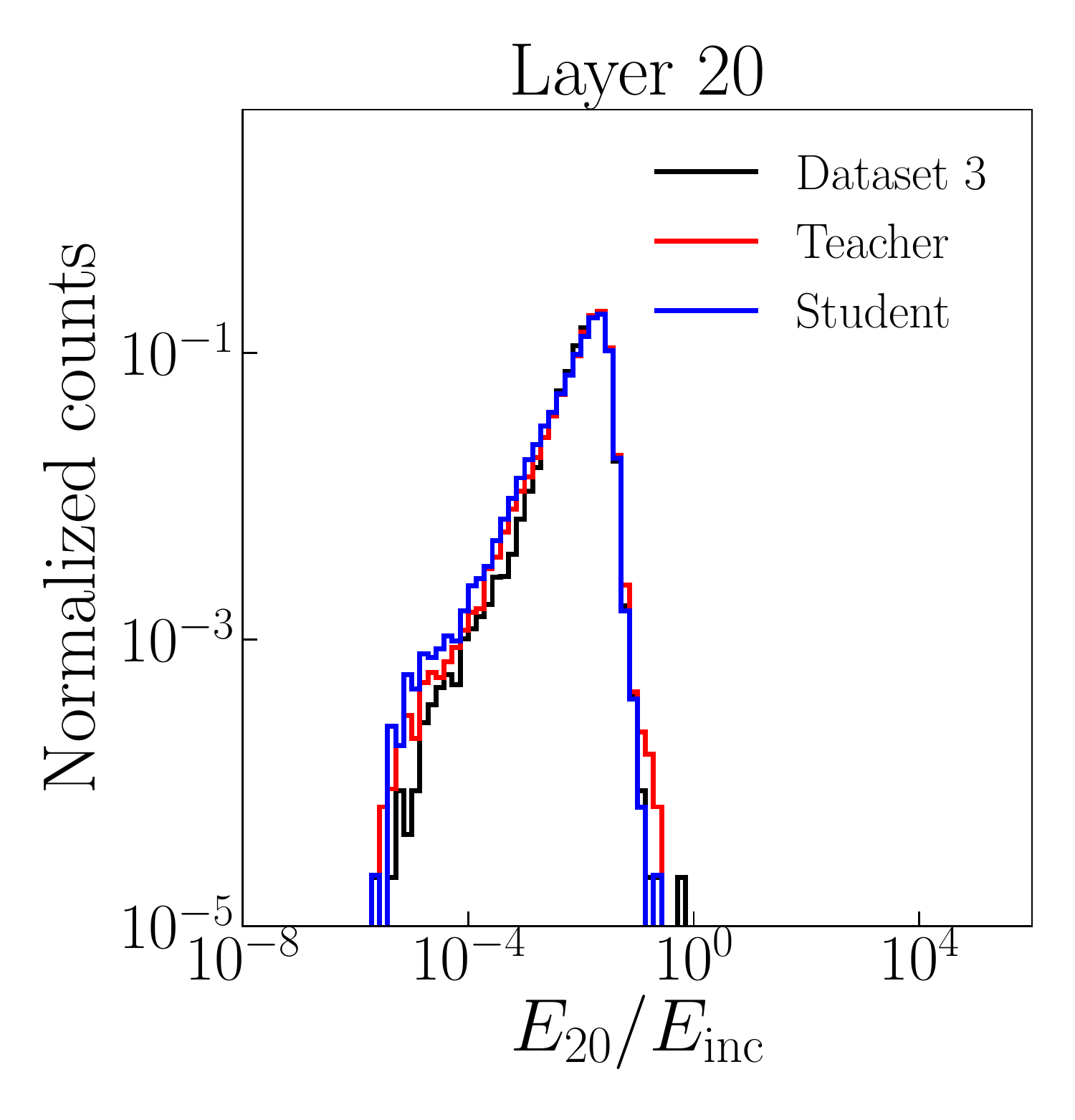}\includegraphics[width=0.5\columnwidth]{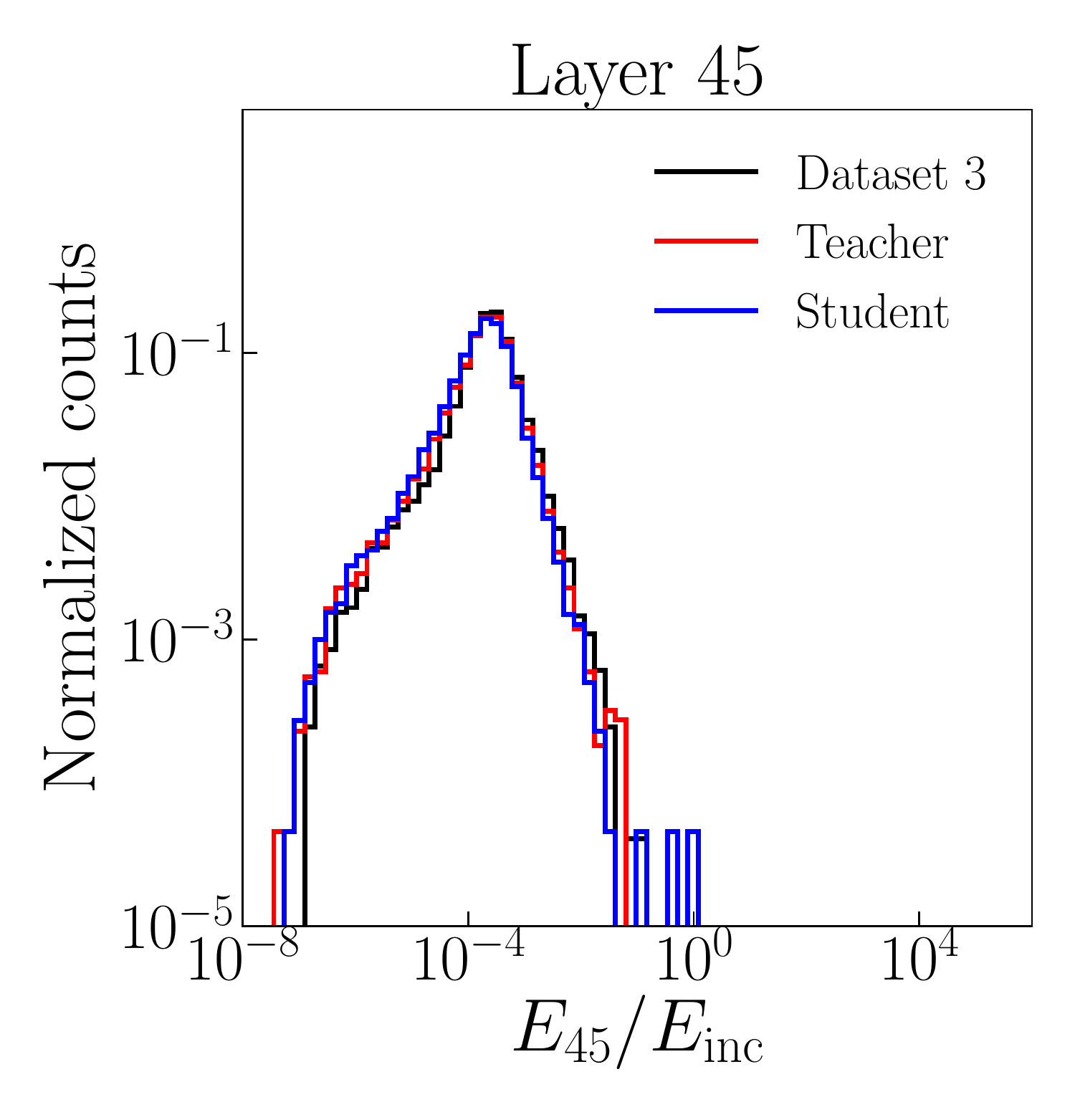}

\caption{Histograms of the ratio of total energy deposited and incident energy in a layer $i$ ($E_i$), for $i=1$, 10, 20, and 45 (from left to right), for Dataset 2 (top row) and Dataset 3 (bottom row). Distribution of \geant\ data is shown as black lines, and that of the \icalo\ teacher (student) trained on Dataset 2 or 3 (as appropriate) in red (blue)\label{fig:edep_einc_layers}}
\end{figure*}

\begin{figure*}[ht]
\includegraphics[width=0.7\columnwidth]{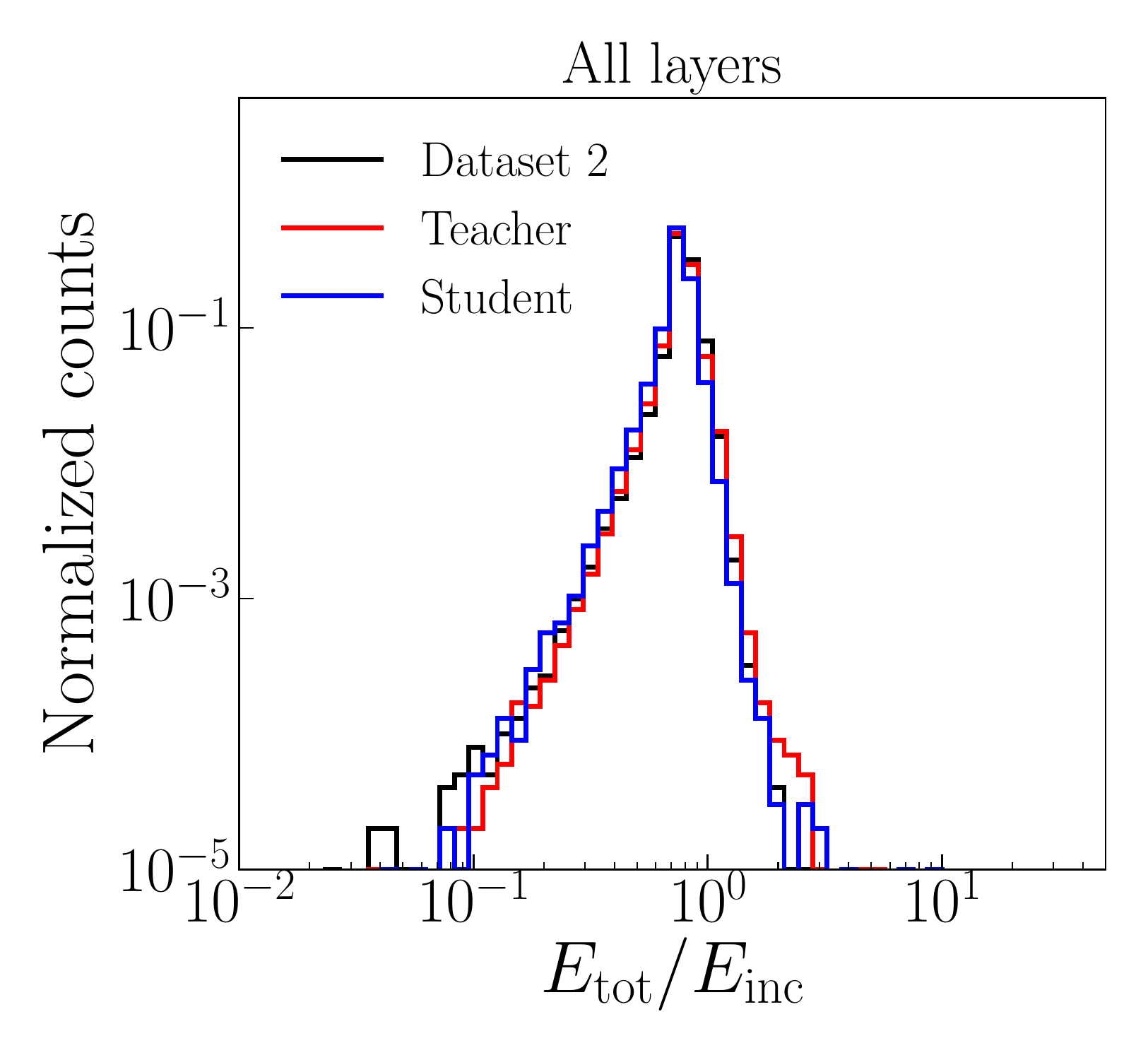}\includegraphics[width=0.7\columnwidth]{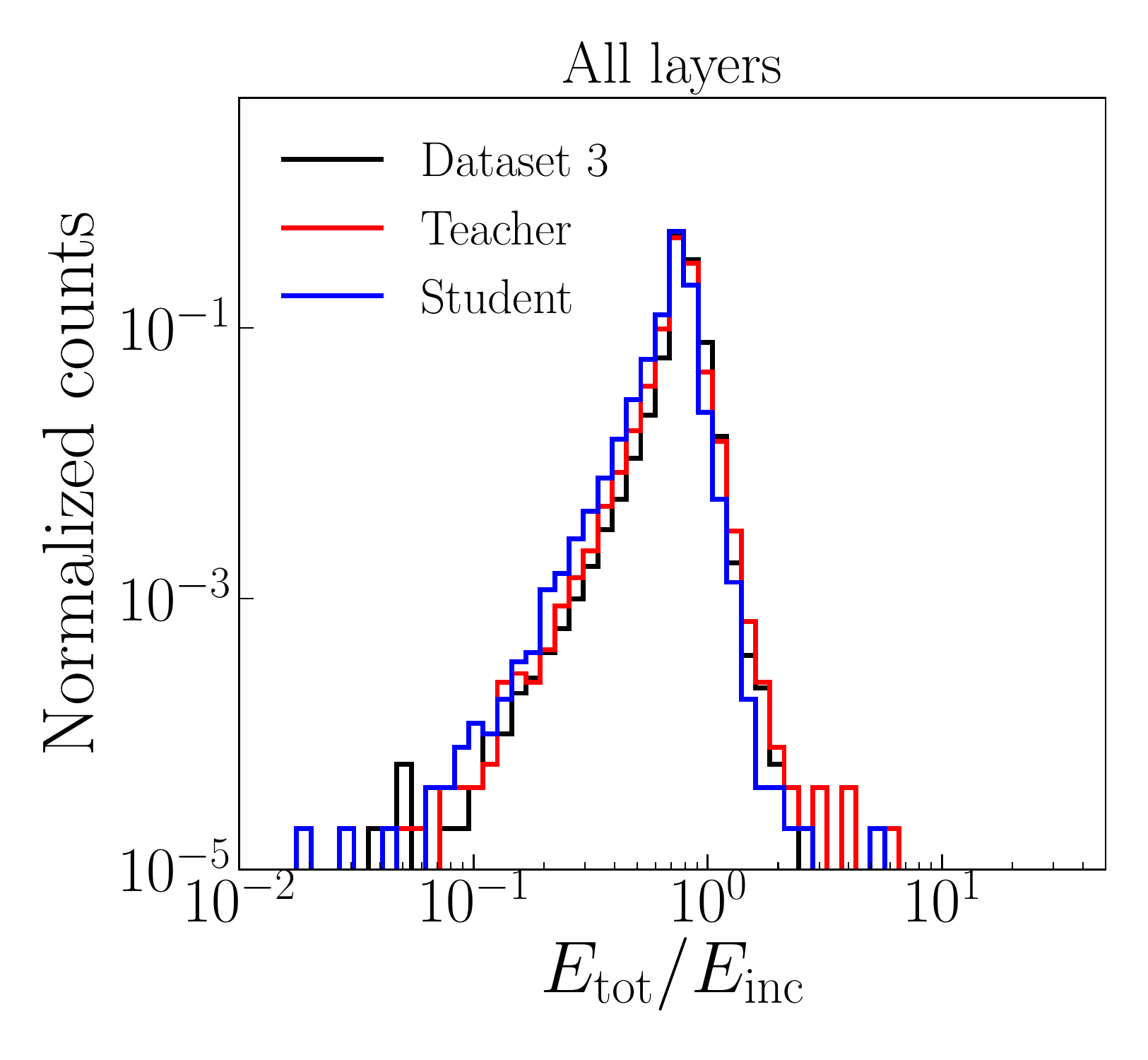}
\caption{Histograms of the ratio of total energy deposition (all layers) and incident energy for Dataset 2 (left) and Dataset 3 (right). 
Distributions of \geant\ data are shown as black lines, and those of \icalo\ teacher (student) trained on Dataset 2 or 3 (as appropriate) in red (blue). \label{fig:total_edep_einc}}
\end{figure*}

We show the ratio of energy deposited in a layer to the incident beam energy in Figure~\ref{fig:edep_einc_layers}, and the ratio of all deposited energy to the incident beam energy in Figure~\ref{fig:total_edep_einc}. There is good overall agreement between \icalo{} and \geant{} events despite some excess at low ratios in the earlier layers. We observed that the showers in the low ratio excess in Layer 1 in these figures are the same showers found in the low energy excess in Layer 1 shown in  Figure~\ref{fig:flowone_generated}. These showers are characterized by having mostly zero energy voxels with a few bright voxels, this makes it difficult for the flow to learn the underlying distribution.

\begin{figure*}[ht]
\includegraphics[width=0.5\columnwidth]{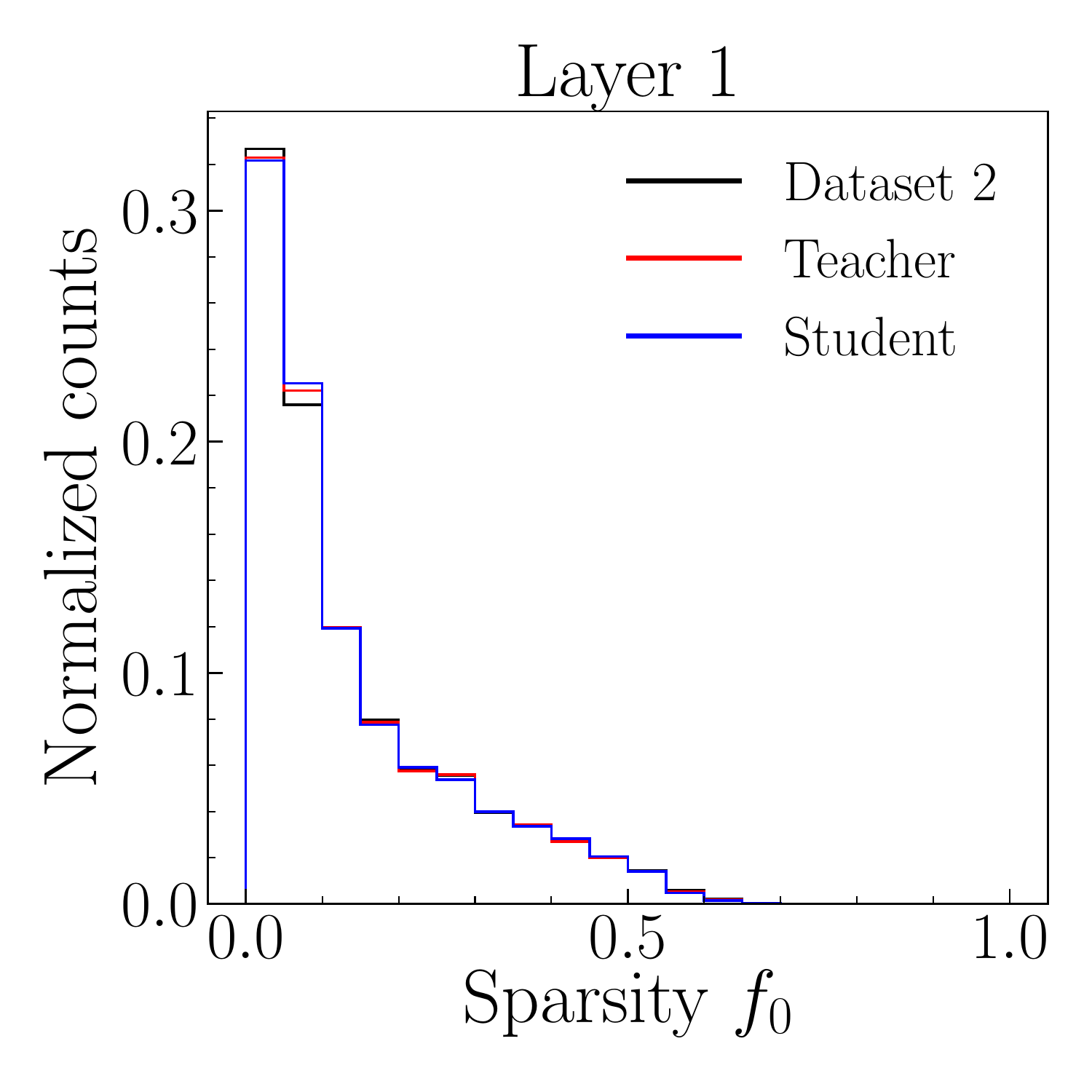}\includegraphics[width=0.5\columnwidth]{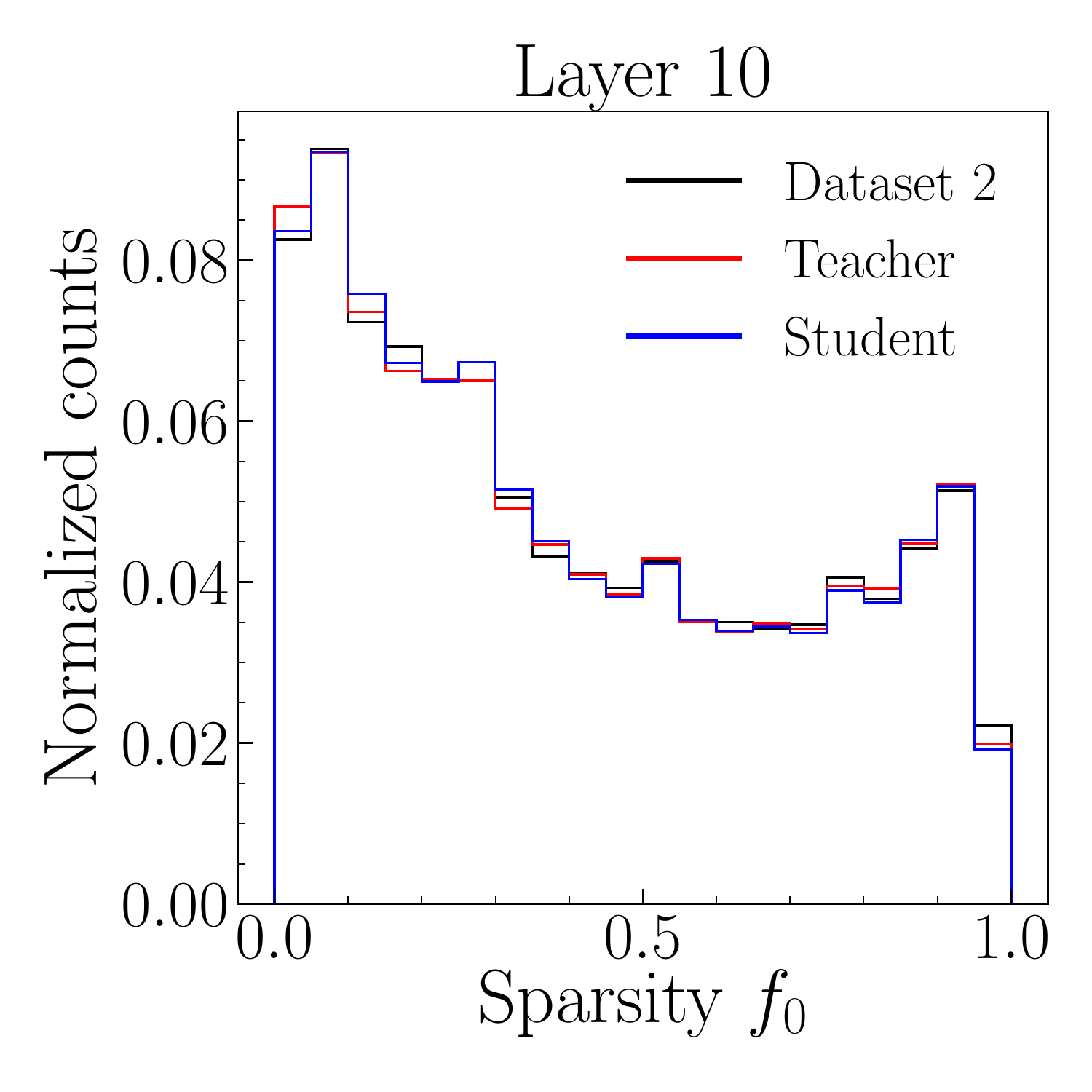}\includegraphics[width=0.5\columnwidth]{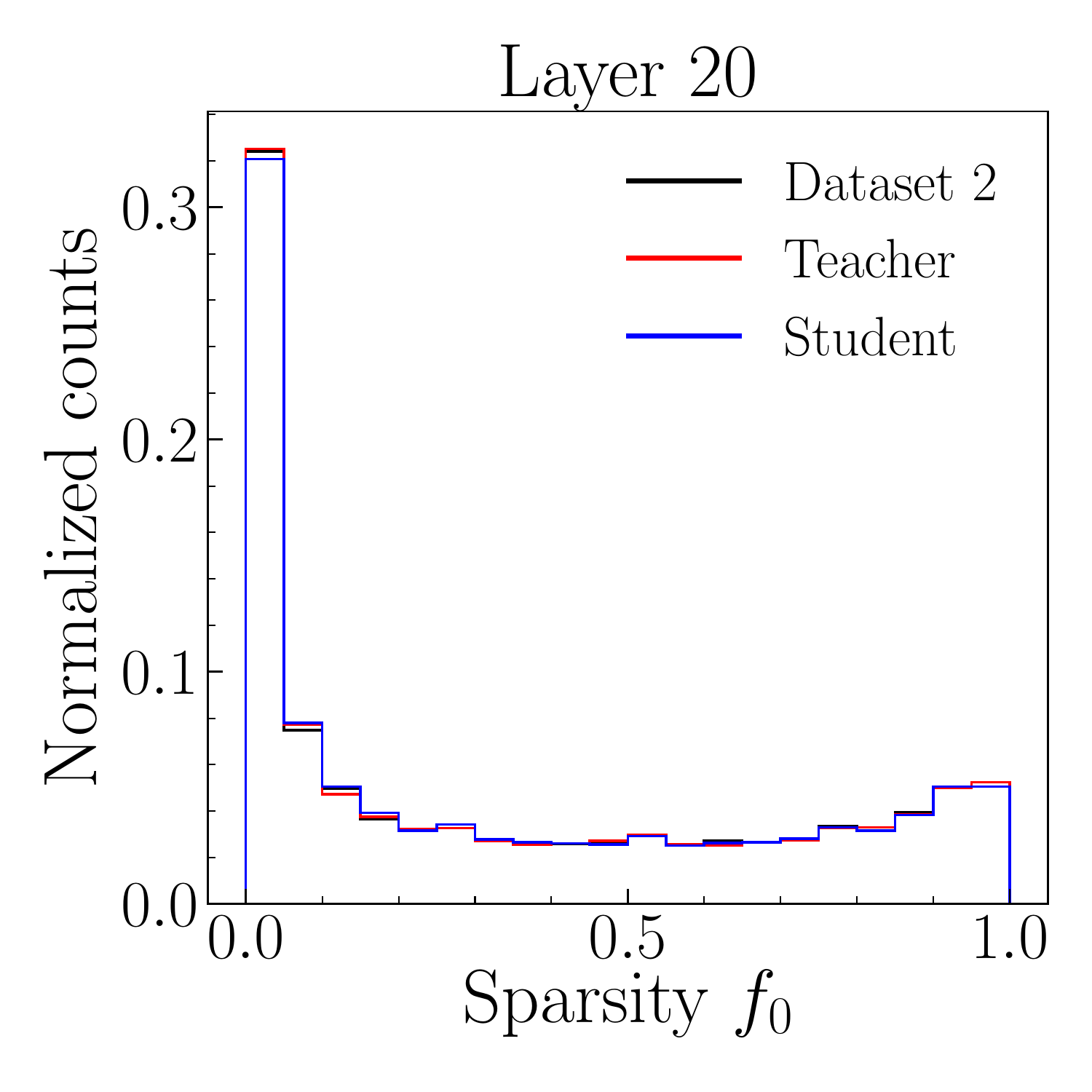}\includegraphics[width=0.5\columnwidth]{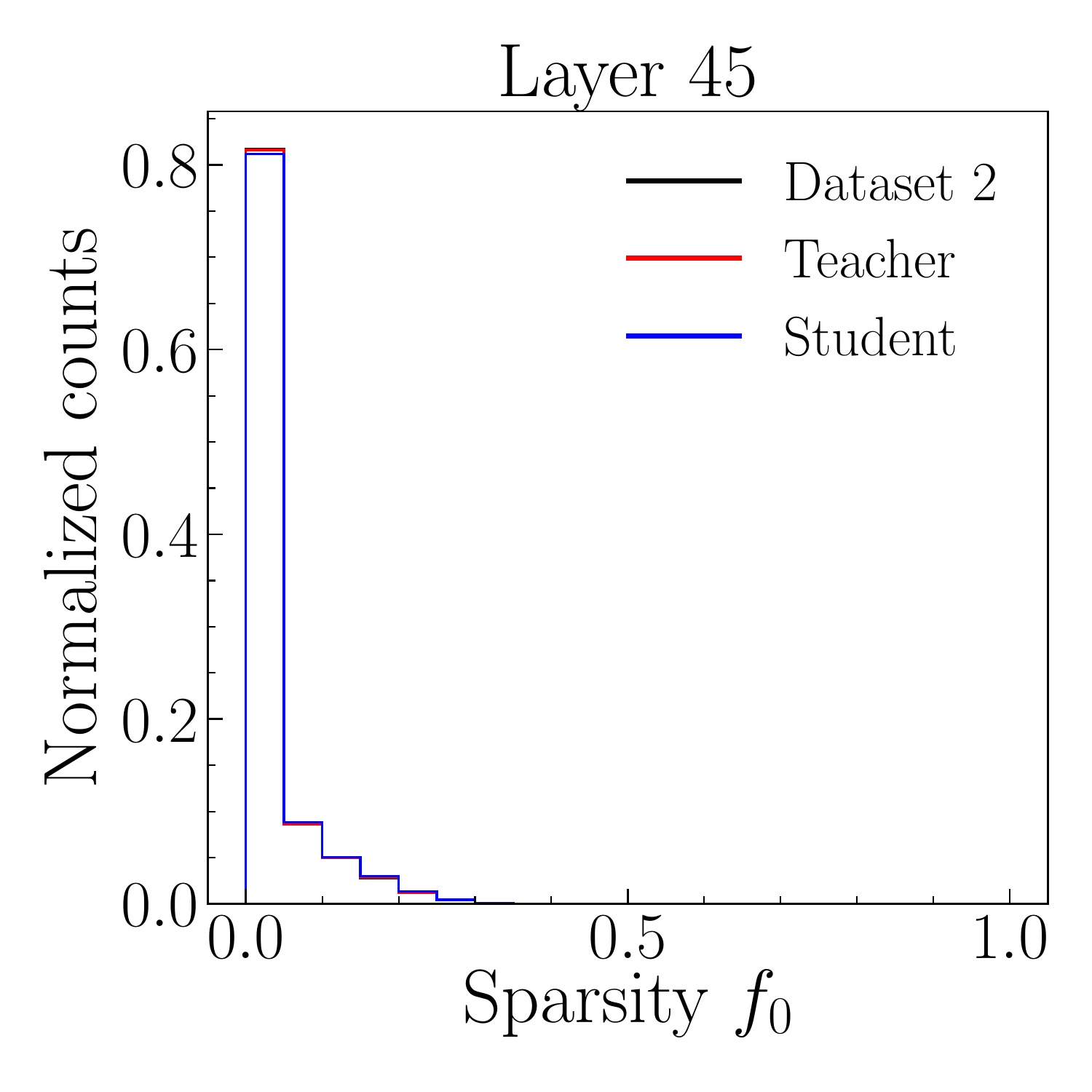}

\includegraphics[width=0.5\columnwidth]{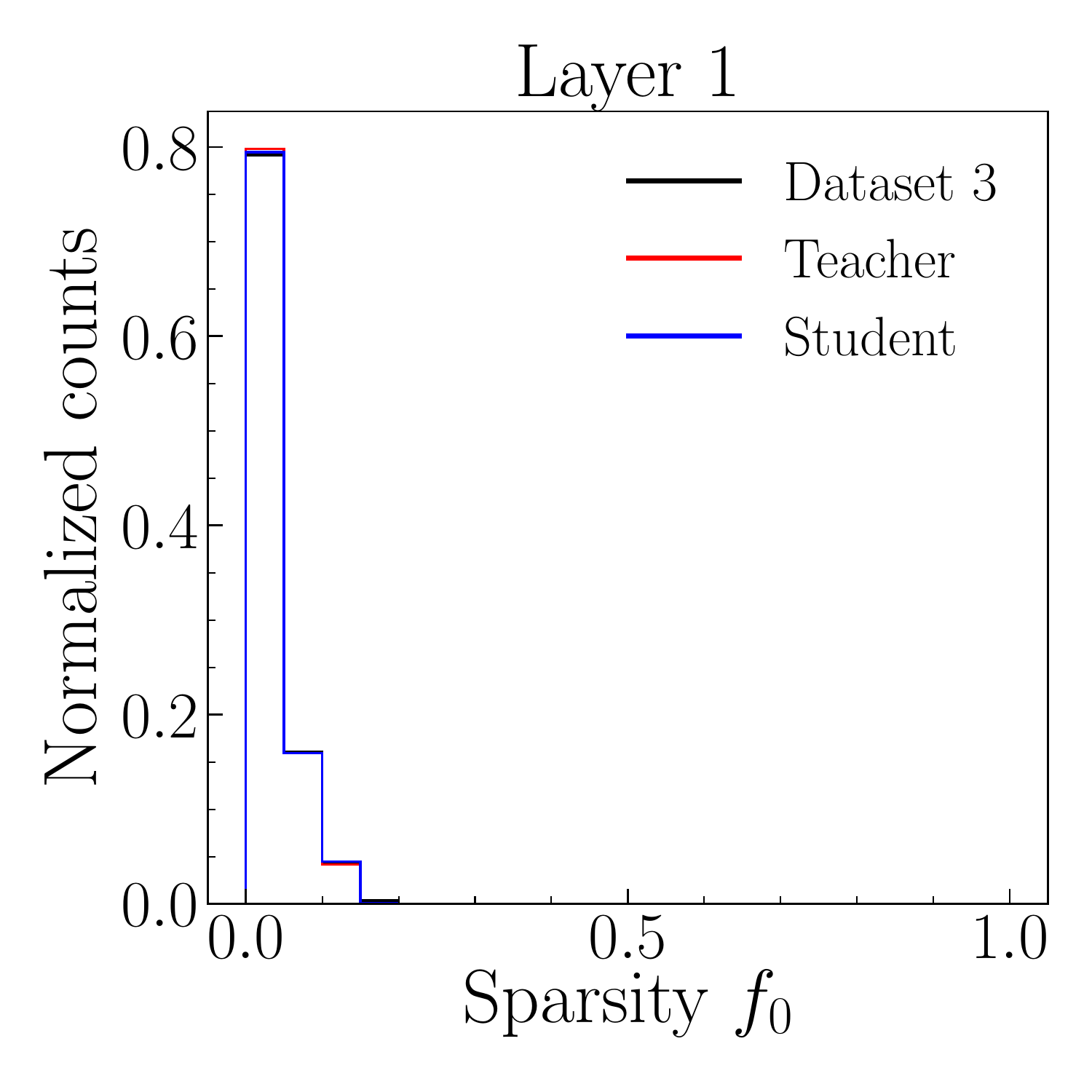}\includegraphics[width=0.5\columnwidth]{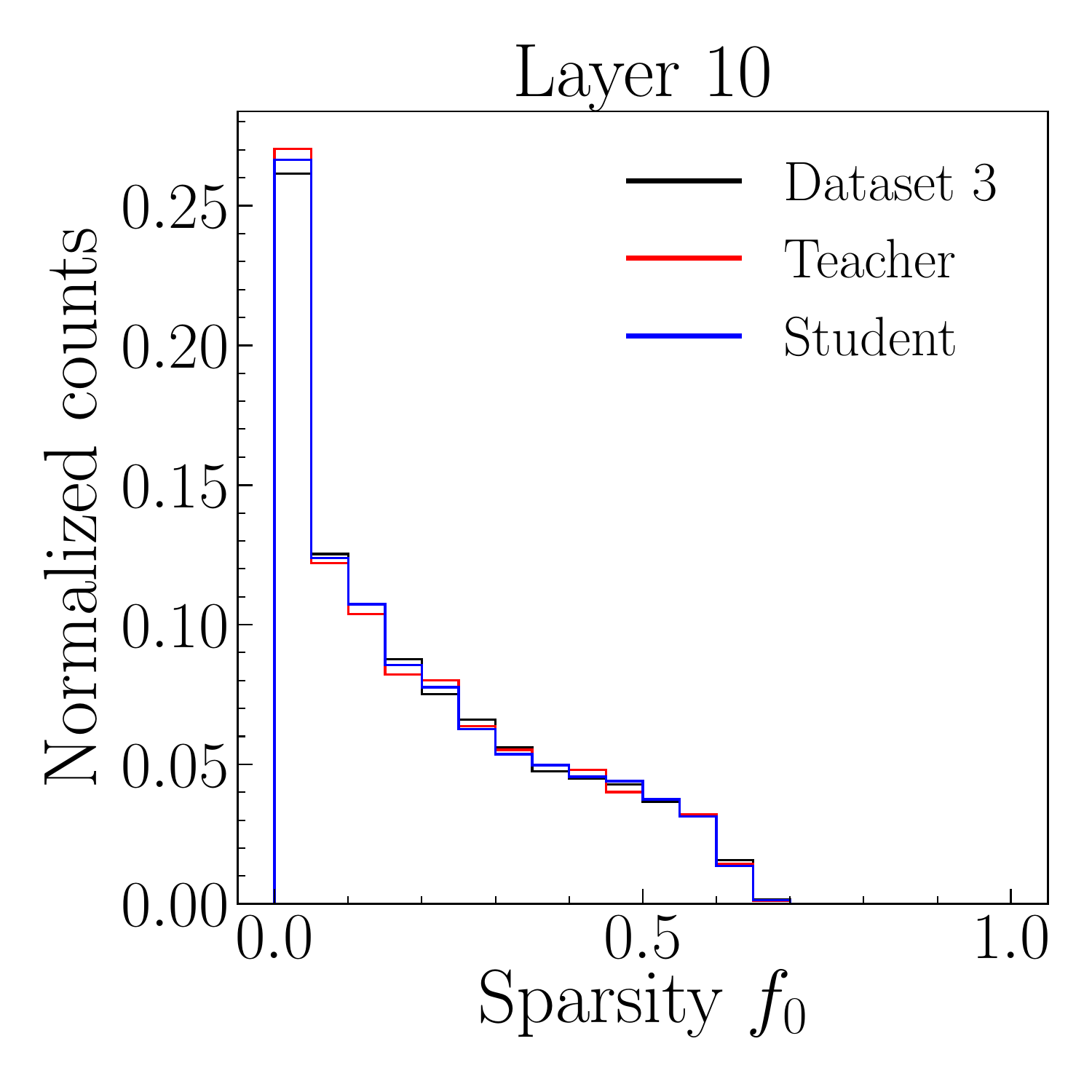}\includegraphics[width=0.5\columnwidth]{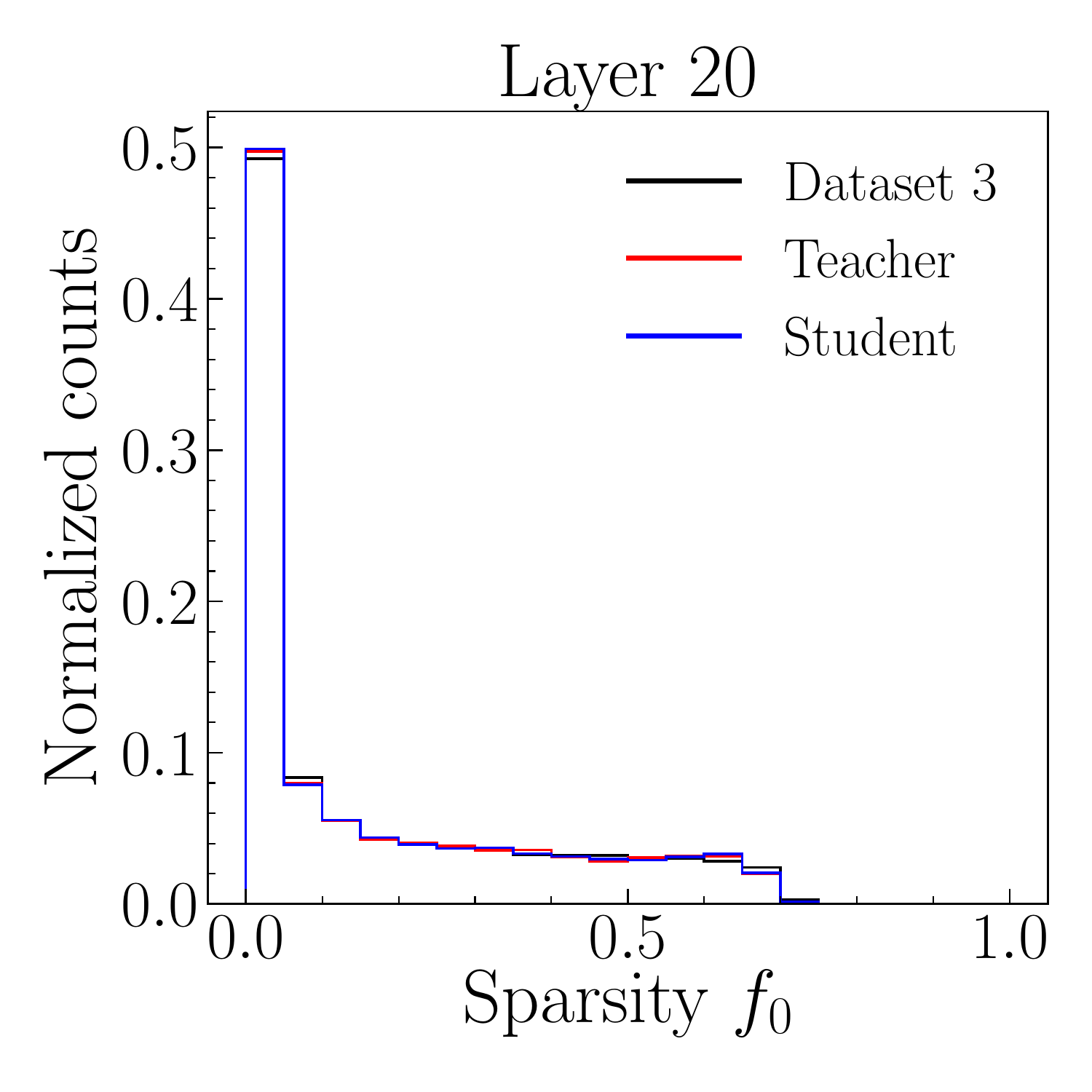}\includegraphics[width=0.5\columnwidth]{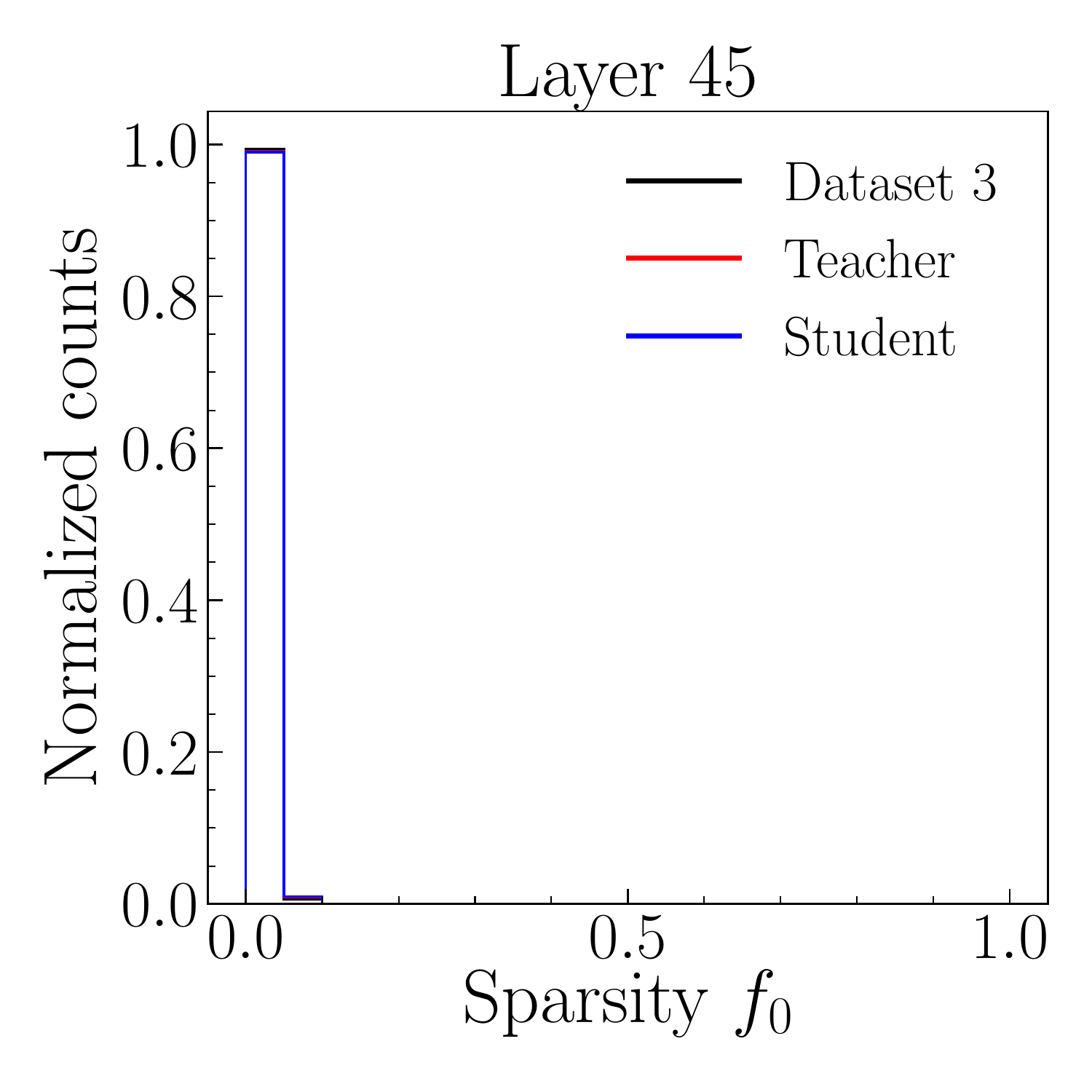}

\caption{Histograms of fraction of voxels in layers 1, 10, 20, and 45 (from left to right) which have non-zero energy deposition $f_0$ for Dataset 2 (upper row) and Dataset 3 (lower row). 
Distributions of \geant\ data are shown as black lines, and those of the \icalo\ teacher (student) trained on Dataset 2 or 3 (as appropriate) in red (blue). \label{fig:layer1_sparsity}}
\end{figure*}

\begin{figure*}[ht]
\includegraphics[width=0.75\columnwidth]{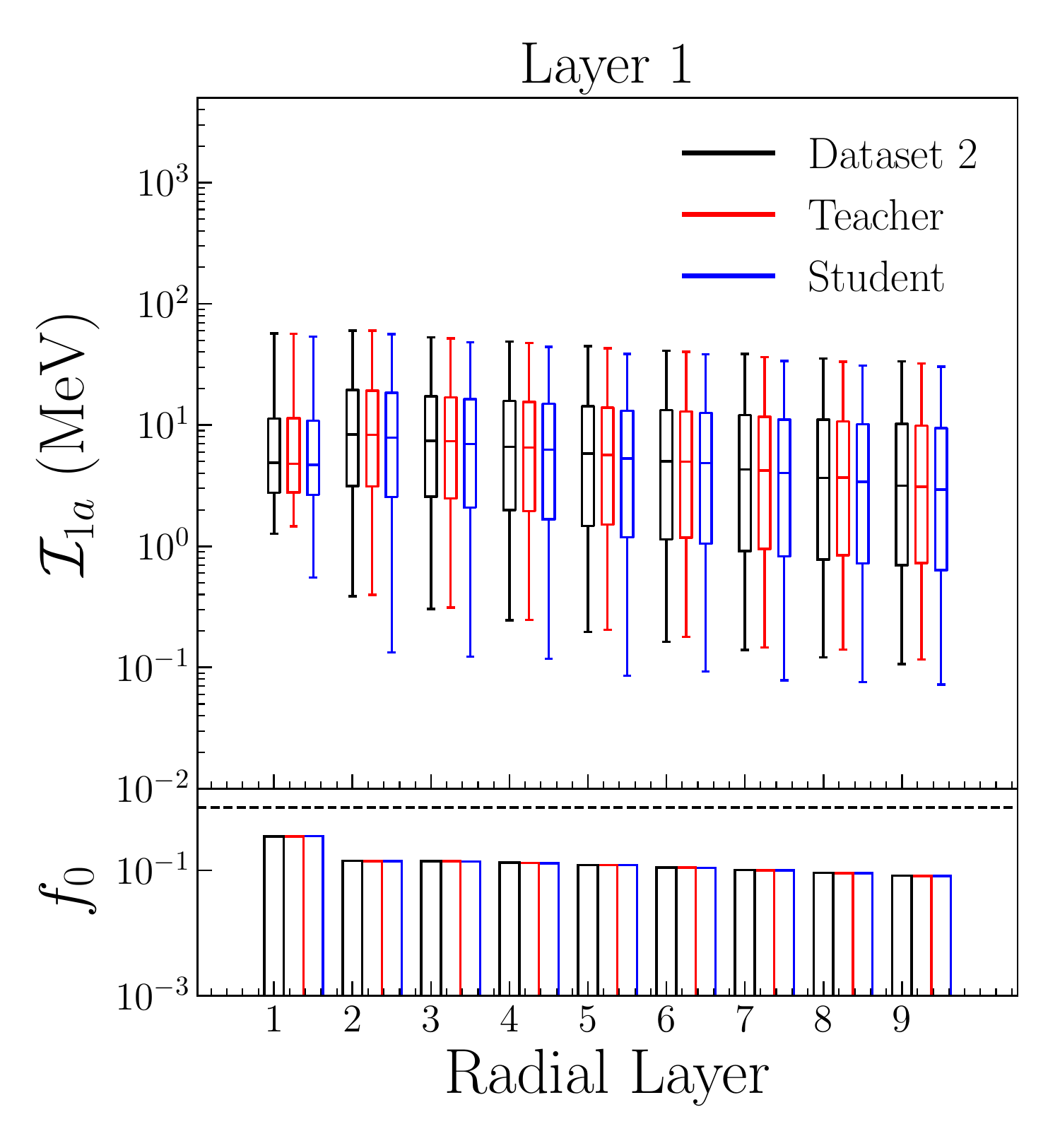}\includegraphics[width=0.75\columnwidth]{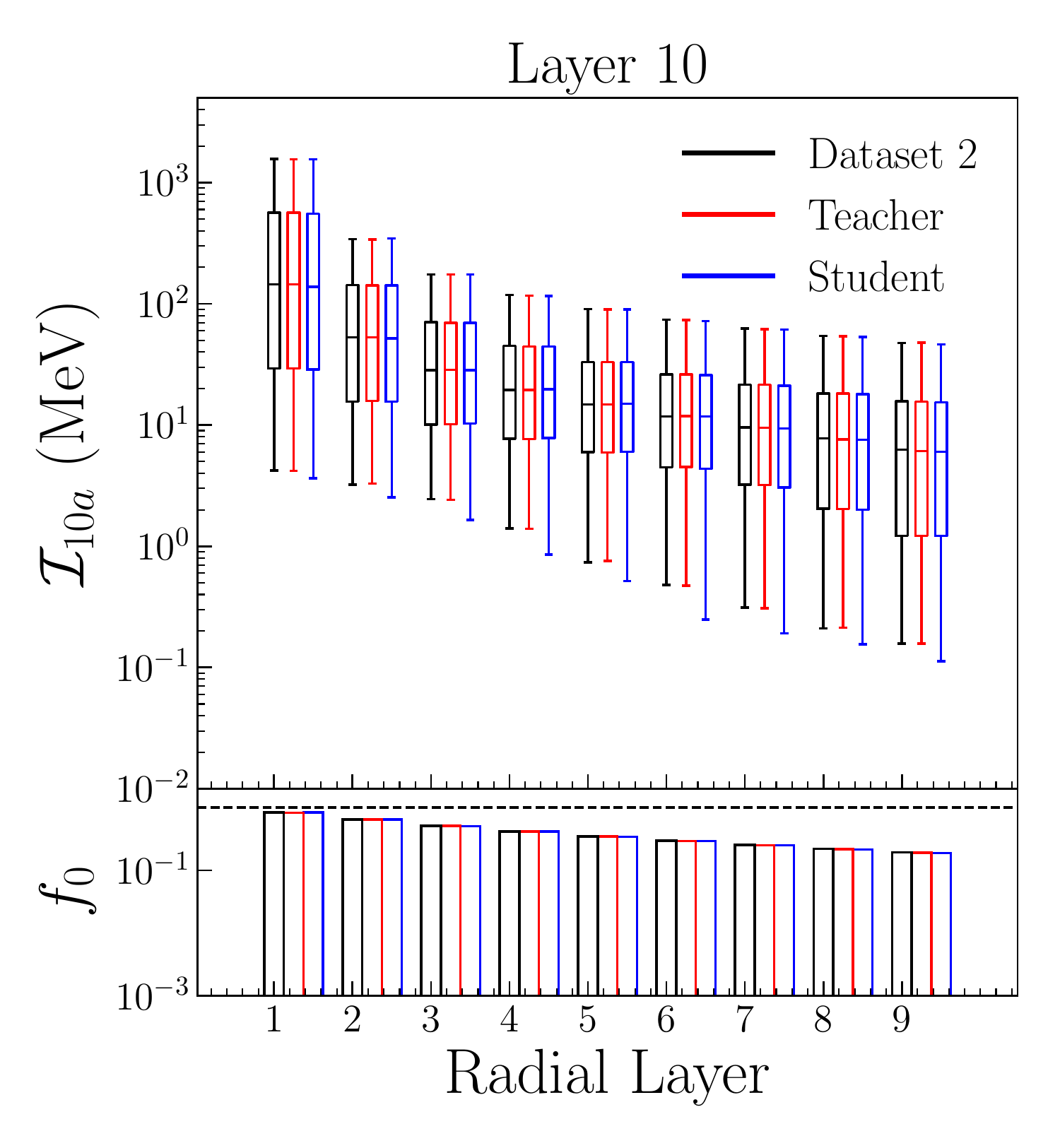}

\includegraphics[width=0.75\columnwidth]{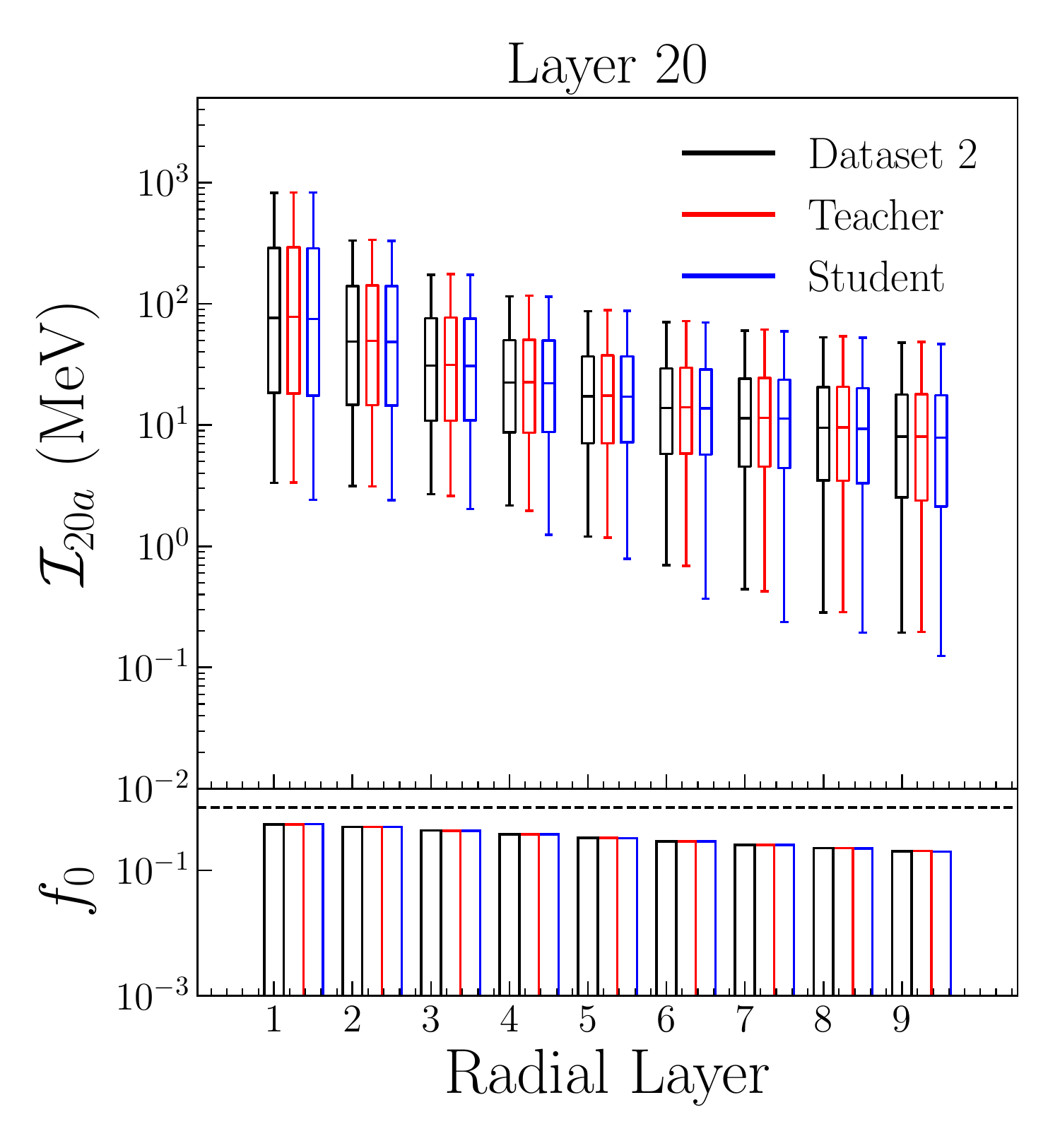}\includegraphics[width=0.75\columnwidth]{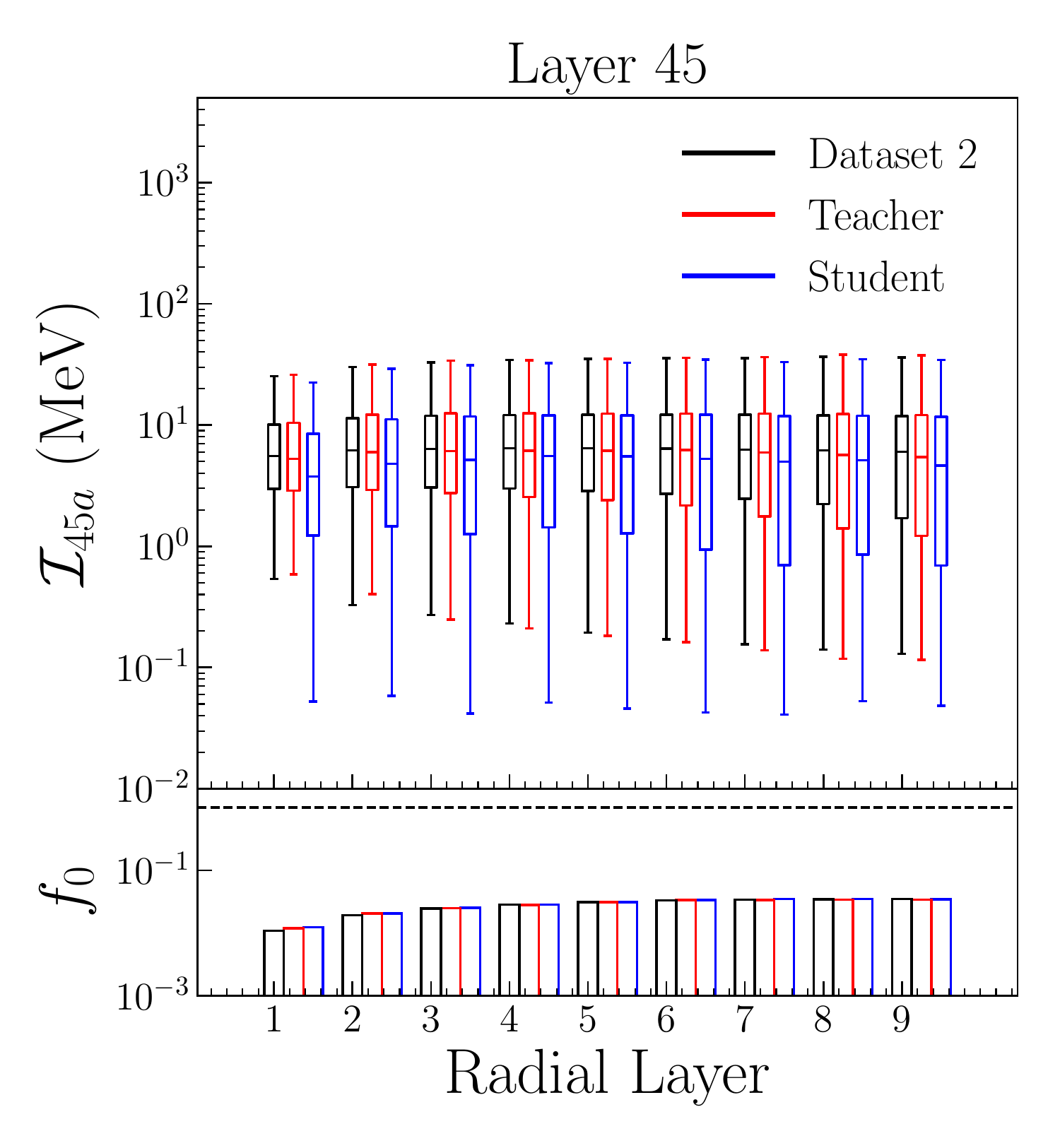}

\caption{Box and whisker plots showing the distribution of energy deposited in each ring of voxels at fixed radial distance from the beam line (9 such rings for Dataset 2) in layers 1, 10, 20, and 45 of Dataset 2. \geant{} data are shown in black, \icalo{} teacher (student) events in red (blue). Each box extends from the first quartile of energies greater than zero to the third quartile. The whiskers extend from the $5^{\rm th}$ to $95^{\rm th}$ percentile of the non-zero energy deposition. Lower subplots show $f_0$, the average fraction of voxels in each radial ring with zero energy deposition.} \label{fig:ds2_boxwhisker}
\end{figure*}

\begin{figure*}[ht]
\includegraphics[width=\columnwidth]{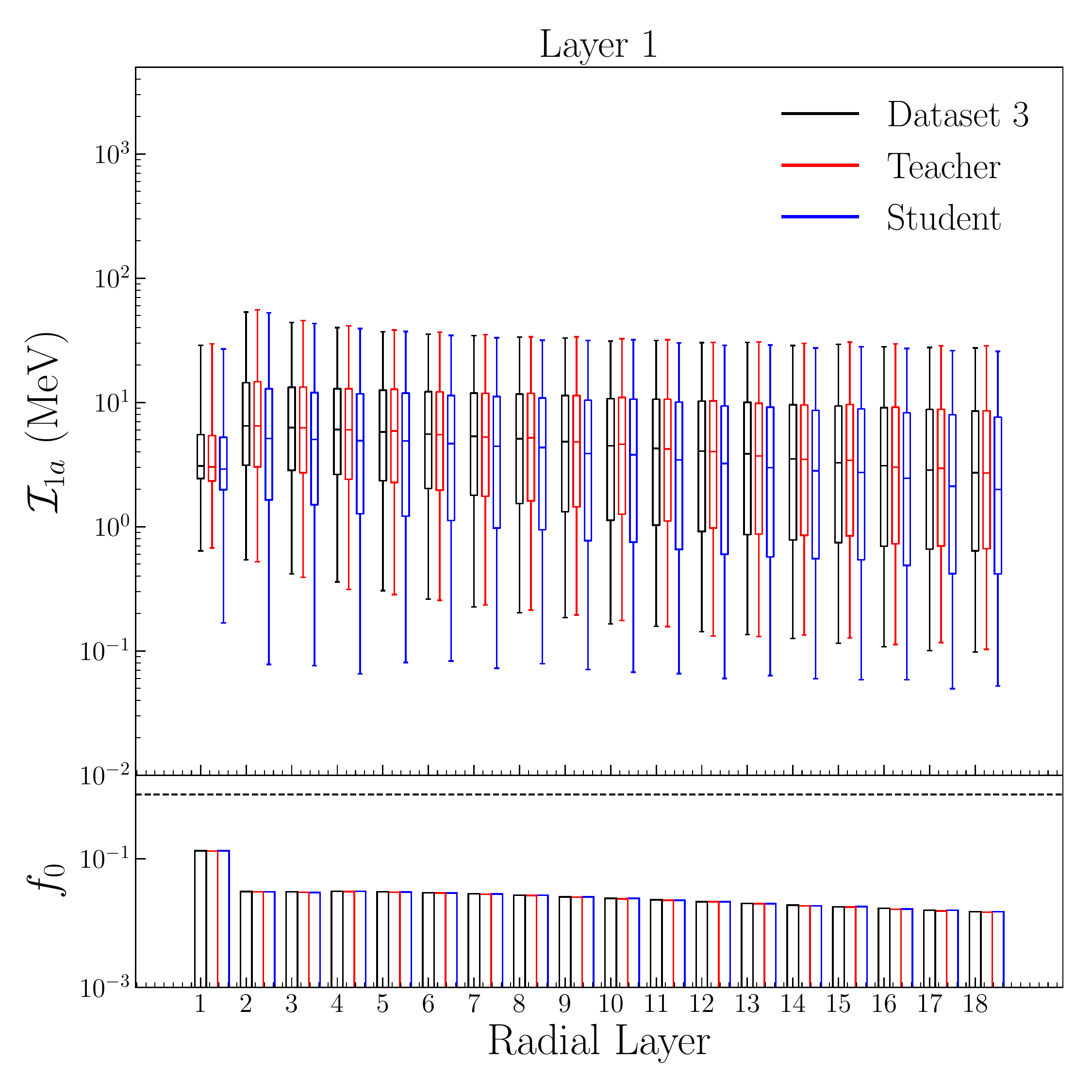}\includegraphics[width=\columnwidth]{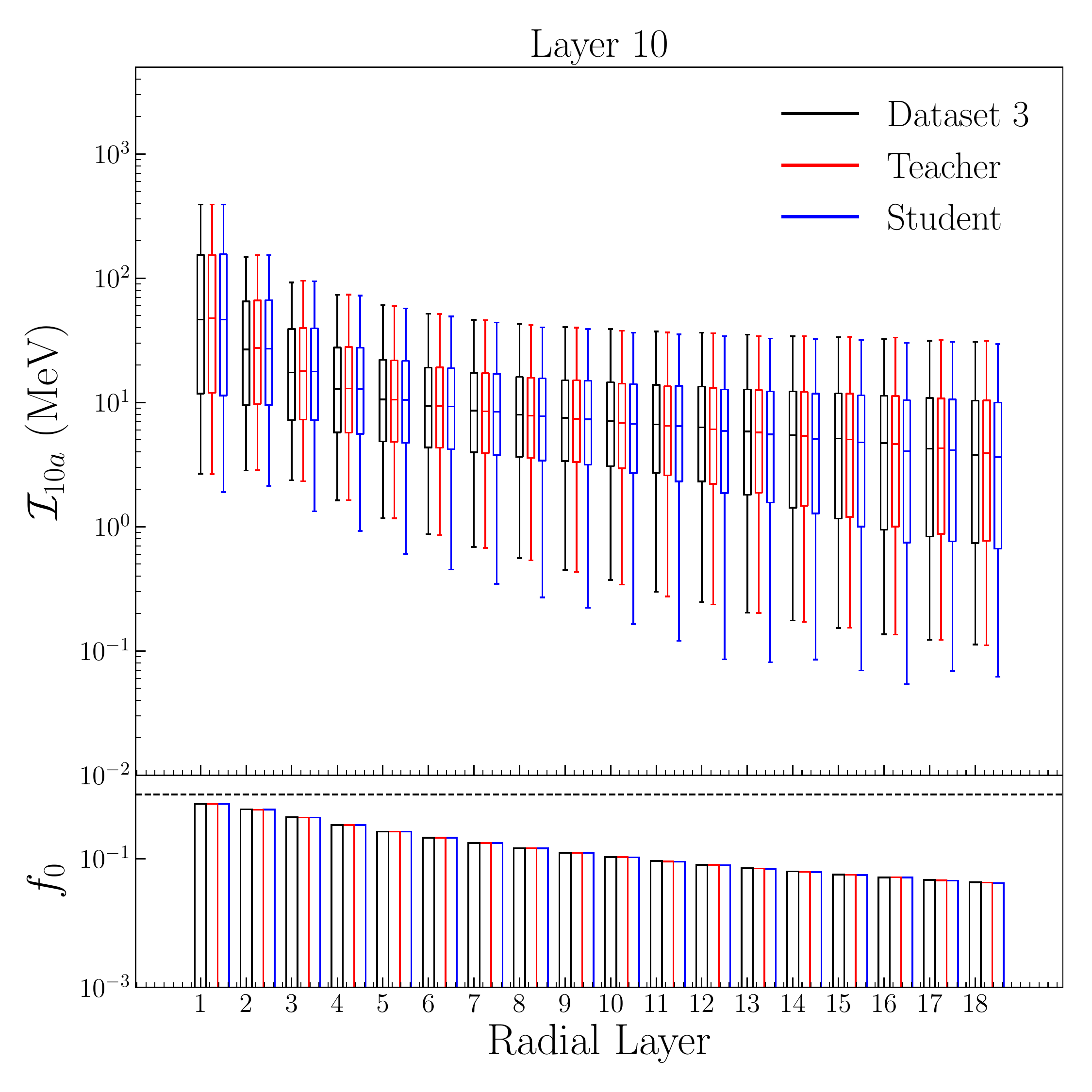}

\includegraphics[width=\columnwidth]{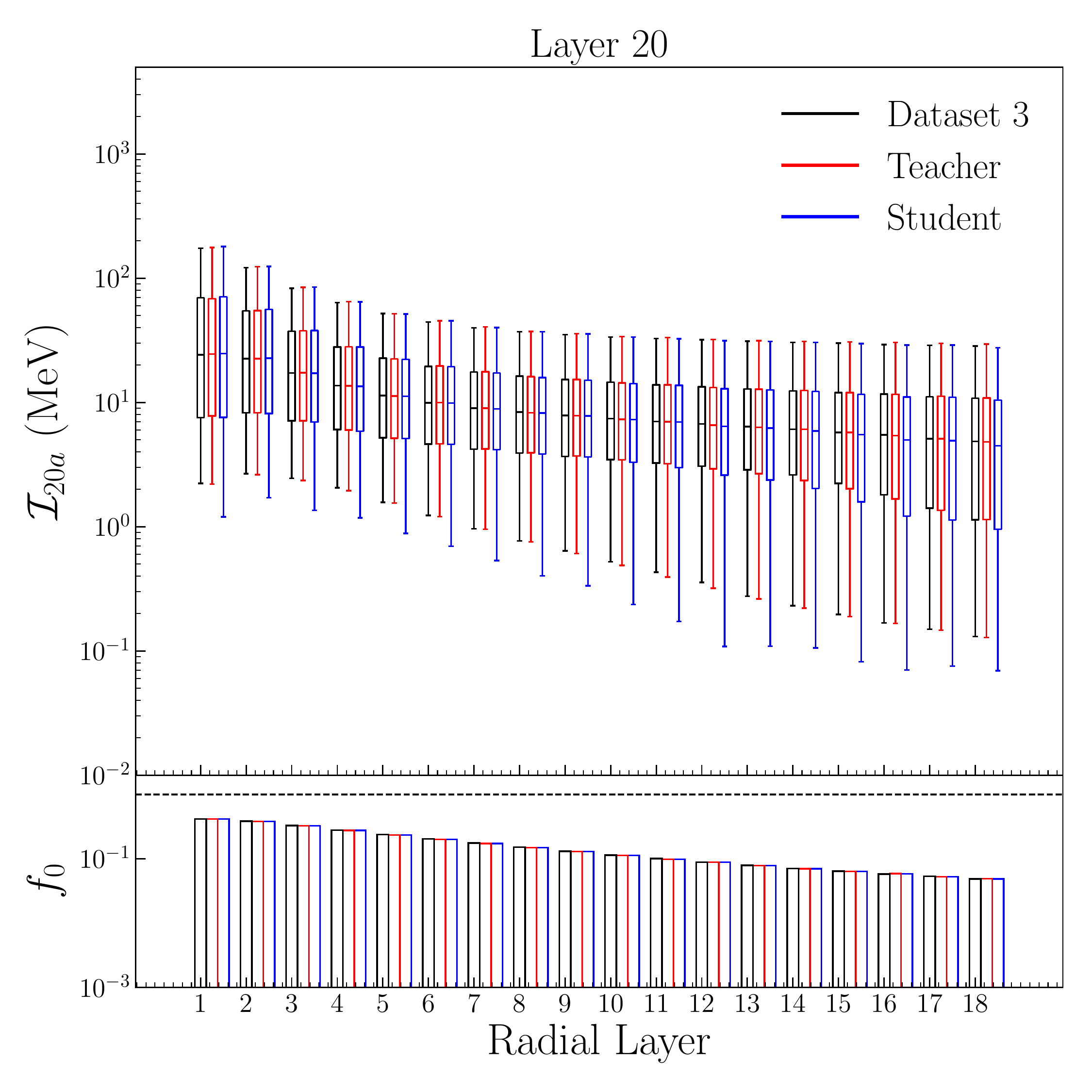}\includegraphics[width=\columnwidth]{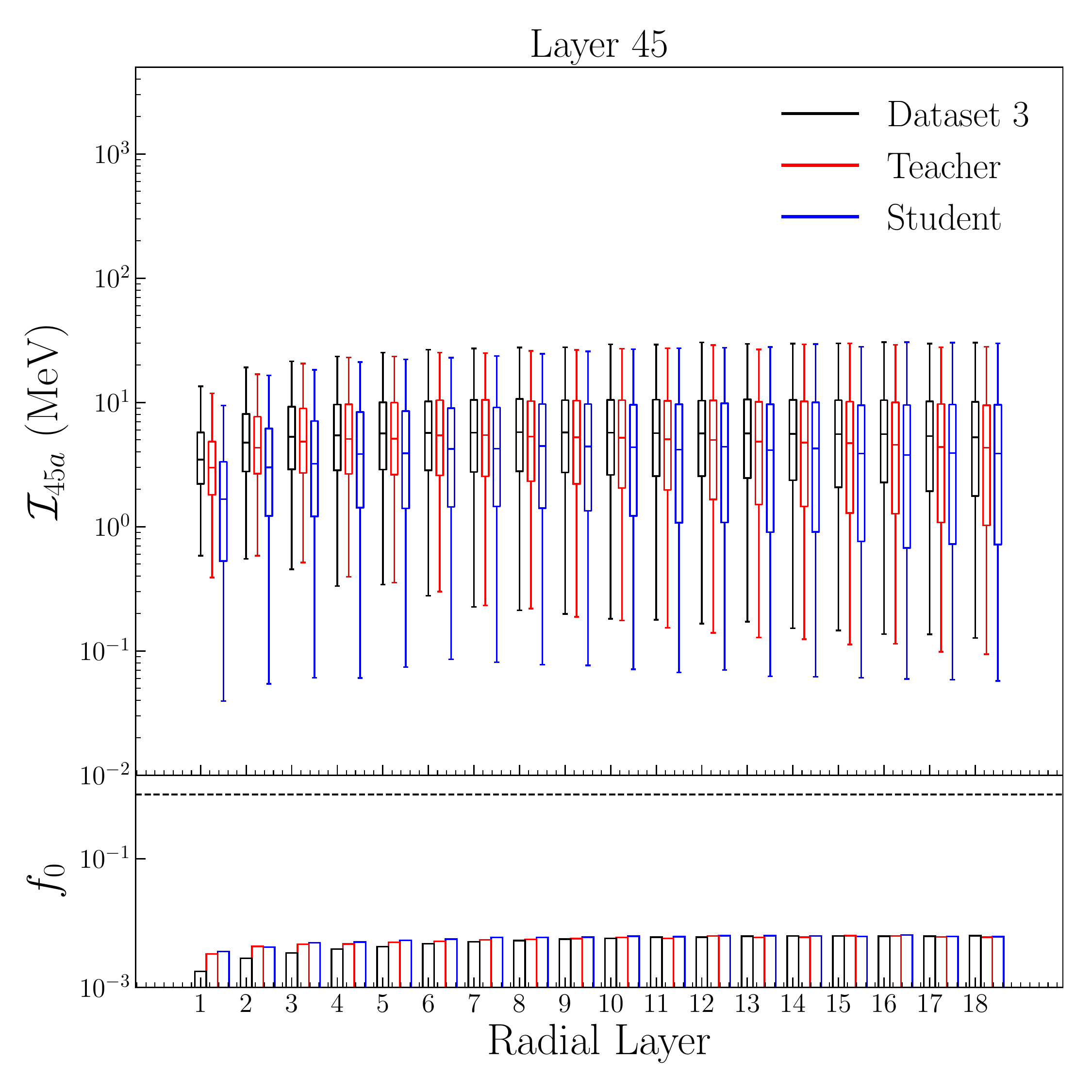}

\caption{Box and whisker plots showing the distribution of energy deposited in each ring of voxels at fixed radial distance from the beam line (18 such rings for Dataset 3) in Layers 1, 10, 20, and 45 of Dataset 3. \geant{} data are shown in blue, \icalo{} events in red. Each box extends from the first quartile of energies greater than zero to the third quartile. The whiskers extend from the $5^{\rm th}$ to $95^{\rm th}$ percentile of the non-zero energy deposition. Lower subplots show $f_0$, the average fraction of voxels in each radial ring with zero energy deposition. }\label{fig:ds3_boxwhisker}
\end{figure*}

We turn now to other aspects of the pattern of energy deposition within each layer. First, we consider the sparsity $f_0$ for each layer, defined as the fraction of voxels that have non-zero energy deposition and shown in Figure~\ref{fig:layer1_sparsity}. We note that the lowest and highest layer numbers have larger fraction of zero energy voxels. It appears to be more difficult for \icalo{} to learn distributions in these layers with small $f_0$ as evident from the deviations found in distributions discussed in this section. In Figures~\ref{fig:ds2_boxwhisker} (Dataset 2) and \ref{fig:ds3_boxwhisker} (Dataset 3), we show box-and-whisker plots of the distribution of energy deposited within all voxels in each radial ring within the layer (9 such rings in the geometry of Dataset 2, each with 16 voxels. Dataset 3 has 18 rings in increasing radius, each with 50 voxels). In the main plots of Figures~\ref{fig:ds2_boxwhisker} and \ref{fig:ds3_boxwhisker}, we show the average distribution of all voxels with non-zero energy deposition. Beneath, we show the averaged sparsity of the voxels within the ring.  These distributions have good agreement between the \geant{} and \icalo{} events. 

\begin{figure*}[ht]
\includegraphics[width=0.5\columnwidth]{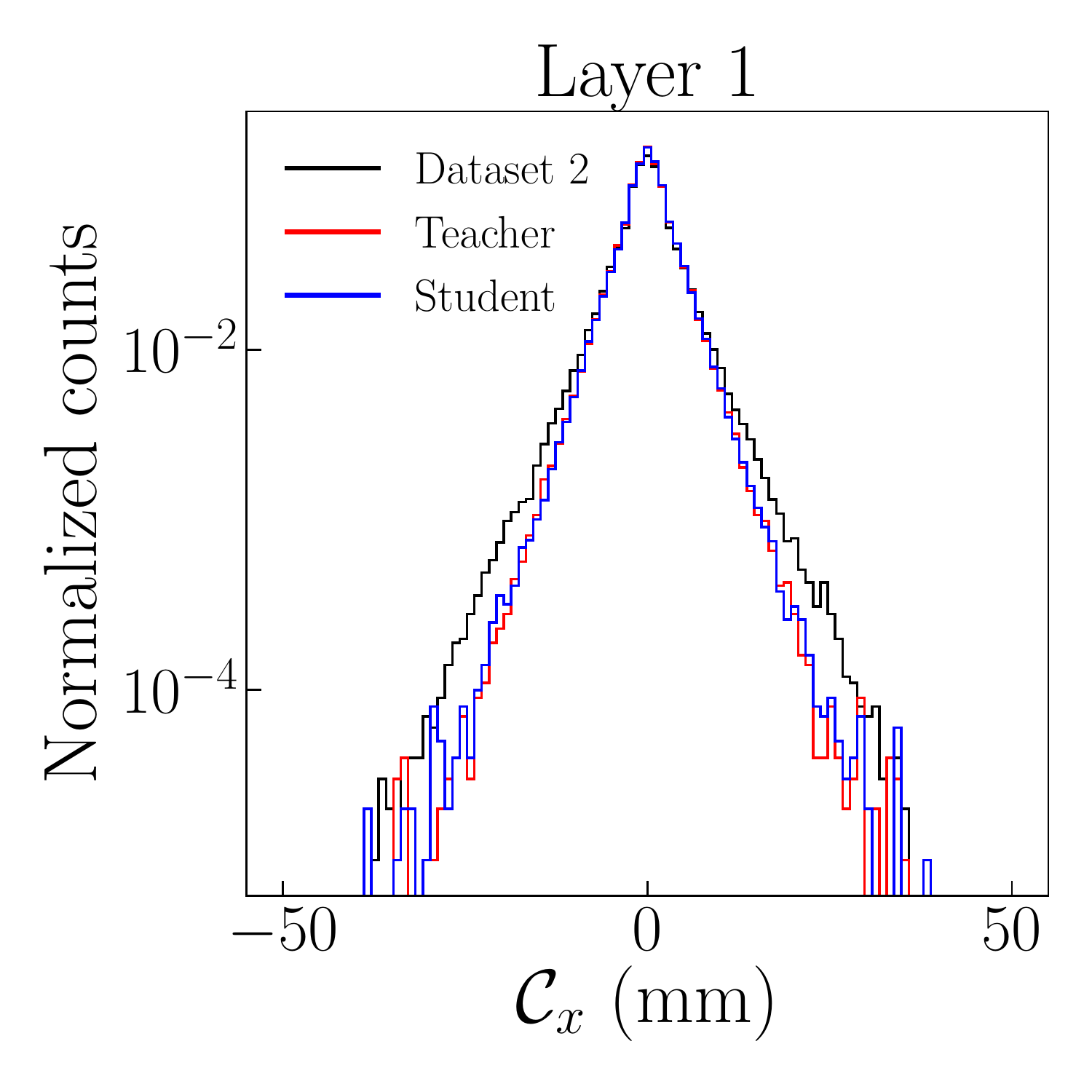}\includegraphics[width=0.5\columnwidth]{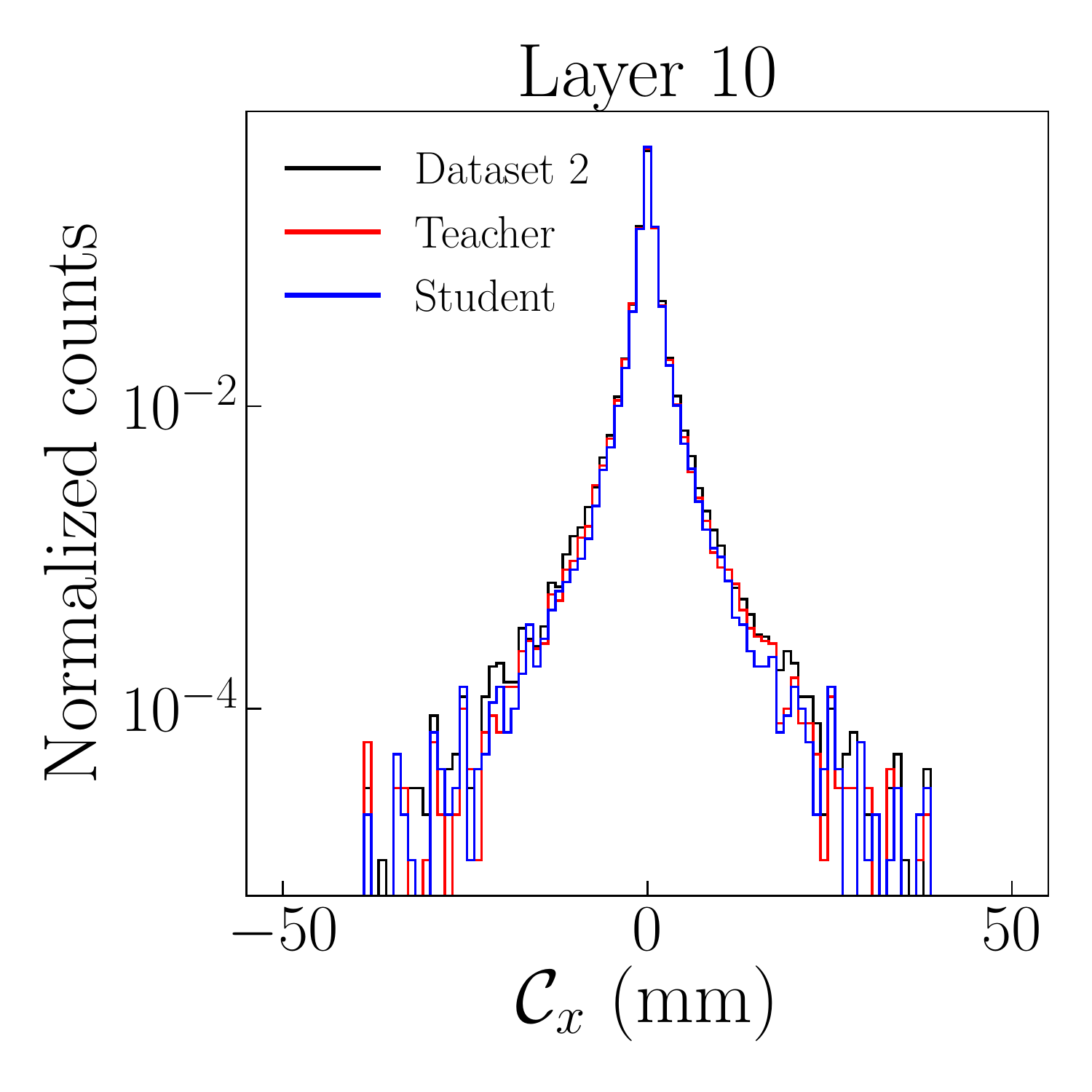}\includegraphics[width=0.5\columnwidth]{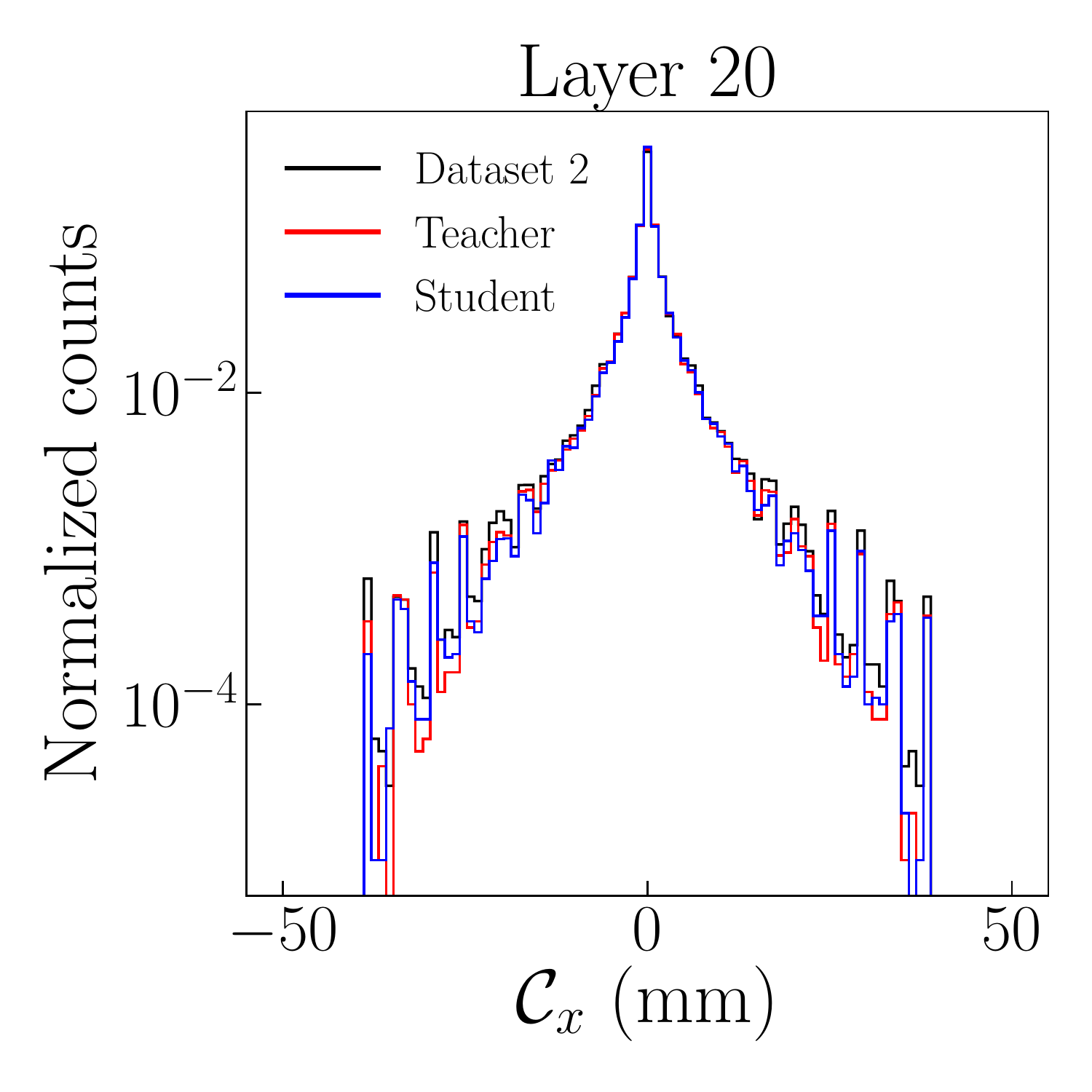}\includegraphics[width=0.5\columnwidth]{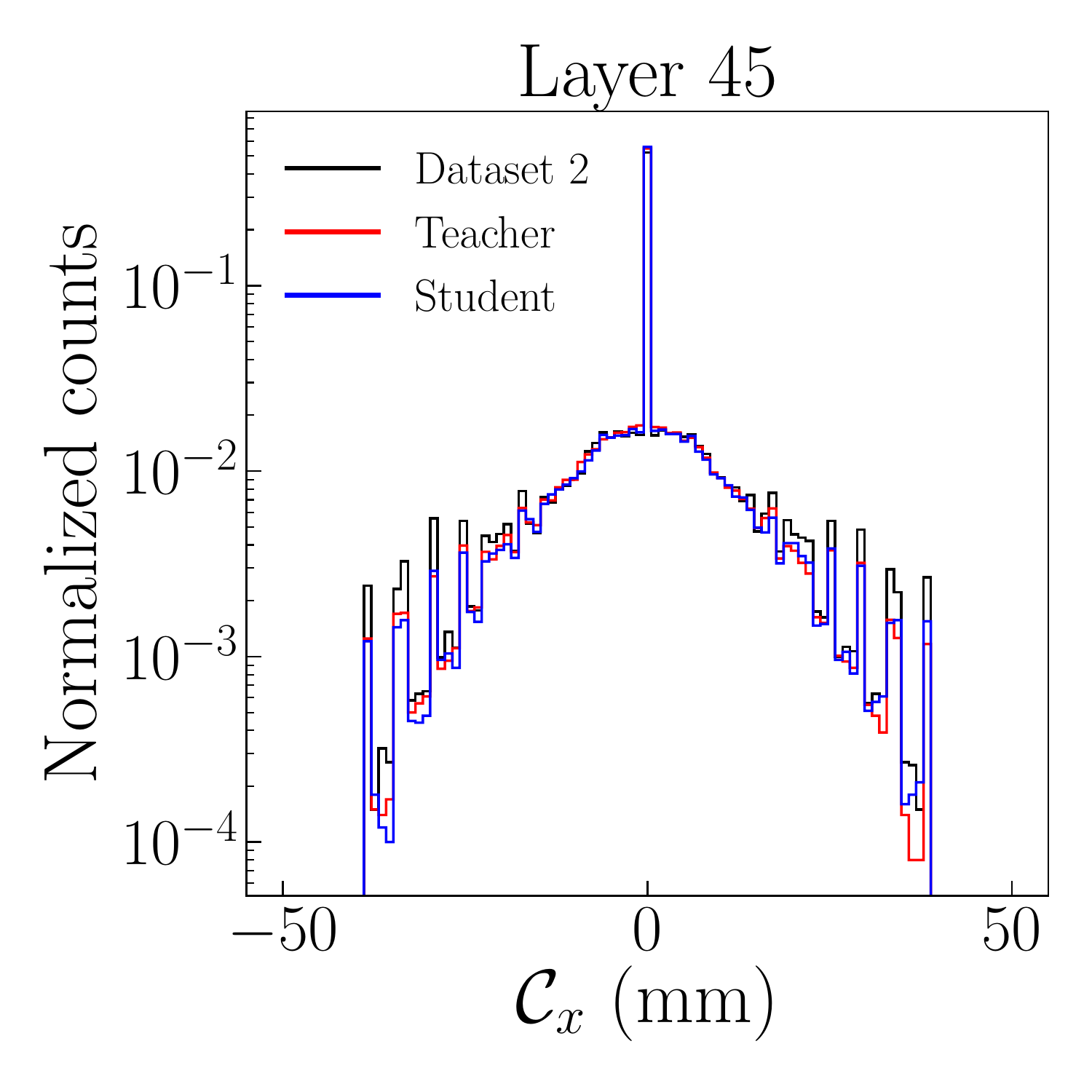}

\includegraphics[width=0.5\columnwidth]{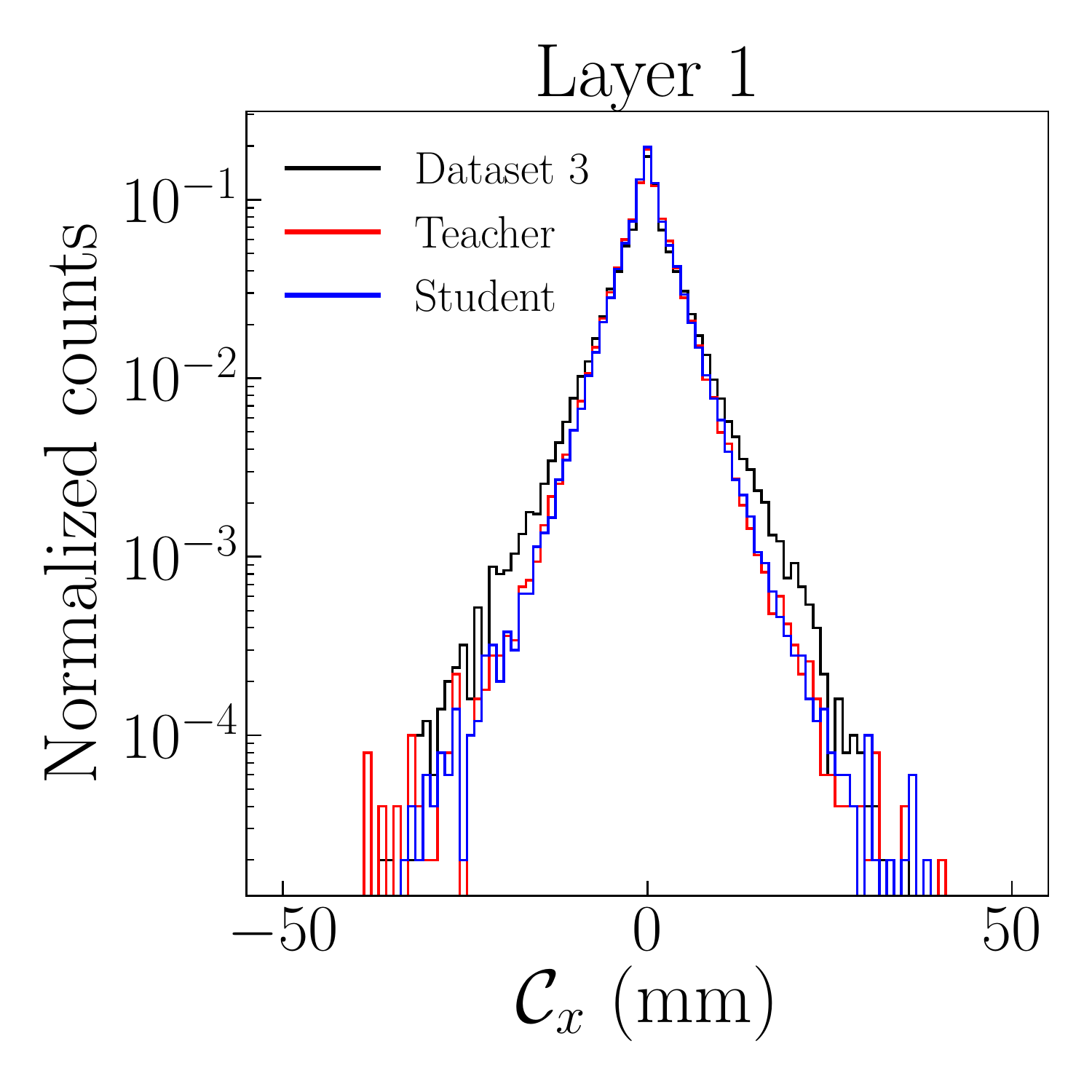}\includegraphics[width=0.5\columnwidth]{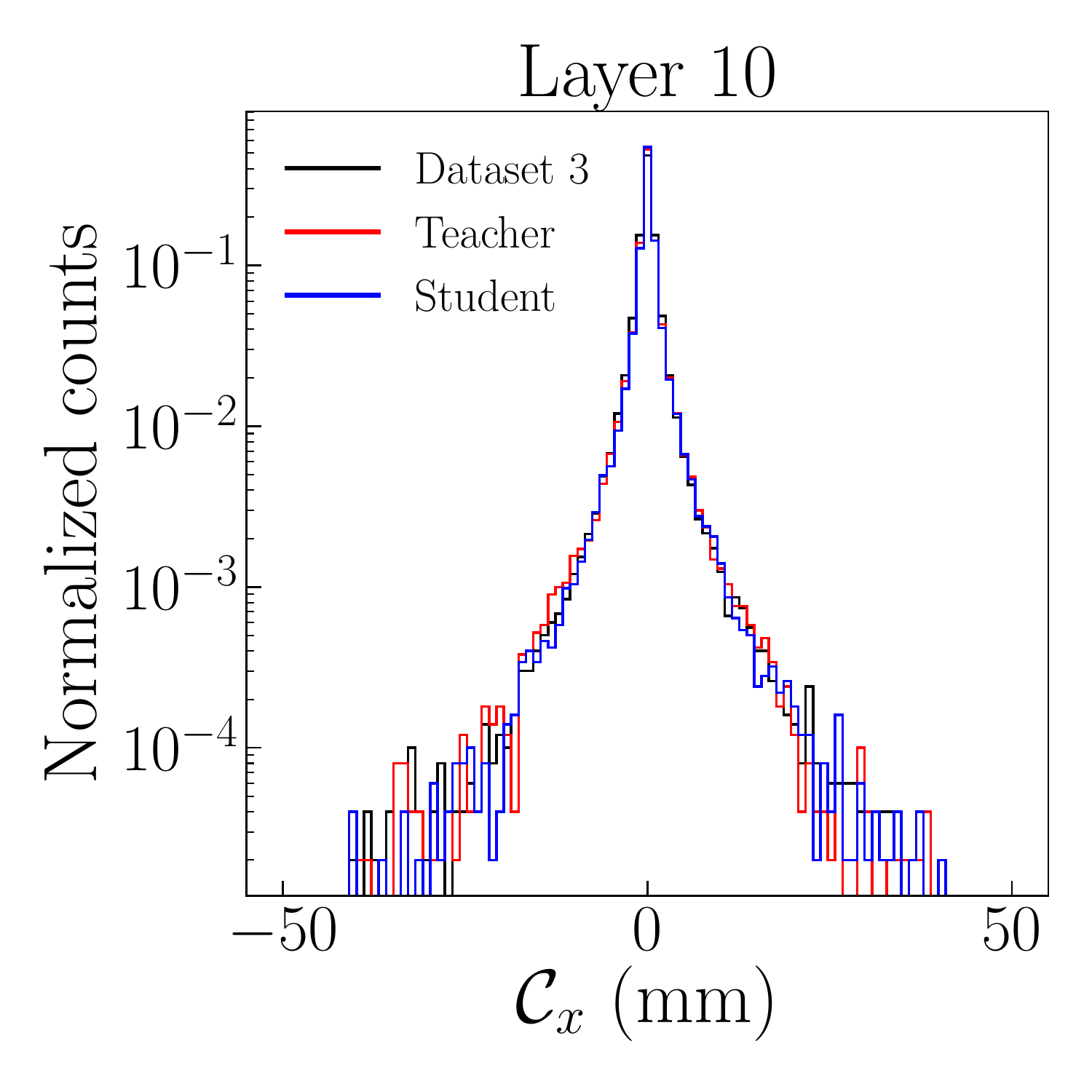}\includegraphics[width=0.5\columnwidth]{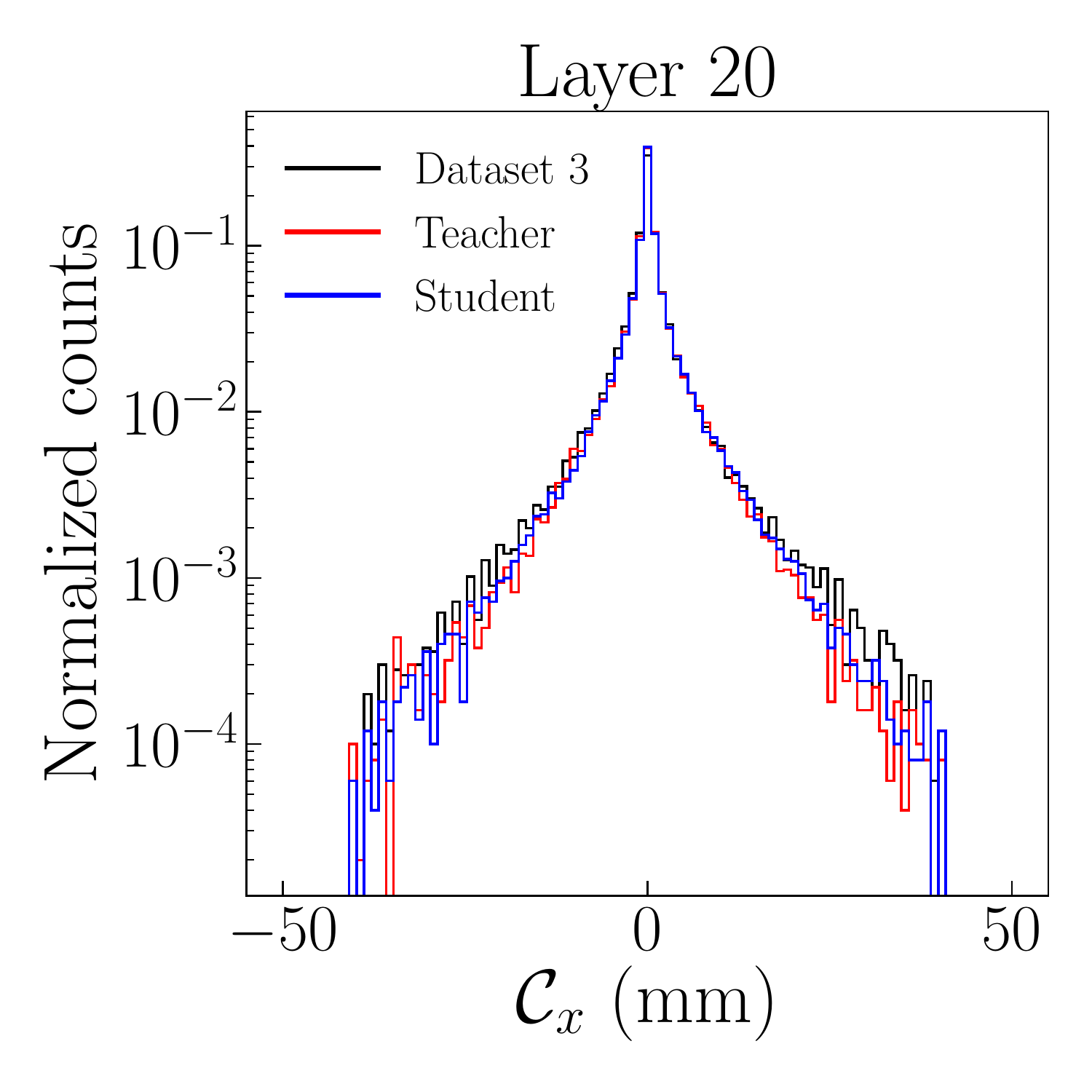}\includegraphics[width=0.5\columnwidth]{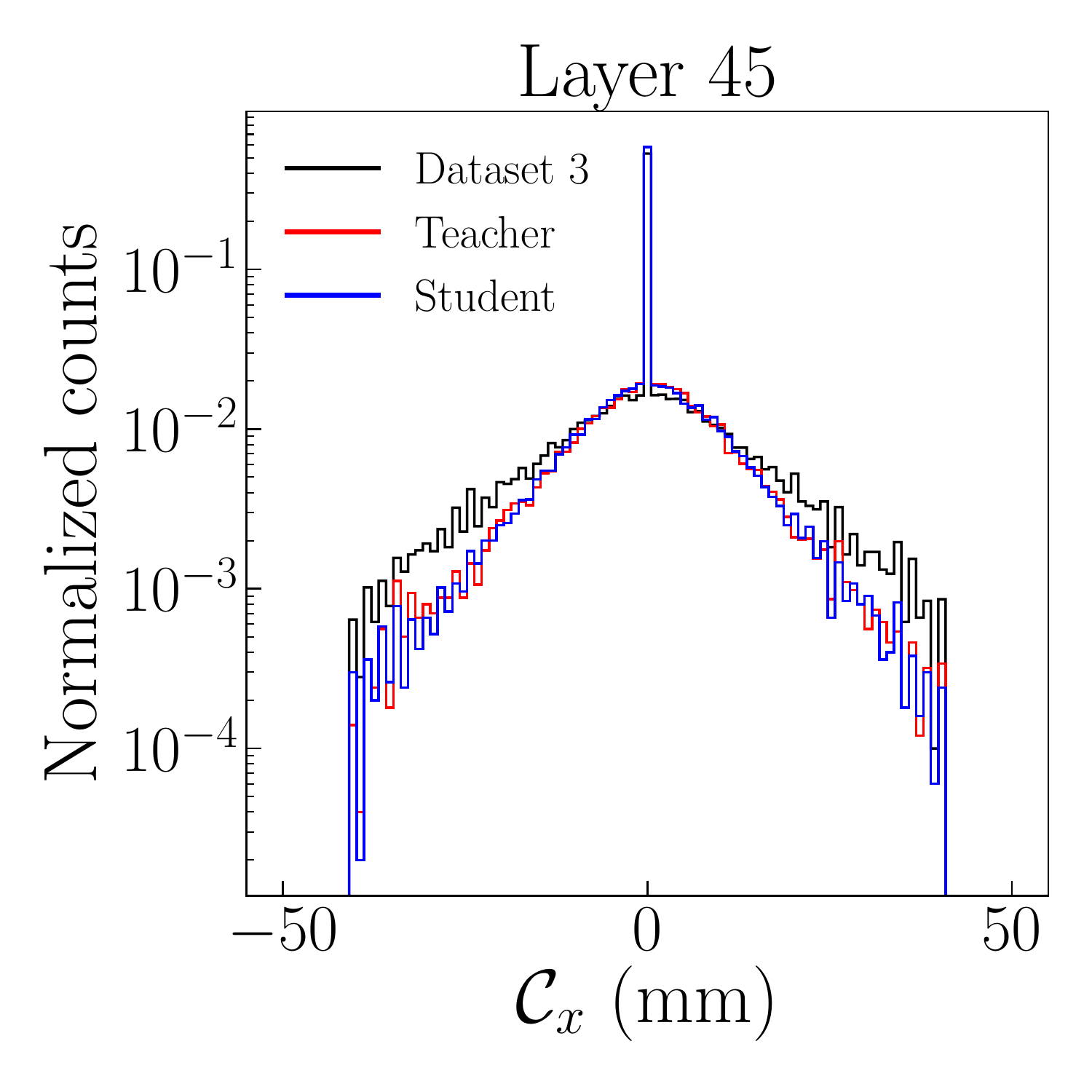}

\caption{Histograms of the centers of energy along the $x$-axis, ${\cal C}_x$, for Dataset 2 (left) and Dataset 3 (right). 
Distributions of \geant\ data are shown as black lines, and those of the \icalo\ teacher (student) trained on Dataset 2 or 3 (as appropriate) in red (blue). Due to the symmetry of the detector and incident beam, the distributions for the centers of energy in the $y$ direction (${\cal C}_y$) are statistically identical to ${\cal C}_y$. \label{fig:layer1_coe}}
\end{figure*}

As a final comparison of the pattern of energy distribution, in Figure~\ref{fig:layer1_coe} we show the distribution of the centers of energy along the $x$ axis of the detector for our representative layer. The center of energy ${\cal C}_j$ along a coordinate ($j=x$ or $y$) is defined as the sum of the energy deposited in each voxel times the voxel's coordinate distance from the origin, normalized by the total energy deposited. We show only ${\cal C}_x$, as the distribution of ${\cal C}_y$ is statistically identical due to the symmetry of the detector around the incident beam. Again, we see the largest deviations in the tails of the centers of energy for the early and late layers of Dataset 3.

While these plots capture only a limited set of diagnostic criteria for the generated events as compared to the \geant{} data, it appears that many of the distributions match well. The most glaring and important exceptions are at low deposited energies, most notably in early layers (see Figure~\ref{fig:flowone_generated} for example).

\subsection{Classifier Metrics}\label{sec:classifier}

The histograms and averaged deposition patterns of the generated events as compared to the \geant\ data suggest that \icalo\ can match many of the properties of the data. Given the complexity of the events however, these simple distributions may not capture correlations within events, thereby giving a misleading impression of the accuracy of the sampler. We wish to more quantitatively determine if the generated probability distribution $p_{\rm generated}$ is identical to that of the data $p_{\rm data}$.

To answer this question, we follow the conventions of \cite{Krause:2021ilc,Krause:2021wez}, and use a binary classifier trained to distinguish the \geant\ and generated events~\cite{2016arXiv161006545L}. (Such binary classifiers can also be used to reweight generative models to improve their fidelity, see~\cite{Diefenbacher:2020rna}.) Table~\ref{tab:cls_results} shows the results of 10 independent classifier runs on generated events from both datasets. ``Low-level features" refers to all energy depositions per voxel (normalized with $E_\text{inc}$ and multiplied by a factor 100) and $E_\text{inc}$ itself (preprocessed as $\log_{10} E_\text{inc}$). ``High-level features" are the incident energy (preprocessed as $\log_{10}{E_{\text{inc}}}$), the energy deposited in each of the layers (preprocessed as $\log_{10}{(E_{i}+10^{-8})}$), the center of energy in the $x$ and $y$ directions (normalized with a factor 100), and the width of the $x$ and $y$ distributions (normalized with a factor 100). More details on the architecture and trainings procedure can be found in Appendix~\ref{app:classifier}. 

In Table~\ref{tab:cls_results}, the results of the classifier runs are presented as AUC and JSD scores.  According to the Neyman-Pearson lemma, we expect the AUC to be 0.5 if the true and generated probability densities are equal. The AUC is 1 if the classifier is able to perfectly distinguish between generated and true samples. The second metric, JSD $\in [0, 1]$, is the Jensen-Shannon divergence which also measures the similarity between the two probability distributions. The JSD is 0 if the two distributions are identical and 1 if they are disjoint. 

In our study, we find that \icalo{} is able to produce generated samples of sufficiently high fidelity for Dataset 2 to fool the classifier. This is evident by the Dataset 2 teacher and student classifier scores being clearly less than unity in Table~\ref{tab:cls_results}. 

In contrast, the AUC scores for our Dataset 3 teacher and student models are all $>0.9$, which indicates lower fidelity. Also, the high-level scores are greater than the low-level scores for our Dataset 3 models. This is likely due to the low-level classifier having insufficient capacity to separate the \icalo{} and \geant{} Dataset 3 samples. It would be interesting to explore more sophisticated architectures to improve the classification performance of the low-level classifier, but we save this for future work. 

In general, the teacher models performed better than the student models at the classifier tests. This is to be expected since the students require a second distillation step, and it is not surprising that some fidelity is lost in the process. 

Finally, we note that the authors of \textsc{CaloScore} \cite{Mikuni:2022xry} have produced results for Datasets 2 and 3 using a score-based diffusion model. They performed a similar classifier test on their samples and obtained an AUC of 0.98 for both datasets. 

\renewcommand{\arraystretch}{1.5}
\begin{table}[!ht]
\begin{center}
\begin{tabular}{|c||c|c|c|c|}
\hline
& \multicolumn{2}{|c|}{low-level features} &  \multicolumn{2}{c|}{high-level features}\\
& AUC & JSD & AUC & JSD \\
\hline \hline
DS2 teacher & \; 0.797(5) \; & \; 0.210(7) \; & \; 0.798(3) \; & \; 0.214(5) \; \\
\hline
DS2 student & \; 0.840(3) \; & \; 0.286(5) \; & \; 0.838(2) \; & \; 0.283(4) \; \\
\hline
DS3 teacher & 0.911(3) & 0.465(6) & 0.941(1) & 0.561(3) \\
\hline
DS3 student & 0.910(8) & 0.462(18) & 0.951(1) & 0.601(5) \\
\hline
\end{tabular}
\caption{Mean and standard deviation of 10 independent classifier runs.}
\label{tab:cls_results}
\end{center}
\end{table}
\renewcommand{\arraystretch}{1}

\subsection{Timing}

\begin{table}
\centering
\resizebox{\columnwidth}{!}{%
\begin{tabular}{|c|c|c|c|}
\hline
\textbf{Model} & \textbf{Batch size} & \textbf{GPU times (ms)} & \textbf{CPU times (ms)} \\
\hline
\hline
\multirow{5}{*}{DS2 teacher} & 1 & $7.12 \times 10^4$ & $8.84 \times 10^4$ \\
\cline{2-4}& 10 & $7.10 \times 10^3$ & $1.21 \times 10^4$ \\
\cline{2-4}& 100 & $7.75 \times 10^2$ & $3.47 \times 10^3$ \\
\cline{2-4}& 1000 & $1.65 \times 10^2$ & - \\
\cline{2-4}& 10000 & $1.21 \times 10^2$ & - \\
\hline
\hline
\multirow{5}{*}{DS2 student} & 1 & $1.06 \times 10^3$ & $4.08 \times 10^3$ \\
\cline{2-4}& 10 & $2.79 \times 10^2$ & $4.08 \times 10^2$ \\
\cline{2-4}& 100 & $2.99 \times 10^1$ & $4.93 \times 10^1$ \\
\cline{2-4}& 1000 & $2.18$ & - \\
\cline{2-4}& 10000 & $\mathbf{1.34}$ & - \\
\hline
\hline
\multirow{5}{*}{DS3 teacher} & 1 & $4.76 \times 10^5$ & $3.86 \times 10^6$ \\
\cline{2-4}& 10 & $5.28 \times 10^4$ & $3.31 \times 10^5$ \\
\cline{2-4}& 100 & $9.17 \times 10^3$ & - \\
\cline{2-4}& 1000 & $5.17 \times 10^3$ & - \\
\cline{2-4}& 5000 & $4.57 \times 10^3$ & - \\
\hline
\hline
\multirow{5}{*}{DS3 student} & 1 & $1.29 \times 10^3$ & $3.36 \times 10^3$ \\
\cline{2-4}& 10 & $1.33 \times 10^2$ & $6.22 \times 10^2$ \\
\cline{2-4}& 100 & $3.58 \times 10^1$ & - \\
\cline{2-4}& 1000 & $8.84$ & - \\
\cline{2-4}& 5000 & $\mathbf{5.87}$ & - \\
\hline
\end{tabular}}
\caption{Average time taken to generate a single shower event by \icalo{} teacher and student models for Datasets 2 and 3. The timing was computed for different generation batch sizes on our Intel Core i9-7900X CPU and our TITAN V GPU. We were not able to generate shower events on the CPU for large batch sizes due to memory constraints.}
\label{tab:timing}
\end{table}

Shower generation timings for different generation batch sizes are shown in Table~\ref{tab:timing}. 
The speedups going from teacher to student are the expected ${\mathcal O}(d)$, where $d$ is the dimensionality of a single layer (144 for DS2 and 900 for DS3). On the GPU, \icalo{} is generally faster than \geant{}, which takes approximately 100~s per event, except for Dataset 3 teacher with a generation batch size of 1. On the CPU, \icalo{} is still faster than \geant{} for both Dataset 2 models and also Dataset 3 student. With the student models, we were able to generate a single shower on the GPU as quickly as 1.34 ms (5.87 ms) for Dataset 2 (3) --- $\mathcal{O}(10^5)$ faster than \geant{}.

\section{Conclusions} \label{sec:conclusions}

In this paper, we have introduced a new approach to generating highly-granular calorimeter showers with normalizing flows. This is the first flow-based approach that has been applied to calorimeter shower data with dimensionality as large as Dataset 3 of {\it CaloChallenge2022}. Instead of learning the joint distribution of all voxels, we focus on the distribution per calorimeter layer and use an inductive algorithm to generate the sample. This approach is inspired by the physical nature of the shower, which starts to develop close to the entry point of the incoming particle and evolves deeper into the detector. 

Previous approaches used separately trained flows for each layer~\cite{Diefenbacher:2023vsw}, while we use a single flow for the step of generating layer $i$ based on the shower in layer $i-1$ (treating $i=1$ separately), in the spirit of a mathematical proof by induction. The advantage of this approach is in its much more efficient setup. Instead of 45+1 normalizing flows, we need only three. This efficiency (along with the inductive approach) would be preserved even if we increased the size of \flowthree{} to be much larger than \flowtwo{} and the flows of \cite{Diefenbacher:2023vsw}.
However, the downside of our inductive approach is that training \flowthree{} cannot be trivially parallelized across multiple GPUs (as the equivalents were in \cite{Diefenbacher:2023vsw}). Also, sharing the same flow across all layers is presumably less expressive than using one flow for each layer.

In both \LtoLF\ and \icalo, event generation must occur sequentially through each layer. On top of this, the slow sampling from the MAF strongly motivates the added step of distilling them into student IAF networks. Compared to the teachers, the students speed up event production by a factor of approximately the number of voxels in a layer --- though it is still necessary to iterate through all the layers. \icalo{} is the first work to include a teacher-student pairing in any \cf{}-like setup for higher dimensional calorimeter shower data such as Datasets 2 and 3. 

At this time, no exhaustive comparison with other generative neural networks is possible. The most natural comparison point, the algorithm of \cite{Diefenbacher:2023vsw}, is applied to a different data set. During the completion of {\it CaloChallenge2022}, \icalo{} will be compared against other codes, and a more complete understanding of its strengths and weaknesses will be available. 

A possible limitation to this inductive algorithm is generalizing it to non-regular calorimeter geometries. In \icalo{}, \flowthree{} requires each layer to be identical in structure, allowing the efficient weight-sharing and training that are the major advantages of the algorithm. Significant modifications would be required to adapt to layer-dependent voxelization. While this might not be possible for arbitrary geometries, some variations (for example, breaking a large voxel in one layer into groups of small voxels in other layers) could admit an inductive flow approach. 

Even with these caveats, the efficiency, event generation speed (when using the IAF students), and relatively high fidelity of \icalo{} make it a competitive architecture for fast calorimeter event generation. One promising future direction is to realize a similar inductive setup based on coupling-layer flows, which are equally fast in density estimation and generation. This would presumably reduce training time as we would not have to train both teacher and student flows.

\section*{Acknowledgements}
We thank Gopolang Mohlabeng for discussion in the early stages of this project. This work was supported by DOE grant DOE-SC0010008. CK would like to thank the Baden-W\"urttemberg-Stiftung for support through the program \textit{Internationale Spitzenforschung}, project \textsl{Uncertainties --- Teaching AI its Limits} (BWST\_IF2020-010).  In this work, we used the {\tt NumPy 1.16.4} \cite{harris2020array}, {\tt Matplotlib 3.1.0} \cite{4160265}, {\tt scikit-learn 0.21.2} \cite{scikit-learn}, {\tt h5py 2.9.0} \cite{hdf5}, {\tt pytorch 1.11.1} \cite{NEURIPS2019_9015}, and {\tt nflows 0.14} \cite{nflows} software packages. 

\appendix 

\section{Pre- and Postprocessing}\label{sec:preprocessing}

The incident energy is a conditional for all three flows. In all cases, it is transformed from the physical range (1~GeV to 1~TeV) to normalized log-space:
\begin{equation}
    E_{\rm inc} \to \log_{10} \frac{ E_{\rm inc}}{10^{4.5}~{\rm MeV}} \; \in [-1.5, 1.5].
    \label{eq:Einc.pre}
\end{equation}

For \flowone{}, the total energy deposited in layer $i$ ($E_i$) is obtained from summing all voxels of the layer. As the flow has difficulty learning a distribution where many elements are exactly zero, uniform noise in the range $[0,5]$~keV is added to $E_i$~\cite{Krause:2021ilc}. We then divided by $65$ GeV (this is slightly larger than the maximum energy deposition in any layer in any of the events of the dataset):
\begin{equation}
    E_i \to  x_i \equiv (E_i + {\rm rand}[0,5~{\rm keV}])/65~\text{GeV}.    \label{eq:layer_norm}
\end{equation} 
As a final step, we apply a logit transformation:
\begin{equation}
    y_i = \log \frac{u_i}{1-u_i},\quad u_i\equiv \alpha+(1-2\alpha)x_i, \label{eq:logit}
\end{equation}
where the offset $\alpha \equiv 10^{-6}$ ensures that the boundaries $x_i = 0$ and $1$ map to finite numbers.

For both \flowtwo{} and \flowthree{}, we apply the same processing steps for the energy depositions per voxel, $\mathcal{I}_{ia}$. First, we add uniform noise in the range $[0,5]$~keV to every voxel. Then, we normalize the voxel by the sum of all voxels of the given layer and then apply the logit transformation of Eq.~\eqref{eq:logit} to it:
\begin{eqnarray}
\mathcal{I}_{ia} & \to & \mathcal{I}_{ia}+{\rm rand}[0,5~{\rm keV}]\nonumber \\
\mathcal{I}_{ia} & \to & \mathcal{I}_{ia}/\sum_b \mathcal{I}_{ib} \equiv \hat{\mathcal{I}}_{ia} \nonumber \\
u_{ia} & = & \alpha+(1-2\alpha)\hat{\mathcal{I}}_{ia} \nonumber \\
y_{ia} & = & \log \frac{u_{ia}}{1-u_{ia}}.
\label{eq:I.pre}
\end{eqnarray}
When generating new events from our trained flows, the preprocessing steps are inverted, and normalized showers are transformed back to physical space using the proxy output of \flowone. Voxel energies below the detector threshold of 15~keV are set to zero.

\flowtwo{} is conditioned on the energy deposited in the first layer, $E_1$. This is preprocessed as in Eqs.~\eqref{eq:layer_norm} and~\eqref{eq:logit}, with the additional step of dividing $y_1$ by 4:
\begin{eqnarray}
    E_1 & \to &  x_i \equiv (E_i + {\rm rand}[0,5~{\rm keV}])/65~\text{GeV} \nonumber \\
    u_1 & \equiv & \alpha+(1-2\alpha)x_1 \nonumber \\
    y_1 & = &  \frac{1}{4}\log \frac{u_1}{1-u_1}.  \label{eq:flow2_proprocess}
\end{eqnarray}
This final division by 4 is to bring the range of $y_1$ down to ${\cal O}(1)$.

The conditional inputs for \flowthree{} are preprocessed as follows: $\mathcal{I}_{(i-1)a}$ follows the steps of Eq.~\eqref{eq:I.pre}. $E_i$ and $E_{(i-1)}$ follow Eqs.~\eqref{eq:layer_norm} and~\eqref{eq:logit} (without the factor 4 of \flowtwo). The layer number $i$ is one-hot encoded in a vector of length 44.

When sampling new events from \icalo{}, these transformations are inverted. As a final step, the detector energy threshold is applied on the generated voxel energies by setting all voxels with ${\cal I}_{ia}<15\,{\rm keV}$ to zero. As noted previously, when generating new patterns of energy deposition from \flowtwo{} and \flowthree{}, the generated $\hat{\cal I}_{ia}$  are not enforced to sum to one. 
Forcing them to sum to one, as was the choice made in the previous iterations of \cf{}~\cite{Krause:2021ilc,Krause:2021wez,Krause:2022jna}, here causes problems in the voxel energy histograms in Figures \ref{fig:layer1_energies} and \ref{fig:voxel_energies}. It enhances low-energy voxels above the cut-off threshold, leading to a large excess at the lower end of the voxel energy histograms. This was less of a problem previously, for Dataset 1 and for the CaloGAN dataset, because the normalization was learned better, perhaps because these datasets were lower dimensional. Meanwhile, in \LtoLF{}~\cite{Diefenbacher:2023vsw}, which was trained on a dataset comparable in dimensionality to Dataset 2, the voxel energies were normalized by maximum voxel energy in the dataset, so no renormalization with $E_i$ was needed and this issue was avoided. However, we found that normalizing by maximum voxel energy did not improve our results here.

\section{Classifier} \label{app:classifier}

The performance metric based on the classifier test uses the neural network architecture that was provided by the {\it CaloChallenge2022}~\cite{calochallenge, calochallenge_github} evaluation script. It is based on the classifiers that were used in~\cite{Krause:2021ilc,Krause:2021wez} and was also used (with slightly different hyperparameters) in~\cite{Krause:2022jna}. To be precise, the classifier is a deep, fully-connected neural network with an input and two hidden layers with 2048 nodes each. All activation functions are leaky ReLUs, with default negative slope of $0.01$, except the output layer, which uses a sigmoid activation function for its single output number.  We do not use any regulators such as batch normalization or dropout. 

The input for low- and high-level feature classification is preprocessed as described in the main text. For Dataset 2, we work with the second provided file of 100k showers~\cite{CaloChallenge_ds2}. We split it in train/test/validation sets of the ratio (60:20:20). For Dataset 3, we use the third file provided at~\cite{CaloChallenge_ds3} split in the ratio (40:10) for training and model selection and the full fourth file of~\cite{CaloChallenge_ds3} to obtain the final scores.

All networks are optimized by training 50 epochs with an \textsc{Adam}~\cite{kingma2014adam} optimizer with initial learning rate of $2\cdot 10^{-4}$ and a batch size of 1000 (250) for Datasets 2 (3), minimizing the binary cross entropy. We use the model state with the highest accuracy on the validation set for the final evaluation and we subsequently calibrate the classifier using isotonic regression~\cite{2017arXiv170604599G} of {\tt sklearn}~\cite{scikit-learn} based on the validation dataset before evaluating the test set.

\bibliographystyle{JHEP}
\bibliography{literature}
\end{document}